\documentclass[12pt,preprint]{aastex}

\def\kms{\,\,\,{\rm km\,s^{-1}}}                      
\def\msol{\,{\rm M}_\odot}              
\def\mpc2{\,{\rm M}_\odot\,{\rm pc}^{-2} }       
\def\gcc{\,\,\,{\rm g\,cm}^{-3}}      
\def\c3{\,{\rm cm}^{-3}} 
\def\cc2{\,{\rm cm}^{-2}}  

\def \nbar{\bar n}

\def \ng{{\bar {n_g}}}

\def\ga{\,\hbox{\hbox{$ > $}\kern -0.8em \lower 1.0ex\hbox{$\sim$}}\,}
\def\la{\,\hbox{\hbox{$ < $}\kern -0.8em \lower 1.0ex\hbox{$\sim$}}\,}

\begin{document}

\title{Analytical theory for the initial mass function: III \\
time dependence and star formation rate}

\author
{Patrick Hennebelle }
\affil{Laboratoire AIM, 
Paris-Saclay, CEA/IRFU/SAp - CNRS - Universit\'e Paris Diderot, 91191, 
Gif-sur-Yvette Cedex, France \\
Laboratoire de radioastronomie, UMR CNRS 8112,
 \'Ecole normale sup\'erieure et Observatoire de Paris,
24 rue Lhomond, 75231 Paris cedex 05, France }

\and

\author{Gilles Chabrier$^{1}$}
\affil{\'Ecole normale sup\'erieure de Lyon,
CRAL, UMR CNRS 5574,69364 Lyon Cedex 07,  France\\
School of Physics, University of Exeter, Exeter, UK EX4 4QL}

\altaffiltext{1}{Visiting scientist, Max Planck Institute for Astrophysics, Garching, Germany}

\date{}


\begin{abstract}
The present paper extends our previous theory of the stellar initial mass function (IMF) by including  the time-dependence,
 and by including the impact of magnetic field.
 The predicted 
mass spectra are similar to the time independent ones
with slightly shallower slopes at large
masses and peak locations shifted toward smaller masses by a factor of a few. Assuming 
that star-forming clumps follow Larson type relations, we obtain core mass 
functions in  good agreement with the observationally derived IMF, in particular when
taking into account the thermodynamics of the gas.
The time-dependent theory
directly yields an analytical expression for the star 
formation rate (SFR) at cloud scales. The SFR values agree  well with the observational  determinations of various Galactic molecular clouds.  Furthermore, we show that the SFR does not simply depend linearly on density, as sometimes claimed in the literature, but depends also strongly on the clump mass/size, which yields the observed scatter.
We stress, however, that {\it any} SFR theory depends, explicitly or implicitly, on very uncertain assumptions  like clump boundaries or the mass of the most massive stars that can form 
in a given clump, making the final determinations uncertain by a factor of a few.   
Finally, we derive a fully time-dependent model for the 
IMF by considering a clump, or a distribution of clumps accreting at a constant rate and thus whose
  physical properties evolve with time. In spite of its simplicity, this model 
 reproduces reasonably well various features
observed in numerical simulations of converging flows. Based on this general theory, we present a paradigm for star formation and the IMF.
\end{abstract}


\section{Introduction}

Understanding the origin of the initial mass function (IMF) and inferring 
the star formation rate (SFR) in galaxies are the two main challenges
  of star formation theory. Many attempts
have been made along the years
to try to resolve these two fundamental issues
 (see Hennebelle \& Chabrier 2010
for a brief review regarding the main modern theories of the IMF).
 Within the past three years, we have developed a new analytical
theory, based on the  gravo-turbulent paradigm of star formation (e.g. MacLow \& Klessen 2004),
aimed at explaining the IMF (Hennebelle \& Chabrier 2008, 2009, hereafter papers I and II).
Indeed, in our theory of the IMF, as well as in the Padoan \& Nordlund (2002) one, large-scale supersonic turbulence
is supposed to generate small-scale overdense regions
with respect to the surrounding background,
within which gravity eventually dominates all supports
and triggers the collapse, leading to the formation of the prestellar cores. As in Padoan \& Nordlund's theory, we
assume that there is a direct correspondence between the prestellar core mass function (CMF) and the final stellar
IMF. More preciselly, we assume that there is a good correspondance between the mass reservoir 
out of which the cores form and the IMF. A fact that seems to be supported observationally by the remarkable
similarity between the IMF
and the CMF (see e.g. Andr\'e et al. 2010 for the most recent results concerning this issue)
but also statistically by the rather strong correlation between the CMF and the IMF
inferred from the analysis of numerical simulations aimed at exploring this issue (Chabrier \& Hennebelle 2010). Recently, Hopkins (2011, 2012) derived an IMF theory in a
 similar spirit as the Hennebelle-Chabrier one, based on a different, so-called excursion set formalism, and has extended the results to larger scales, typical of large-scale structures in galactic disks. 
His results are found to agree fairly well with the Hennebelle-Chabrier ones when considered at the star forming clump scale.

The theory developed in papers I and II, which extends to the context of star formation, characterized by non-linear density fluctuations, the formalism developed
in cosmology for linear fluctuations by Press \& Schechter (1974), consists in properly counting  at all scales 
the self-gravitationally bound density fluctuations\footnote{As discussed in \S3 and \S5 of paper I, the HC theory  takes into account the probability for overdense, collapsing structures to be included in larger collapsing ones, and thus properly addresses the so-called "cloud in cloud" problem present in the Press \& Schechter formalism.}.
In the HC theory, these collapsing density fluctuations represent
the overdense regions that isolate themselves from the surrounding
medium and start to contract under the action of gravity at the {\it very initial stages of star formation}, and out of which
prestellar cores, and later on individual stars (or multiple star systems such as binaries) will form (see Chabrier \& Hennebelle 2011 for a simplified explanation of the HC formalism). 
As mentioned in papers I and II, a limitation so far of this theory 
resides in the time independent nature of its formulation. 
The importance of time-dependence for the IMF has been stressed, for instance, by Clark et al. (2007)
who argue that, since massive (low-density) cores are expected to collapse in a time longer 
than low-mass (high-density) ones, the CMF should be flatter than the IMF, if the latter one
is to be inherited from the former one. 
In the Hennebelle-Chabrier theory, however, turbulence-induced velocity dispersion, rather than purely thermal motions,  plays a dominant role in setting-up the mass 
of the {\it massive} cores when they form, so that
the {\it turbulent} Jeans mass,  which entails a Mach dependence,
instead of the thermal Jeans mass, should
be used as the characteristic mass scale. This does not mean that turbulence is acting as a pressure
or even a support in a static sense, as often misunderstood, but rather that on large scales, turbulent motions act to prevent larger amounts of material from immediate gravitational collapse. Eventually, for instance where the flows collide, the density is temporarily enhanced, yielding the  subsequent collapse of these large mass reservoirs, progenitors of massive cores (see Chabrier \& Hennebelle 2011). Under these conditions, as shown in Appendix C of Paper II, the characteristic free-fall timescale depends much more weakly on the star mass ($\tau_{ff}\propto M^{1/4}$), making the aforementioned time problem much less severe.
In the present paper, we develop this argument quantitatively by 
deriving a time-dependent theory of the IMF. We show that, indeed, time-dependence barely affects the slope of the IMF at large masses.
The peak of the mass spectrum, however, is 
shifted by a factor of $\simeq 3$ towards smaller masses, as more low-mass prestellar cores are able to form. This time-dependent
formulation of star formation enables us to derive a star formation rate.
We stress, however, that this SFR is valid at the scale of the clouds and not necessarily at the scale 
of the entire Galaxy (see e.g. Ostriker et al. 2010).

A summary of the present results and comparisons with previously published SFR theories and with some observational results have been presented in Hennebelle \& Chabrier (2011, HC11). 
The present paper presents in details the whole derivation of our time-dependent IMF theory and of the analytical SFR and confronts the results with further recent observational determinations. 
The paper is organized as follows. The extension of our formalism to a time-dependent
derivation, and comparisons between the time-dependent and time-independent theories of the mass spectrum of self-gravitating fluctuations, 
are presented in details in \S2, both for the isothermal and non-isothermal case.  The effect of magnetic field is also considered in this section.
In \S3, we derive the 
 star formation rate and star formation efficiency. We first derive the theoretical expression 
and then compute the SFR values for an ensemble of clump parameters and  explore the dependence of the SFR upon various clump characteristic properties\footnote{In our theory (see papers I and II), star forming "clumps" are identified as overdense
($\nbar \gtrsim 10^2$-$10^3$ cm$^{-3}$) $\sim$1-10pc-size unbound regions within large, diffuse
molecular clouds, within which $\sim$0.1pc-size gravitationally bound prestellar "cores" will preferentially form.}.
 In  \S4, we make comparison with recent SFR determinations of molecular clouds in the Galaxy. We briefly discuss the impact of filaments in star formation upon our formalism.
In  \S5, we present the complete self-consistent time-dependent model, 
by exploring the consequences of our time-dependent theory  on the mass spectrum 
of different evolving clumps. Then, we 
investigate the impact on the mass spectrum of a time-dependent 
clump {\it distribution} instead of a single clump. Section 6
is devoted to the conclusion and our paradigm for star formation and the IMF is presented.

\section{Time dependent theory}

\subsection{Analytical expression of the mass spectrum}

The underlying concept on which relies the theory developed in papers I and II consists to identify, in a random field of density fluctuations, 
the mass $M_R$ which at scale $R$ (physically speaking, $R$ denotes the radius of 
the density fluctuation) is gravitationally unstable, i.e. the mass 
contained in regions within which gravity dominates over all sources of
support. To achieve this, the first step is to determine $M_R$,
 the mass contained in regions whose
density exceeds a {\it scale-dependent} density threshold, $\log \delta^c_R$, determined by
the virial condition. Since all this gas is unstable at scale 
smaller than or equal to $R$, it is expected that it will end up in objects of mass
smaller than or equal to the Jeans mass associated to the density threshold.
Thus the second step is to equal $M_R$ with the mass contained in the structures
of mass equal to or smaller than the associated (turbulent) Jeans mass (see eqns.(28) and (29) of paper I).

This approach implicitly assumes that each part of the flow is initially assigned 
a specific  (scale dependent) Jeans mass, which collapses. While such an assumption
is certainly reasonable in the case of cold dark matter fluctuations within 
the primordial universe, it is a priori not the case in a turbulent flow in 
which fluctuations of scale $R$  are replenished within a few 
crossing times, $\tau _R \simeq 2R / V_{{\rm rms}}(R)$, where $V_{{\rm rms}}(R)\equiv \langle \sigma_R^2\rangle^{1/2}$ denotes the turbulent rms velocity at scale $R$ and $\sigma_R$ the velocity fluctuation over scales smaller than $R$. Therefore, in the case of 
turbulent molecular clumps, it seems necessary to take such {\it dynamical effects} into account.  
This is not obvious by any means, in particular for large scale fluctuations for which, first, a long enough time 
is required and, second, gravity must not entirely freeze the gas motions 
by preventing fluid particles to escape their local Jeans masses. In this 
latter case, gravity will prevent the flow motions and the 
replenishment of the density fluctuations, which have become gravitationally unstable.

On the other hand, if the  clump is accreting at a sufficiently high rate or 
if it contains a large fraction of gas that is not dense enough to be 
locally gravitationally unstable (i.e. if the Jeans length at the gas density 
is  comparable to or larger than the size of the clump), dissipation of large scale turbulence leads to turbulent compression which
 can continuously generate density fluctuations at all scales
and trigger, for the densest ones, gravitational collapse. In this case, one must
take into account the fact that, during the lifetime of the cloud, $\tau_0$, the fluctuations
at scale $R$ have been replenished a number of time equal to $\tau_0/\tau_R$.
This implies that eq.~(31) of paper I must be modified as

\begin{eqnarray}
{ M_{\rm tot}(R) \over V_c} =
 \int ^{\infty} _ {\delta_R^c} \bar{\rho} \exp(\delta)   {\cal P}_R(\delta)  ({\tau_0 \over \tau_R}) d\delta
= \int _0 ^ {M_R^c} M' \, {\cal N} (M') \,   P(R,M')\, dM'. 
\label{bal_mass}
\end{eqnarray}
where $\delta=\log(\rho/{\bar \rho})$ denotes the (logarithm of) density fluctuation, $\delta^c_R$ the threshold density at scale $R$,
 ${\cal P}(\delta)$ is the density distribution (PDF)\footnote{We stress that, in our general formalism, 
${\cal P}$ is not necessarily a lognormal, as emphasized in Schmidt et al. (2010).}
and $V_c\sim L_c^3$ denotes the clump's volume.

Apart from the term ${\tau_0 / \tau_R}$, this equation is the same as the 
one derived in paper I. It uses the formulation developed by 
Jedamzik (1995) in the context of dark matter halos (see also Yano et al. 1996). 
The first equality for $M_{\rm tot}(R)$ stems from the fact
that the  mass contained within structures of mass $M< M_R^c$
 is equal to the mass of the gas which, smoothed at scale $R$, has a (logarithmic)
density larger than a critical threshold $\delta_R^c$.
The second expression arises from the fact that 
the number-density of structures of mass $M< M_R^c$  is $ {\cal N}(M') P(R,M')  dM'$.
Here, ${\cal N}(M') dM'$ is the number-density of  structures of  mass
between $M'$  and $M'+dM'$, while
$P(R,M')$ is the probability to find a gravitationally unstable structure 
of  mass $M'$ embedded inside a structure of gas which at scale $R$ has a (logarithmic) density
larger than $\delta_R^c$. $P(R,M')$ is assumed to be equal to 1 (see Appendix D of Paper I for further justification).
The time ratio $\tau_0 / \tau_R$ thus simply illustrates
the fact that the flow fluctuations  at scale $R$ have been 
rejuvenated $\tau_0 / \tau_R$ times.

\noindent Taking the derivative of eq.~(\ref{bal_mass}) with respect to  $R$,
we obtain for the {\it number-density mass spectrum of gravitationally bound structures}, ${\cal N} (M)=d(N/V_c)/dM$:

\begin{eqnarray}
\label{n_general}
 {\cal N} (M_R^c)  &=& 
  { \bar{\rho} \over M_R^c} 
{dR \over dM_R^c} \,
\left( -{d \delta_R ^c \over dR} ({\tau_0 \over \tau_R}) \exp(\delta_R^c) {\cal P}_R( \delta_R^c) + \int _  {\delta_R^c}^\infty \exp(\delta) {d  \over dR}[ ({\tau_0 \over \tau_R}){\cal P}_R]\, d\delta \right)
\label{spec_mass1}
\end{eqnarray}
As shown in paper I, the first term in this expression is the most important one and dominates over the second one
except when the scale $R$ becomes comparable to the injection scale, $L_i$, which is basically the size of the cloud (see the interesting formulation of Hopkins (2012) to avoid this large-scale limitation of the present formalism).

Once the expression of the critical density threshold, $\delta_R^c=\ln (\rho(R)/{\bar \rho})$, is specified from the Virial condition, $\langle V_{rms}^2(R)\rangle + 3C_S^2 < E_{{\rm pot}}(R)/M $ (see \S4.2 of Paper I),
the mass spectrum of the self-gravitating pieces of fluid can be inferred from
\begin{eqnarray}
M_R=C_m\,  \rho(R)\, R^3,  
\label{masse}
\end{eqnarray}
where $C_m$ is a geometrical factor, typically of the order of $4 \pi/3$.

Before proceeding further, it is worth stressing what is exactly selected by 
our procedure. As seen from eq.~(\ref{bal_mass}), the integration 
is performed from 0 to $M_R^c$. This means that the smallest Jeans masses, corresponding 
to the densest pieces of gas, are first accounted for and removed from the available 
gas mass. Then the larger Jeans masses are progressively taken into account 
and removed. By doing so, {\it we properly take into account the fact that there are 
small Jeans masses embedded into larger ones}. This happens, in particular,
when the PDF significantly varies  locally, i.e. when turbulence is strong, which 
implies that in that case the scale must be large. Therefore, strictly speaking, we do not 
identify well defined bound cores but rather coherent mass reservoirs in the density field that isolate themselves from the surrounding
medium under the action of gravity. For sufficiently small scales, the density becomes reasonably uniform within the mass reservoir
to have a clear correspondence between this latter and a well identified "core".
For large scales, which will lead eventually to the formation of
 massive stars, however, the one-to-one core-reservoir 
correspondence becomes more blurry (as the potential well
associated with the mass reservoir becomes itself more shallow).
 In that case, the reservoir 
of mass out of which the most massive cores/stars will form
 corresponds to what is left
once all the self-gravitating {\it small-scale} density fluctuations
embedded in the reservoir have been properly accounted for and
 "taken away".
In that sense, {\it for the largest scales,  thus the most 
massive cores}, our formalism rejoins in some sense the 
so-called "competitive" accretion process 
(see e.g. Smith et al. 2009). 
Therefore, in principle, the turbulent fragmentation process
is properly described in our formalism, apart from the fact that, 
for sake of simplicity, the conditional probability 
$P(R,M)$ is taken to be equal to 1. What has {\it not} been taken
 into
account so far, however, is the gravitational fragmentation
 that occurs 
during the collapse. Although such fragmentation may occur, its 
importance is likely to remain limited because of the impending roles 
of magnetic field (Machida et al. 2005, 2008, 
Hennebelle \& Teyssier 2008, Price \& Bate 2008, 
Hennebelle et al. 2011) and radiation (Krumholz et al. 2007, Bate 2009,
 Commercon
et al. 2010) particularly when both are present 
(Commercon et al.
2011), an issue which still needs to be properly quantified. 
Observations of massive cores indeed suggest that fragmentation 
is rather limited, most of the mass of the core ending up in one 
or just a few smaller cores (Bontemps et al. 2010, 
Longmore et al. 2011,  Bressert et al. 2010, Palau et al. 2013).

\subsection{Influence of the time-dependence on the CMF/IMF: 
isothermal case}
In this section, we examine the impact of time-dependence on the CMF 
in the simple isothermal case. The barotropic
 case will be examined
in  \S\ref{baro-mag}. We first discuss the density PDF, then 
the crossing time.
 Finally, we derive the analytical expression and 
compare the results with the time-independent ones.

\subsubsection{Density probability function}
Numerical simulations
 (V\'azquez-Semadeni 1994, Padoan et al. 1997, Passot \& V\'azquez-Semadeni 1998,  
Kritsuk et al. 2007, Federrath et al. 2008, Schmidt et al. 2009)  have revealed that 
the turbulence-induced density distribution, ${\cal P}(\delta)$, is reasonably well described by a lognormal distribution
\begin{eqnarray}
{\cal P}(\delta) = {1 \over \sqrt{2 \pi \sigma_0^2}} 
\exp\left(- { (\delta - \bar{\delta})^2 \over 2 \sigma_0 ^2} \right),\nonumber\\
 \delta = \ln (\rho/ \bar{\rho} ), \,\, 
 \bar{\delta}=-\sigma_0^2/2, \;
 \; \sigma_0^2=\ln (1 + b^2 {\cal M}^2).
\label{base}
\end{eqnarray}
In this expression, ${\cal M}$ is the characteristic cloud Mach number and $b$ a non-dimensional coefficient 
that depends on the turbulence forcing (see Federrath et al. 2010). It typically 
varies from 0.25 when the forcing is purely solenoidal to almost 1 when 
the forcing is applied only on compressible modes. 
Such a 3D lognormal shape for density fluctuations has received observational support from its 2D observed projection, namely
the power spectrum column density of molecular clouds, measured from dust extinction maps (Kainulainen et al. 2009,  Brunt et al. 2010).
The scale-dependence of the variance of the distribution reads, in 3D (see Paper I)
\begin{eqnarray}
\sigma^2(R) = \int _{2 \pi / L_c} ^{2 \pi / R} \widetilde{\delta} (k)^2 4 \pi k^2 dk
= \sigma_0^2 \left( 1 - \left({R \over L_c} \right)^{n'-3} \right),
\label{power_spec}
\end{eqnarray}
where $L_c$ is the cloud's size, $\widetilde{\delta} (k)^2 \propto k^{-n'}$ is the 
power spectrum of $\log (\rho)$, of 3D index $n'$. This latter is found  in isothermal, shock-dominated hydrodynamical and MHD simulations to be very similar to the index $n$ of the {\it velocity} power spectrum, 
with a typical value $n^\prime \sim n \sim 3.8$, between  11/3 (Kolmogoroff limit) and 4 (Burgers limit) (see e.g. Kritsuk 2007, Federrath et al. 2008, 
Schmidt et al. 2009). Then (see paper I) 
\begin{eqnarray}
{\cal P}(\delta) = {1 \over \sqrt{2 \pi \sigma(R)^2}} \exp\left(- { [\delta + 
 {\sigma(R)^2 \over 2} ]^2 \over 2\, \sigma(R)^2} \right).
\label{Pr}
\end{eqnarray}

\subsubsection{The crossing time, $\tau_R$}
\label{crossing_sect}
The crossing time at scale $R$ is given by $\tau_R\equiv \tau_{ct}(R)=2 \alpha_{ct} R/V_{ct}$,
where $V_{ct}$ is the relevant velocity and $\alpha_{ct}$
a dimensionless coefficient of the order of a few. At large scales, $V_{ct}$ is 
typically equal to  the one-dimensional velocity dispersion $V_{\rm rms} ^{1D}=V_{\rm rms}/{\sqrt 3}$,
where  $V_{\rm rms}\equiv V_{\rm rms} ^{3D}$ designates the 3D velocity dispersion all along this paper, 
while at small scales, below the sonic length, $V_{ct} \simeq C_s$. 
This crossing time is the typical time that is necessary for 
the density field to be significantly modified at scale $R$, 
implying that a new set of fluctuations, statistically 
independent of the former one, has been processed.  

As mentioned earlier and discussed in papers I and II, in the present context, we select the pieces of
 gas which are self-gravitating, i.e. such that their 
internal gravitational, kinetic and thermal energies obey the condition: 
$-E_{\rm grav} > 2 E_{\rm kin} + 3 P_{th}$. 
At large scales, this implies that $\alpha_g G M / R > V_{\rm rms}^2$,
where $\alpha_g$ is a dimensionless coefficient, equal to $3/5$ for
a uniform density fluctuation.  
We thus get
\begin{eqnarray}
\tau_R ={2 \alpha_{ct} R \over V_{\rm rms} ^{1D} } = 2 \alpha_{ct} 
\sqrt{ 24 \over \pi^2 \alpha_{g}} \tau_{ff}(R),
\label{crossing}
\end{eqnarray}
where $\tau_{ff}(R)=\sqrt{3 \pi \over 32 G \rho(R) }$ is the free-fall time of the density fluctuation of density $\rho(R)$.
A similar expression is obtained below the sonic length. In the following 
we thus define the crossing time of a collapsing density fluctuation of scale $R$ as
\begin{eqnarray}
\tau_R &=& \phi_t \tau_{ff}(R),\\
{\rm with}\,\,\,\,\phi_t &=& 2 \alpha_{ct} \sqrt{ 24 \over \pi^2 \alpha_{g}}\simeq 3.
\label{crossing2}
\end{eqnarray}

\noindent Defining $\tau_{ff}^0$ as the free-fall time at the clump's mean density, 
$\tau_{ff}^0 =\sqrt{3 \pi \over 32 G \bar{\rho}}\simeq 1.07({\mu\over 2.33})^{-1/2}({\nbar\over 10^3{\rm cm}^{-3}})^{-1/2}$ Myr (where $\mu=2.33$ is the mean molecular weight for a cosmic H$_2$/He composition), we get
\begin{eqnarray}
{\tau_{R}\over \tau_{ff}^0 }= \phi_t \sqrt {{\bar{\rho}\over \rho}}.
\label{time}
\end{eqnarray}
Note that the choice of $\tau_{ff}^0$ is not consequential at this stage 
as it simply modifies the value of ${\cal N}$ uniformly without
affecting its shape.   
Combining eqs.~(\ref{spec_mass1}),~(\ref{Pr}), and~(\ref{crossing2}) and dropping the second term in eq.~(\ref{spec_mass1}), as mentioned earlier, 
and assuming that $\tau_{ff}^0\simeq \tau_0$, we obtain (see paper I)
\begin{eqnarray}
{\cal N}(\widetilde{M}) = -{1 \over \phi_{t}} 
{ \bar{\rho} \over M_J^{0} \widetilde{M}}
 \left( { \widetilde{M}_R^c \over \widetilde{R} ^3 } \right)^{1/2}
 {d \widetilde{R} \over d \widetilde{M}_R^c}  {d \delta_R^c\over d\widetilde{R}}  {1 \over \sqrt{2 \pi \sigma^2} } \exp \left( -{(\delta_R^c)^2 \over 2 \sigma^2} + 
{\delta _R^c \over 2} - {\sigma^2 \over 8}  \right), 
\label{mass_spec}
\end{eqnarray}
where 
$M_R^c$ is the critical mass at scale $R$ (see eq.(\ref{masse})) while
\begin{eqnarray}
\delta_R^c = \ln \left( { \widetilde{M}_R^c \over \widetilde{R}^3} \right),
\label{dens_crit}
\end{eqnarray}
where $\widetilde{R}= R / \lambda_J^0$,
$\widetilde{M} =  M / M_J^0 $ and $M_J^0$, 
$\lambda_J^0$ denote the usual mean thermal Jeans mass and Jeans length, respectively:
\begin{eqnarray}
M_J^0&=& {a_J\over C_m}\,{ C_s^3 \over \sqrt{G^3 \bar{\rho}}}\approx 0.8\,\, ({a_J\over C_m}) \,({T \over 10\,{\rm K}})^{3/2}\,
({\mu \over 2.33})^{-2}\,({{\bar n} \over 10^4\,{\rm cm}^{-3}})^{-1/2}\, \msol 
\label{mjeans} \\
\lambda_J^0&=& \left( {\pi^{3/2} \over C_m} \right)^{1/3}{C_s\over \sqrt {G{\bar \rho}}}\approx 0.2\,\left( {\pi^{3/2} \over C_m} \right)^{1/3} \,({T \over 10\,{\rm K}})^{1/2}\,
({\mu \over 2.33})^{-1}\,({{\bar n} 
\over 10^4\,{\rm cm}^{-3}})^{-1/2}\,\, {\rm pc},
\label{ljeans}
\end{eqnarray}
where $a_J$ is a dimensionless geometrical factor of order unity.
Taking the standard  definition of 
the Jeans mass as the mass enclosed in a sphere of diameter equal to the Jeans length, one gets  $a_J=\pi^{5/2}/6$.
We now need to know $M_R^c$, and its 
derivative $d M_R^c / dR$, as a function of the scale, $R$. It is determined 
by the physical processes at play in the cloud, as examined in the next 
sections.

\subsubsection{Analytical expression}
We now need to specify the density threshold, $\delta_R^c $, which is determined from the virial theorem. 
As shown in paper I (\S4.3),
the condition for collapse, which simply reads $\delta>\delta_R^c $ or equivalently $M > M_R^c=M_J(R)$, yields after calculations
\begin{eqnarray}
M > a_J { \Bigl[ (C_s(\rho))^2 + ({V_{rms}^2 \over 3})  \Bigr]^{3/2}   \over \sqrt{G^3 \bar{\rho} \exp(\delta) }  },
\label{cond_tot}
\end{eqnarray}

We  assume that the effective dispersion velocity obeys the Larson's (1981) relationship:
\begin{eqnarray}
\langle V_{\rm rms}^2\rangle &=&  V_0^2 \times \left( {R \over  1 {\rm pc}} \right) ^{2 \eta},\nonumber\\
{\cal M} &\equiv& {\cal M}(R) = {\langle V_{\rm rms}^2\rangle^{1/2}\over C_s}.
\label{larson}
 \end{eqnarray}
As mentioned in paper I,  the Mach number ${\cal M}$ represents "effective" values which include both the hydrodynamical and magnetic contributions,
i.e. $V_{\rm rms}=\{(V_{{\rm rms}})_{\rm hydro}^2+V_A^2/2\}^{1/2}$ where  $V_A=B/(4\pi\rho)^{1/2}$
denotes the Alfv\'en velocity.  The coefficient $\eta$ is related to $n$, the index of the velocity power spectrum, by the relation (see eq.(24) of Paper I):
\begin{eqnarray}
 \eta=\frac{n-3}{2}.
\label{eta}
\end{eqnarray} 
As mentioned earlier, 3D numerical simulations of compressible turbulence (e.g. Kritsuk et al. 2007) suggest a value $n\simeq3.8$, yielding  $\eta\simeq 0.4$, as indeed found in observations.

With eq.(\ref{masse}), eq.(\ref{cond_tot}) implies (see Paper I)
\begin{eqnarray} 
M > M_R^c = a_J^{2/3} 
\left( {  C_s^2 \over G    } R + {V_{\rm rms}^2  \over 3\, G  }\right), 
\label{crit_Mtot}
\end{eqnarray}  

After normalisation, eq.~(\ref{crit_Mtot}) becomes (see \S5.4 Paper I)
\begin{eqnarray}
\widetilde{M}_R^c =  M / M_J^0 = \widetilde{R}\,
(1+ {\cal M}^2_* \widetilde{R}^{2 \eta}),
\label{mass_rad}
\end{eqnarray}
where  ${\cal M}_*$ is given by:

\begin{eqnarray}
{\cal M}_* = { 1  \over \sqrt{3} } { V_0  \over C_s}\left({\lambda_J^0 \over  
 1\, {\rm pc} }\right) ^{ \eta}
\simeq (0.8-1.0) \,\left({\lambda_J^0\over 0.1\,{\rm pc}}\right)^{\eta}\,\left({C_s\over 0.2\kms}
\right)^{-1},
\label{mass_star}
\end{eqnarray}
and thus illustrates the impact of turbulence induced velocity fluctuations at the {\it Jeans scale} (see paper I).

Equations~(\ref{mass_spec}) and~(\ref{mass_rad}) finally yield for the mass spectrum of gravitationally bound prestellar cores:
\begin{eqnarray}
{\cal N} (\widetilde{M} ) &=&  {2\ \over \phi_{t}}  {\cal N}_0 \, { 1 \over \widetilde{R}^6} \,
{ 1 + (1 - \eta){\cal M}^2_* \widetilde{R}^{2 \eta} \over
[1 + (2 \eta + 1) {\cal M}^2_* \widetilde{R}^{2 \eta}] }
\times    \left( {\widetilde{M} \over \widetilde{R}^3}  \right) ^{-1 -   {1 \over 2 \sigma^2} \ln (\widetilde{M} / \widetilde{R}^3) }
\times {\exp( -\sigma^2/8 ) \over \sqrt{2 \pi}\, \sigma },
\label{grav_tot2}
\end{eqnarray}
where ${\cal N}_0=  \bar{\rho} / M_J^0$. 
Equation~(\ref{grav_tot2}) is the time-dependent generalization of eq.~(44) of paper I.
The time dependence appears explicitly through the factor $1/\phi_t$ but also through the modification of the exponent,
$-1 -{1 \over 2 \sigma^2} \ln (\widetilde{M} / \widetilde{R}^3)$, instead of $-3/2 -   
{1 \over 2 \sigma^2} \ln (\widetilde{M} / \widetilde{R}^3)$,  
which arises from the  time correction, proportional to $\sqrt{\rho} \propto (M/R^3)^{1/2}$.
As expected from the discussion in the introduction, the time dependence thus affects the slope of the CMF.
We  quantify this effect in the next sections.

\setlength{\unitlength}{1cm}
\begin{figure} 
\begin{picture} (0,8)
\put(0,0){\includegraphics[width=7.5cm]{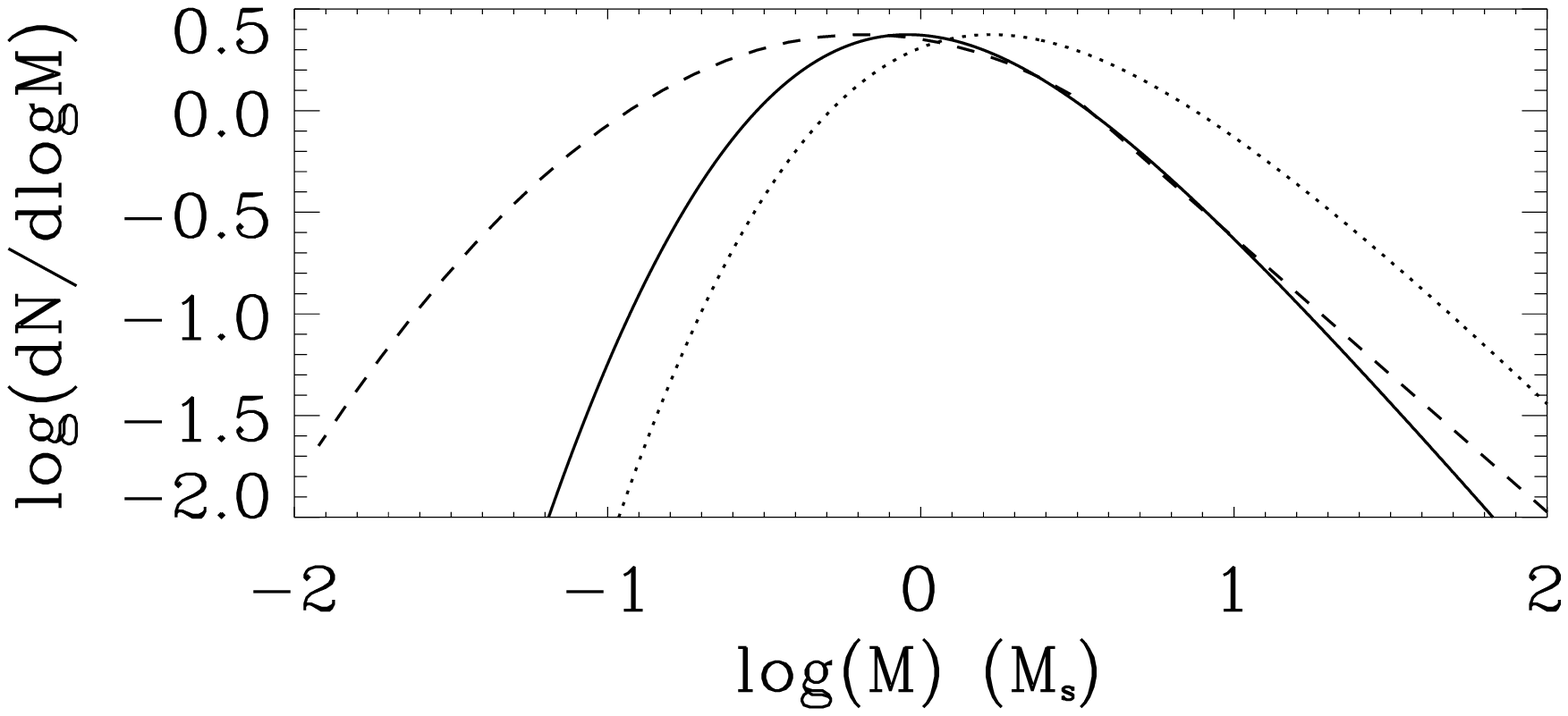}} 
\put(0,4){\includegraphics[width=7.5cm]{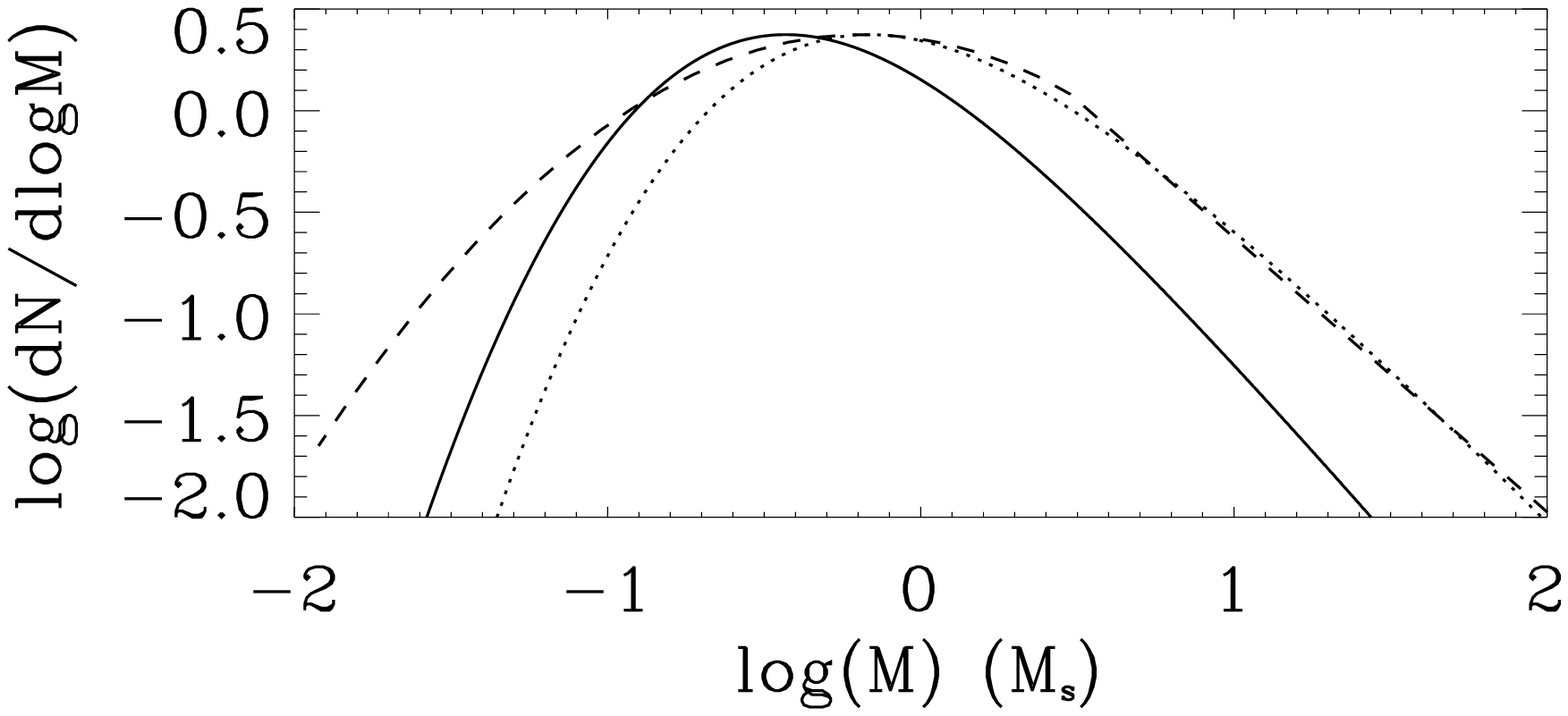}} 
\end{picture}
\caption{Isothermal case. 
Core mass function, $dN/d\log M$, for ${\cal M}=6$, ${\cal M}_* =\sqrt{2}$ and mean
clump  density ${\bar n}=3 \times 10^4$ cm$^{-3}$ (top panel) and ${\bar n}=5 \times 10^3$ cm$^{-3}$ (bottom panel).
The solid line corresponds to the time-dependent model 
while the dotted line represents the time independent one. 
The dashed line is the Chabrier IMF shifted upward in mass by a factor 3 (see text).}
\label{fig_isotherm}
\end{figure}

\subsubsection{Results and comparison with the time independent model}
\label{timedep}
Before comparing the   time-dependent and 
time-independent distributions, we determine the position of the peak of the CMF. As discussed in paper I (\S7.1.4),
 the peak occurs in the thermally dominated regime,
which corresponds to ${\cal M}_*=0$. The derivative of eq.~(\ref{grav_tot2}) with respect to 
the mass yields in that case
\begin{eqnarray}
 \widetilde{M}_{\rm peak} = {M_{\rm peak}\over M_J^0} =  \exp ( -  \sigma ^2 ) = { 1 \over (1 + b^2 {\cal M}^2) }, 
\label{mpeak}
\end{eqnarray}
whereas in the case of a time independent distribution we have (eqn.(46) Paper I)
\begin{eqnarray}
 \widetilde{M}_{\rm peak} = \exp ( - {3 \over 4 } \sigma ^2 ) = { 1 \over (1 + b^2 {\cal M}^2) ^{3/4}}.
\label{mpeak_indept}
\end{eqnarray}
This implies that, for a given Mach number, the peak of the distribution is 
shifted toward smaller masses when time dependence is 
taken into account. This is intuitively expected from the fact that more 
small-scale objects, with short free-fall times, are continuously produced during the clump evolution
in a time-dependent collapse.  It is thus important to stress that in the (both time-dependent and time-independent) HC theory of gravo-turbulent fragmentation, {\it the peak (i.e. the characteristic) mass of the IMF depends not only on the clump's mean thermal Jeans mass, but also strongly ($\propto {\cal M}^{-2}$) on the characteristic Mach number.}

Another important quantity is the exponent of the high-mass tail of the distribution.
Indeed, in the limit  $\widetilde{M} \simeq {\cal M}_*^2 \widetilde{R}^{2 \eta +1}$,
which corresponds to the turbulence dominated regime  (see eq.~(\ref{mass_rad})),
eq.~(\ref{grav_tot2}) tends to a power law plus a lognormal, such as ${\cal N}\propto M^{-(1+x)}$ with
\begin{eqnarray}
x = {3 \over 2 \eta +1} - 6 { 1 - \eta \over (2 \eta + 1)^2 \sigma^2 } \ln ( {\cal M}_*),
\label{mslope}
\end{eqnarray}
whereas in the time-independent case, this coefficient reads (see eq.~(43) of paper II)
\begin{eqnarray}
x = {2 + \eta \over 2 \eta +1} - 6 { 1 - \eta \over (2 \eta + 1)^2 \sigma^2 } \ln ( {\cal M}_*).
\label{mslope_indep}
\end{eqnarray}
As expected, the distribution becomes slightly steeper when the time dependence is included in the derivation of the CMF/IMF.
This is due to the fact that some massive clumps fragment into smaller pieces during the cloud collapse.
 Typical values for $x$ range from 1.1 to 1.5, depending 
on $\eta$, $\sigma$ and ${\cal M}_*$, bracketing the Salpeter value, $x=$1.35.

Figure~\ref{fig_isotherm} displays the core
mass function, $dN / d \log M$ for $\eta=0.45$,
${\cal M}=6$, ${\cal M}_* =\sqrt{2}$ and two typical clump densities, $\bar{n}=5000$ cm$^{-3}$
and $\bar{n}=3 \times 10^4$ cm$^{-3}$.
Solid lines display  the time dependent results, 
dotted lines the time independent ones. For reference, 
dashed lines
represent the Chabrier system IMF (Chabrier 2003),
shifted upward in mass along the x-axis by a factor 3 to account for the observed shift 
between the CMF and the IMF. In the rest of the paper, we will refer to this IMF as the "shifted 
Chabrier IMF" (SCIMF). As mentioned above,
the time-dependent distribution peaks
at lower masses and has a slightly steeper
high-mass slope than  the time-independent one.
In spite of these differences, 
both distributions match  well the SCIMF above about the mean Jeans mass.
At low masses,
both distributions appear to be too narrow. As already discussed in 
papers I and II,
 this is  essentially due to the isothermal approximation for
the equation of state of the gas (see Paper II and below), but also to the
assumed exact correspondence between the initial core mass and the final star mass. 
Indeed, as shown in Chabrier \& Hennebelle (2010), 
taking into account some (expected) dispersion between
the mass of the parent core and the one of the final star, i.e. between the CMF and the IMF,
naturally leads to a broadening of the  latter one compared with the former one in the low-mass regime. At last,
gravity may also broaden the PDF.

\subsection{Barotropic equation of state and magnetic field}
\label{baro-mag}
Here we compare the time dependent and independent models 
when a barotropic equation of state is used. We also 
introduce the magnetic field in our model and investigate 
its impact on the mass spectrum. 

\subsubsection{Formalism}
Again, assuming that the pieces of fluid which collapse are 
gravitationally dominated, eq.~(\ref{cond_tot}) becomes
\begin{eqnarray}
M > a_J { \Bigl[ (C_s(\rho))^2 + (V_{rms}^2 / 3) + (V_{a}^2 / 6) \Bigr]^{3/2}   \over \sqrt{G^3 \bar{\rho} \exp(\delta) }  },
\label{cond_tot2}
\end{eqnarray}
where $V_a$ is the mean Alfv\'en speed. Based on observational and 
numerical results, we assume that the magnetic intensity 
correlates with the gas density as 
\begin{eqnarray}
B = B_0 \left( {\rho \over \bar{\rho} } \right)^{\gamma_b}.
\label{mag_rho}
\end{eqnarray}
Typical values for $B_0$ and $\bar{\rho}$ are 10-20 $\mu G$ and 
10$^3$ cm$^{-3}$, respectively (Crutcher 1999). The coefficient 
$\gamma_b$ seems to depend on density. In the diffuse gas, $\gamma_b$
is typically smaller than 0.5 and around 0.1-0.3 (Troland \& Heiles 
1986, Hennebelle et al. 2008) while for higher densities when the gas 
is self-gravitating, it is 
of the order of 0.5 (although see Crutcher et al. 2010).

After normalisation, eq.~(\ref{cond_tot2}) becomes
\begin{eqnarray}
\widetilde{M}_R^c =  M / M_J^0 = \widetilde{R}\,
\left( { C_s(\rho)^2 \over (C_s^0)^2 } + {\cal M}^2_* \widetilde{R}^{2 \eta}
+ (V_a^*)^2 \left( {\rho \over \bar{\rho}} \right)^{2 \gamma_b -1} \right),
\label{mass_rad2}
\end{eqnarray}
where
\begin{eqnarray}
V_a^* = { 1  \over \sqrt{6} } { B_0  \over \sqrt{4 \pi  \bar{\rho}} \,C_s}.
\label{va_star}
\end{eqnarray}

As emphasized in paper II, the thermodynamics of the gas has a drastic impact on the 
low-mass end of the mass function. As in paper II, we use the barotropic eos suggested by Larson (1985, see also Glover \& MacLow 2007)
from observations of molecular clouds, namely:

\begin{eqnarray}
T\propto \rho^{\gamma_1-1}{\hskip 1.cm} \rho<\rho^{crit},\nonumber \\
T\propto \rho^{\gamma_2-1}{\hskip 1.cm} \rho>\rho^{crit},
\label{baro}
\end{eqnarray}
with $\gamma_1 \simeq 0.7$, $\gamma_2 \simeq 1-1.1$, and $\rho^{crit}$ corresponds to the density above which dust grains
 become thermally coupled with the gas and thus dust cooling becomes the dominant cooling mechanism instead of line cooling. Observations suggest that $\rho^{crit}\simeq 10^{-18}\gcc$, i.e. ${\bar n}^{crit}\simeq 2.5\times 10^5$ cm$^{-3}$. We keep the same eos than the one we used in 
paper I, inspired from Larson (1985) and write
\begin{eqnarray}
\nonumber
C_s^2 &=& \left[  \left((C_{s,1}^0)^2 ({\rho  \over \bar{\rho} })^{\gamma_1-1}\right)^m + \left( (C_{s,2}^0)^2 ({\rho  \over \bar{\rho}})^{\gamma_2-1} \right)^m \right]^{1/m} \\
  &=&  (C_{s,1}^0)^2 \left[   \left({\rho  \over \bar{\rho} }\right)^{(\gamma_1-1)m}  + \left({C_{s,2}^0\over C_{s,1}^0}\right)^{2m} 
\left({\rho  \over \bar{\rho}}\right)^{(\gamma_2-1)m}  \right]^{1/m},
\label{prescript}
\end{eqnarray}
where $m$ is a real number of order unity. In the following, 
$m$ will be equal to 3 (see Fig.~4 of paper II). Typically, $\gamma_1 \simeq 0.7$
while $\gamma_2 \simeq 1-1.1$. The critical density, $\rho ^{crit}$, at which the transition 
between the two regimes is occurring, is expected to be about $10^{-18}\gcc$ and 
we define 
\begin{eqnarray}
K_{\rm crit}= \left({C_{s,2}^0 \over C_{s,1}^0}\right)^2 =    \left({\rho^{crit}  
\over \bar{\rho} }\right)^{\gamma_1  -\gamma_2}.
\label{const_K}
\end{eqnarray}

In order to get the mass spectrum from eq.~(\ref{mass_spec}), we need to know
$M_R^c$ and $dM_R^c / d R$ as a function of $R$. While for the first, it is 
not possible to get an explicit relation from eq.~(\ref{mass_rad2}), the  
second can be obtained by deriving eq.~(\ref{mass_rad2}) with respect 
to $\widetilde{R}$. The analytical expressions that we obtain 
are identical to eq.~(38) of paper II except for the extra terms related to the 
magnetic field. We write them explicitly for completeness.

\begin{eqnarray}
{d \widetilde{M}_R^c \over d \widetilde{R} } &=&  { B \over C}, 
\end{eqnarray}

\begin{eqnarray}
\nonumber
B &=&  D - 3 { \widetilde{M} \over  \widetilde{R}^3} 
{d D \over d \widetilde{\rho}} 
+ (2 \eta + 1) {\cal M}_*^2 \widetilde{R}^{2 \eta} , \\
 C &=&  1 -  {  \widetilde{R}^ {-2}}
\label{multi_crit2}
{d D \over d \widetilde{\rho}} , \\
D &=& \left(   { \widetilde{\rho} }  ^{(\gamma_1-1)m}  + 
( K_{\rm crit})^m  { \widetilde{\rho} } ^{(\gamma_2-1)m}
 \right)^{1/m} + (V_a^*)^2 \widetilde{\rho} ^{2 \gamma_b -1}.
\nonumber
\end{eqnarray}

\subsubsection{Comparison between time dependent and independent mass spectra in the barotropic case}
\setlength{\unitlength}{1cm}
\begin{figure*} 
\begin{picture} (0,16)
\put(0,0){\includegraphics[width=7.5cm]{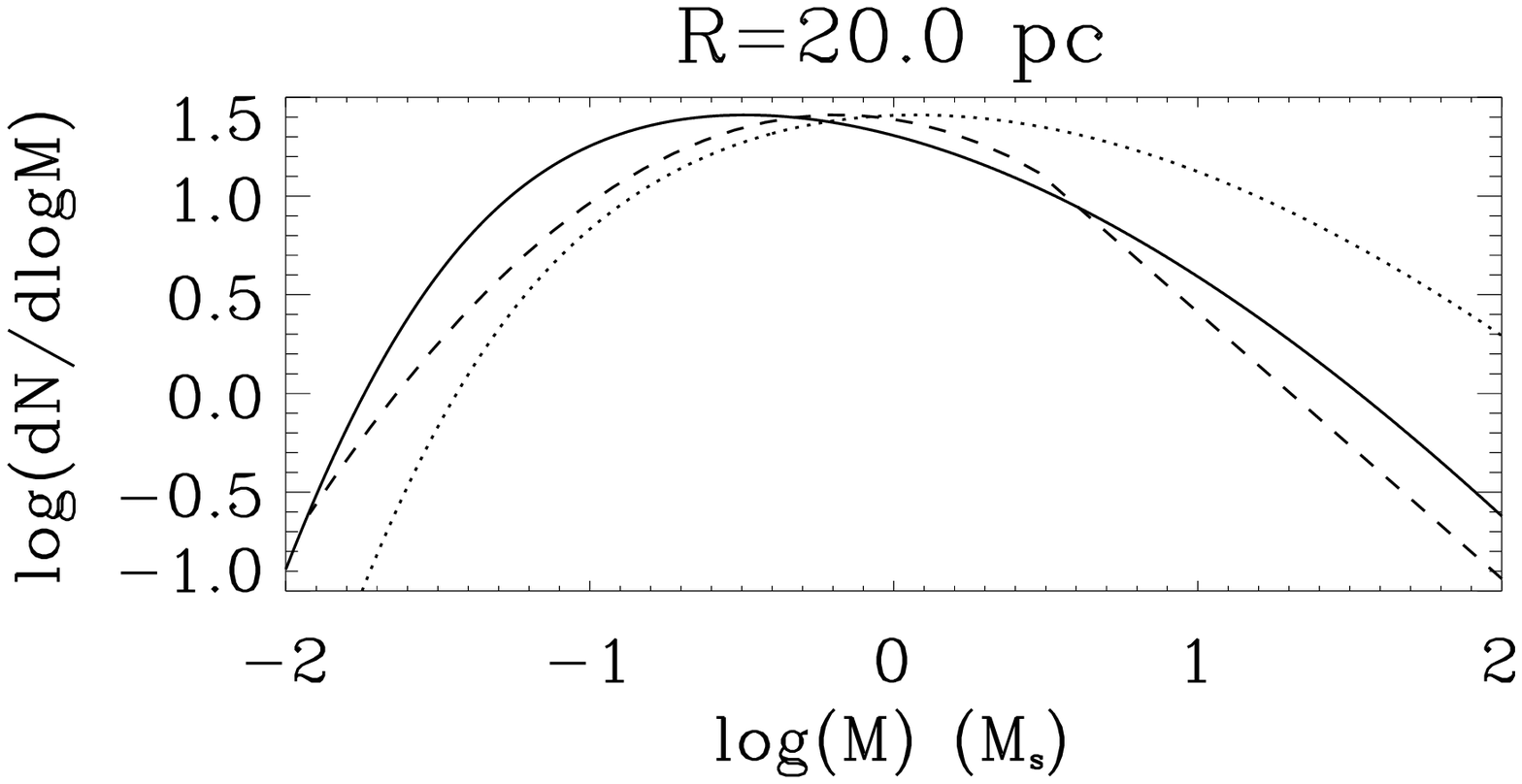}} 
\put(0,4){\includegraphics[width=7.5cm]{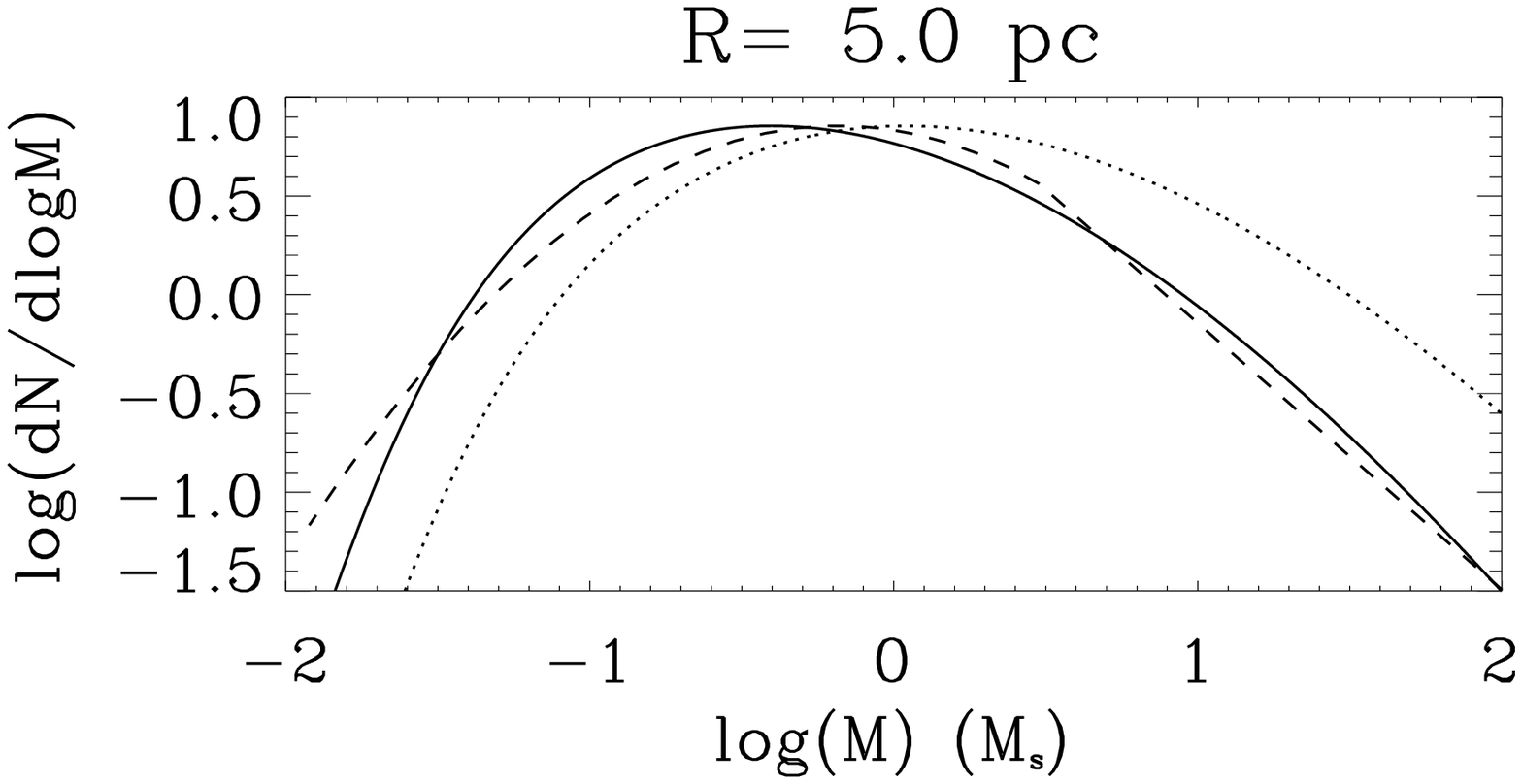}} 
\put(0,8){\includegraphics[width=7.5cm]{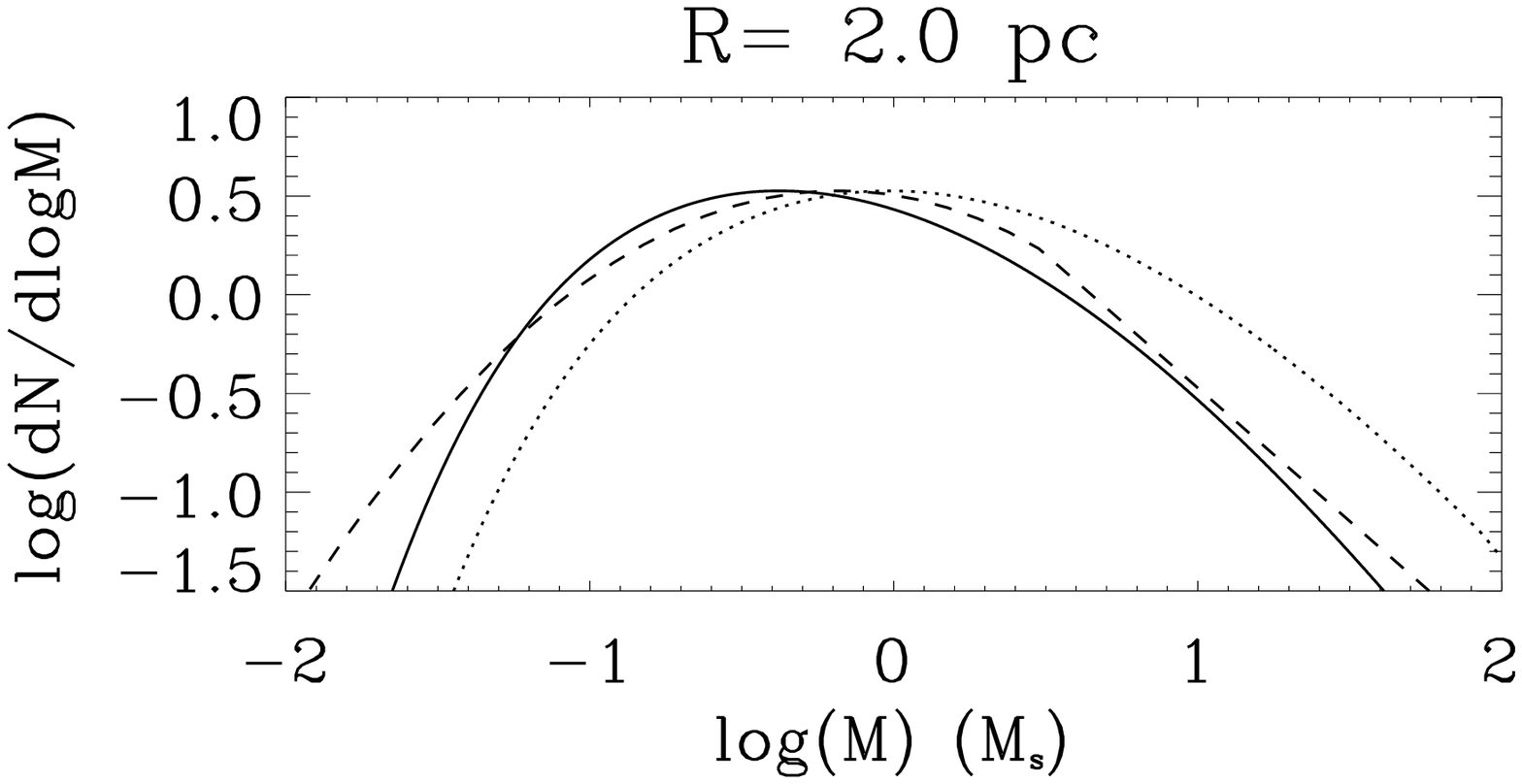}} 
\put(0,12){\includegraphics[width=7.5cm]{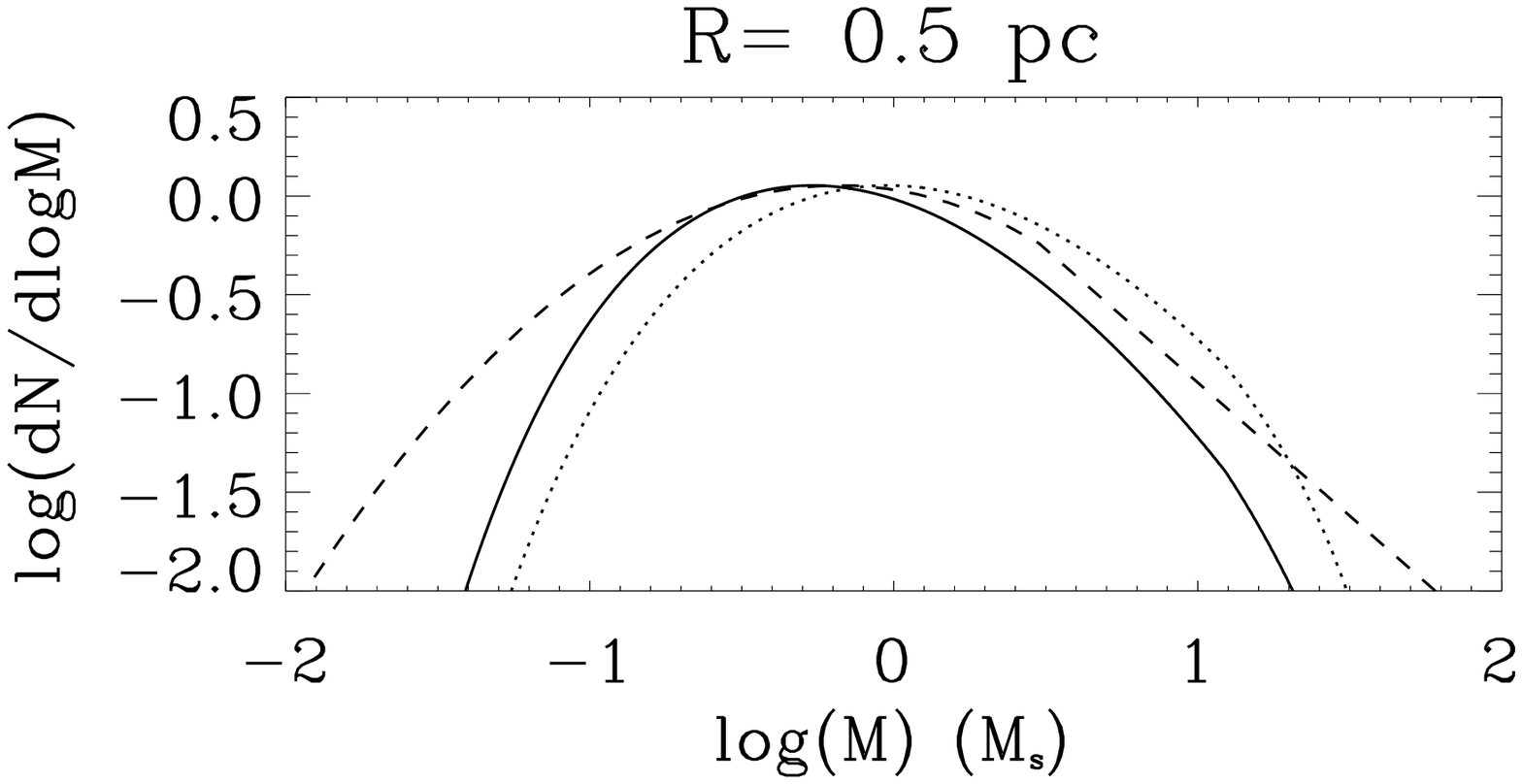}}
\put(8,0){\includegraphics[width=7.5cm]{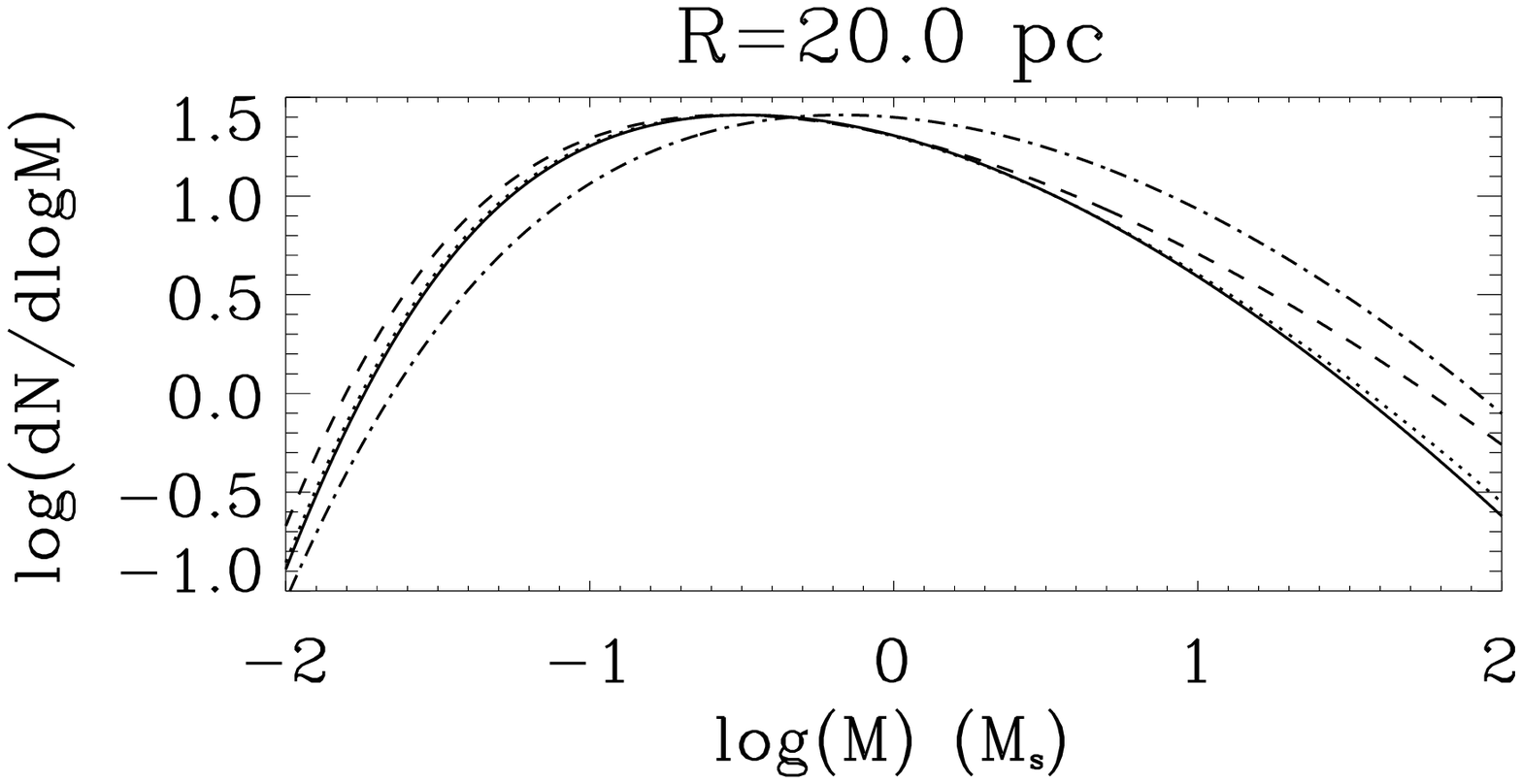}} 
\put(8,4){\includegraphics[width=7.5cm]{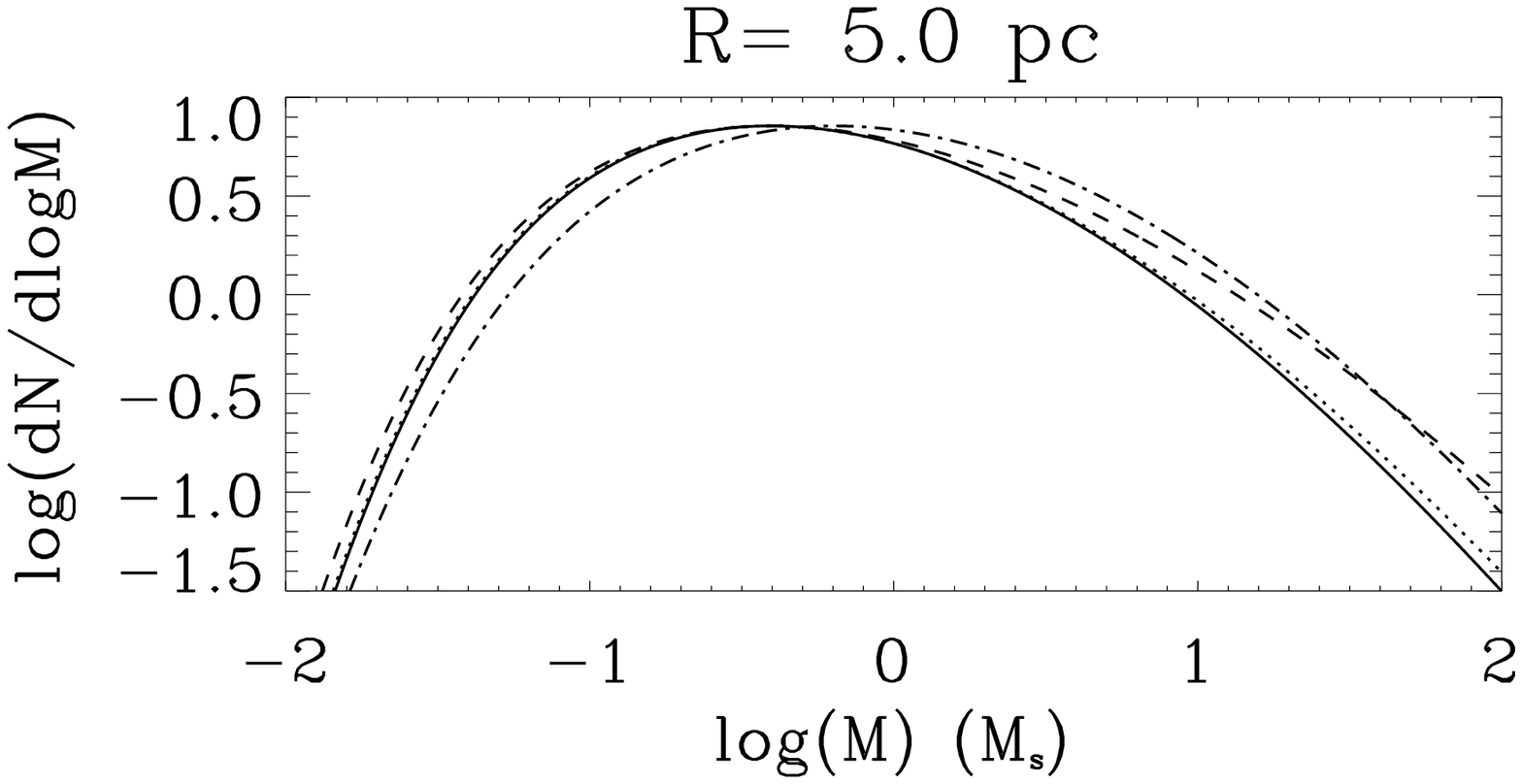}} 
\put(8,8){\includegraphics[width=7.5cm]{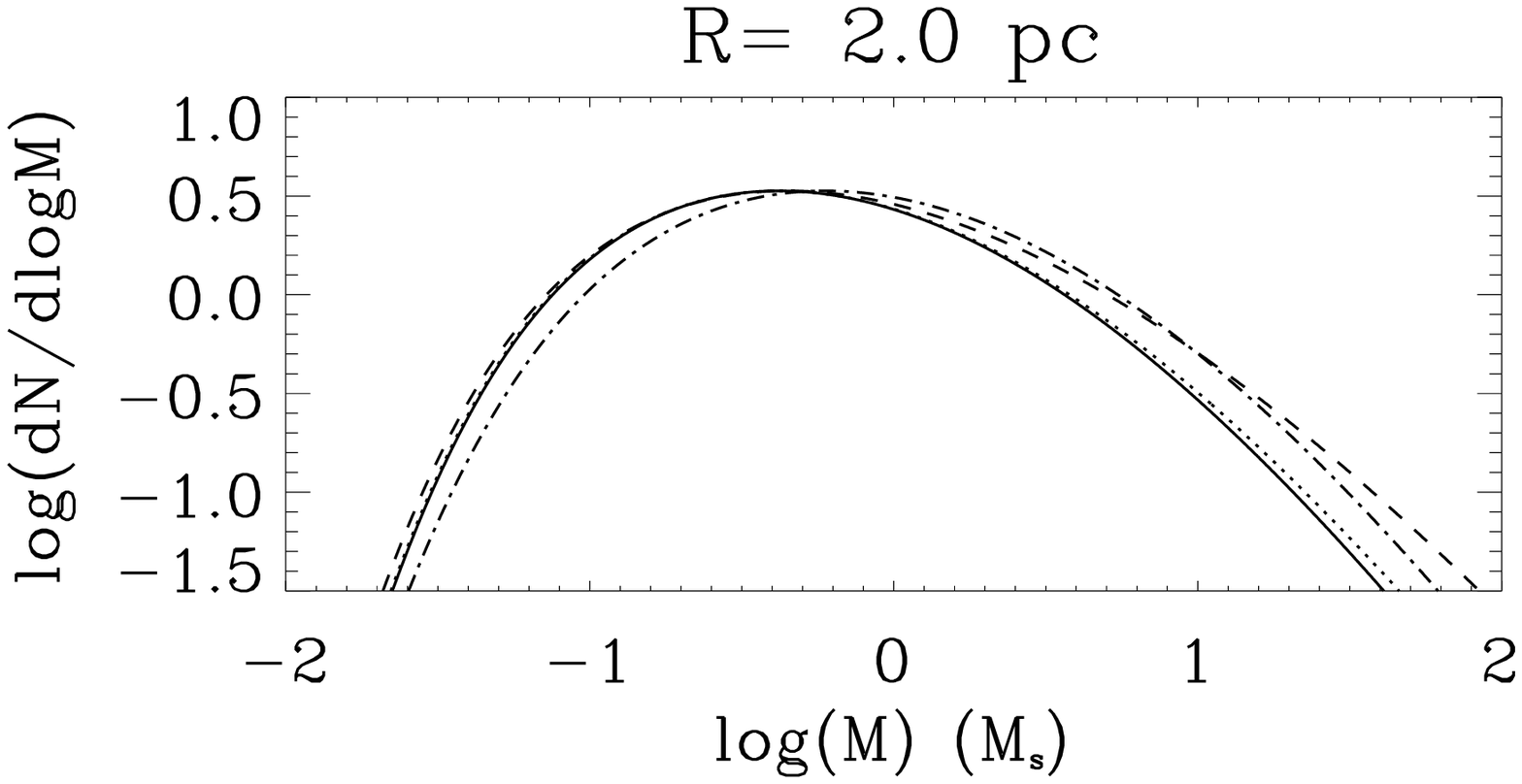}} 
\put(8,12){\includegraphics[width=7.5cm]{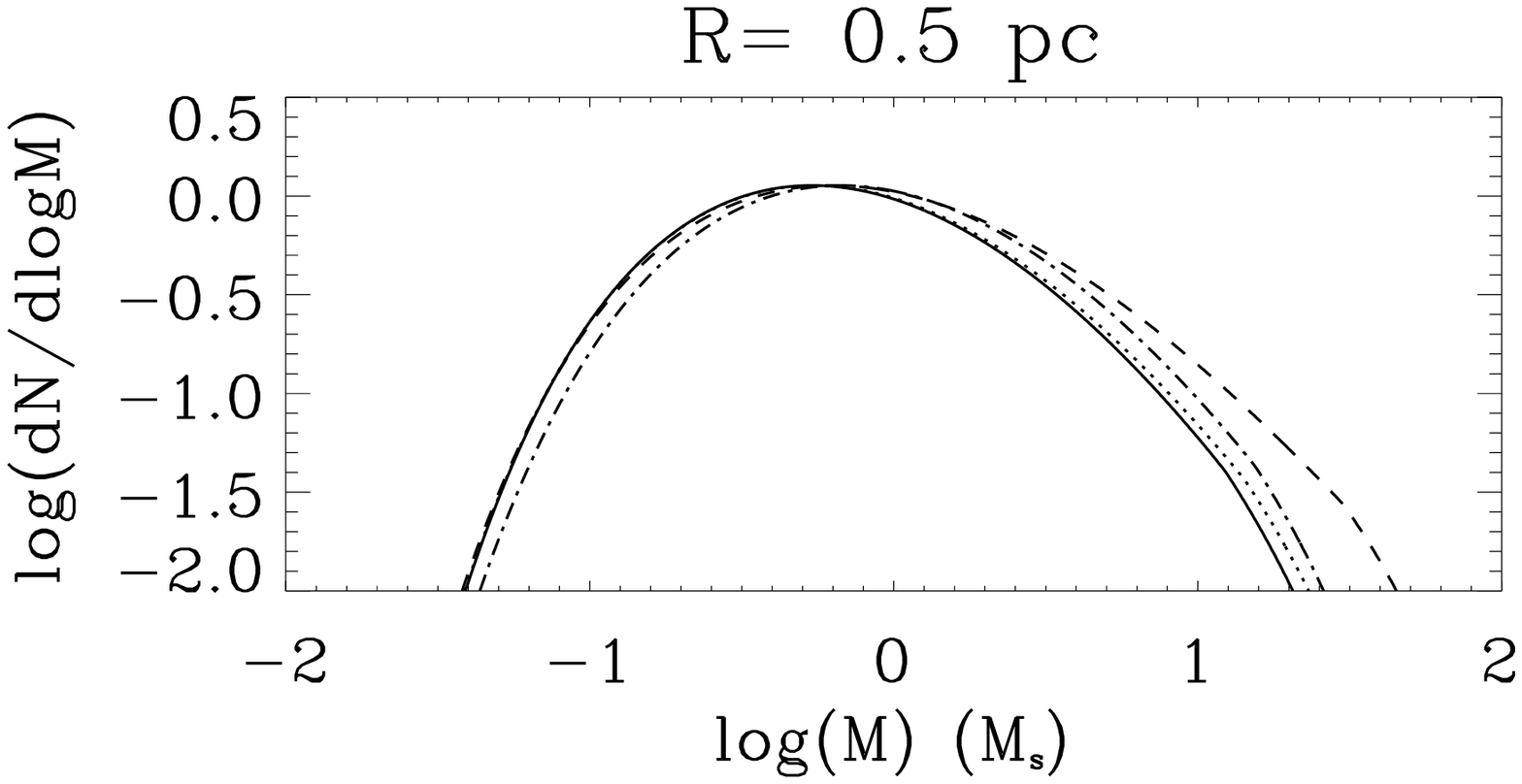}} 
\end{picture}
\caption{Barotropic and magnetized mass spectra for clumps of size $R_c=0.5$, $R_c=2$, $R_c=5$ and $R_c=20$ pc. The 
clump density and Mach numbers are given by Larson-type 
relations (eq.~(\ref{def_dens})) with $u_0=1$ and $d_0=5,4,3,2$ from top to bottom. Left panel:
the solid line corresponds to the time-dependent CMF
while the dotted line represents the time independent one. 
The dashed line is the Chabrier IMF shifted by a factor 3 in mass 
(see text). Right panel: the solid line is identical 
to the left panel. The dotted line corresponds to $V_a^0=1$, 
$\gamma _b=0.1$, the dashed line to $V_a^0=3$, 
$\gamma _b=0.1$ and the dot-dashed one to $V_a^0=1$, $\gamma _b=0.3$.}
\label{fig_non_iso}
\end{figure*}

When a barotropic eos is considered, the mass spectrum depends 
on the clump's mean density not only as a normalisation factor 
 (as for the isothermal eos) but also 
through the eos itself. As a fiducial value, we adopt a critical density (eq.~(\ref{const_K}))  
$\nbar^{crit}=2 \times 10^5$ cm$^{-3}$, i.e. $\rho^{crit}\simeq 10^{-18}\gcc$, as inferred from observations. The density and velocity dispersions are taken to
follow Larson (1981) (see also Falgarone et al. 2004) type relations
\begin{eqnarray}
\nbar &=& (d_0 \times 10^3 \; {\rm cm}^{-3})  \left( { R \over 1 {\rm pc}} \right)^{-0.7},\label{def_dens}\\
V_{\rm rms} &=& (u_0 \times 0.8 \; {\rm km \; s}^{-1}) \left( { R \over 1 {\rm pc}} \right)^{\eta}.
\label{def_dens2}
\end{eqnarray}

\noindent According to eq.~(\ref{larson}), $V_0=u_0 \times 0.8$ km s$^{-1}$.
Figure~\ref{fig_non_iso} (left panel) shows the mass spectrum for four clumps
of size $R_c =$0.5, 2, 5 and 20 pc. 
The value of $u_0$ is taken to be equal to 1 in all cases while 
the top, middle and bottom panels correspond to $d_0=$5, 4, 3 and 2,
 respectively.  The four panels thus correspond, from top to bottom, 
to typical clump masses and mean densities, M$_c$=183, 3564, 21995 and
355607 $\msol$,
 $\nbar=8120, 2460$, 970 and 240$\c3$.
As in Fig.~\ref{fig_isotherm}, solid lines display the time dependent case,
 dotted lines  the time independent one while the dashed 
line corresponds to the SCIMF.
In the four cases, except possibly for the smallest clump (as expected, see \S6.2.2 of paper II and \S\ref{cutoff}a below), the agreement with the SCIMF is very good. Importantly enough, 
the  clump densities found to yield a good agreement with the SCIMF are smaller by a factor of about 2-5
than what has been found in the time-independent case for similar agreement (Fig.~8 of paper II) and thus agree fairly well with the "standard" Larson density normalization values, $d_0\simeq 3$. 
This is a direct consequence of the peak position being shifter toward 
smaller masses, as discussed earlier. Another noticeable improvement is the width of the distribution, 
which is larger than for the isothermal cases presented in Fig.~\ref{fig_isotherm}. As discussed in paper II, this improvement  stems from the larger compressibility of the gas 
for a softer than isothermal eos, promoting small-scale collapsing structures, and from the strong dependence of the peak position on $\gamma$ (see fig.~3 of paper II), which becomes  even more
acute in the time dependent 
CMF, whose peak is shifted toward  smaller masses compared with the time-independent solution.

\subsubsection{Influence of the magnetic field on the core mass function}
\label{magfield}

To study the influence of the magnetic field on the mass spectrum, we 
define 
\begin{eqnarray}
V_a = V_a^0 {B_0^{ref} \over \sqrt{4 \pi \bar{\rho} }} \left( { \bar{n} 
\over 1000 \, {\rm cm}^{-3}} \right)^{\gamma_b},
\label{mass_spect_mag}
\end{eqnarray}
where $B_0^{ref}$ is a reference magnetic intensity equal to 10$\mu$G, $\bar{n}$ 
is the cloud mean particle density and $\bar{\rho}=m_p \bar{n}$.
We consider four cases, hydrodynamical (which serves as a reference), 
$\gamma_b=0.3$ with $V_a^0=1$, and $\gamma_b=0.1$ with $V_a^0=1$ and 
$V_a^0=3$. We investigate primarily values of $\gamma_b$ smaller than 0.5 because 
our analysis consists in analyzing the density fluctuations generated 
by supersonic turbulence and to identify the ones which are self-gravitating 
and that subsequently will be amplified by gravity. Therefore, at the stage 
where the analysis is performed, the correlation between 
magnetic intensity and density is the result of turbulent processes 
rather than gravity,  implying $B \propto \rho^{0.1-0.3}$. 
Note that, as emphasized in Hennebelle \& Chabrier (2008), the 
case $\gamma_b=0.5$ is formally equivalent to the isothermal case and 
all the magnetic results can be obtained by a simple renormalisation  of the rms velocity and Mach number (see eqn.(16)).

The impact of the magnetic field on the mass spectrum can 
  be inferred from a comparison between the various 
magnetized cases  and the hydrodynamical one 
(Figure~\ref{fig_non_iso}, right panel).
When $V_a^0=1$ and $\gamma_b=0.3$ (dot dashed line) the mass spectrum is essentially 
shifted toward larger masses with respect to the purely hydrodynamical 
case. This stems from the magnetic support, as expressed by eq.~(\ref{crit_Mtot}),
which adds up to the thermal pressure. Indeed, for this value of 
$\gamma_b$, the magnetic pressure  density dependence, $P_{mag} \propto \rho^{0.6}$,
 is very similar to the thermal pressure density dependence, in particular because 
the Larson eos has an effective exponent at low density which is equal to 0.7.

When $\gamma_b=0.1$, the impact of the magnetic field is different.
For $V_a^0=1$ (dotted line), the departure from  the hydrodynamical case (solid
line) is only marginal. This is because the magnetic support increases
very slowly with density; thus at high density (which corresponds
to low mass cores), the magnetic support is less important than the thermal one.
At high masses, on the other hand, the support is dominated by turbulence.
For $V_a^0=3$ (dashed line), the distribution 
tends to be shallower at high masses. As discussed in paper II, 
the exponent of the mass distribution at high masses, ${\cal N}(M)\propto M^{-(1+x)}$, when
  only pressure terms are included  (${\cal M}_\star=0$) is given by (see eq.~32 of paper II)
$1+x=(9 - 6 \gamma)/(4 - 3 \gamma)$,  so that if $\gamma <0.2$ the slope becomes shallower than the Salpeter value.

 Overall, for a typical magnetic field intensity {\it in the diffuse (low density) gas} of the order of  $B \simeq 10 \, \mu$G, and a density dependence such that
$\gamma_b < 0.3$, the mass spectrum is not very different from the 
hydrodynamical case. For $\gamma_b = 0.3$, the CMF still resembles the SCIMF but is shifted toward larger masses, a direct consequence of the magnetic support, which shifts the characteristic Jeans mass to larger values.
When the magnetic field intensity is significantly larger than the aforementioned value (e.g. 30 $\mu$G, like in the considered exemple), however,  
the mass spectrum becomes too shallow at large masses and starts departing significantly from a Chabrier/Salpeter IMF in this regime. Therefore, one of the predictions of our theory is that diffuse magnetized environments with magnetic field
intensities largely exceeding about $10 \, \mu$G should have high-mass IMF tails shallower than the Salpeter value.

\section{Star formation rate}
In this section, we derive the SFR from our time-dependent analytical theory  of star formation described in the previous section,
analyze its dependence upon various parameters and clump properties
and compare the results   with  recent observational determinations.

\subsection{Theoretical considerations}
\label{sfr_sec}


Following Krumholz \& McKee (2005, KMK), 
we define the dimensionless 
{\it star formation rate per free-fall time}, $SFR_{ff}$, as 
the fraction of clump mass converted into stars
 per clump {\it mean} free-fall time, 
i.e. the free-fall time defined at 
the clump mean density $\bar{\rho}$:
\begin{eqnarray}
SFR_{ff} = {\dot{M}_*\over M_c} \, \tau_{ff}^0,
\label{sfr_def}
\end{eqnarray}
where $\dot{M}_*$ denotes the {\it total star formation rate} arising from a clump
of mass $M_c$ and volume $V_c\simeq R_c^3$. The {\it star formation efficiency} (SFE),
is defined as the global mass fraction of a clump converted into stars during the lifetime $\tau_0$ of the clump, which can last a few free-fall times (e.g. Murray 2011):
\begin{eqnarray}
SFE = {{M}_*\over M_c}=SFR_{ff} \times ( {\tau_0\over \tau_{ff}^0}).
\label{sf}
\end{eqnarray}

According to these definitions, $SFR_{ff}$ is thus given by the integral of the mass spectrum specified by 
eq.~(\ref{spec_mass1}), as
\begin{eqnarray}
\nonumber
SFR_{ff} &=&
 \epsilon \, \int _0 ^{M_{cut}}  { M {\cal N} (M) dM \over \bar{\rho}}
\\
&=& \epsilon \int _0 ^{M_{cut}} dM  
{dR \over dM} \,
\left( -{d \delta_R \over dR} {\tau^0_{ff} \over \tau_R} \exp(\delta_R) {\cal P}_R( \delta_R) + 
\int _ {\delta_R}^\infty {d \over dR} [{\cal P}_R  ({\tau^0_{ff} \over \tau_R})]\, e^{\delta_R} d\delta \right)
\label{sfr}
\end{eqnarray}
In this expression $\epsilon$ is the (supposedly uniform) efficiency with which the mass within 
the collapsing prestellar cores is converted into stars\footnote{Note the typo in eq.(7) of Hennebelle \& Chabrier (2011) where it should simply be $dM$ and not ${dM \over M}$ in the integrand.}. Indeed, during the collapse, a substantial fraction of the mass initially within the core
is blown away by outflows and jets. Calculations (e.g. Matzner \& McKee 2000, 
Ciardi \& Hennebelle 2010), as well as observations, suggest that 
\begin{eqnarray}
\epsilon \simeq 0.3-0.5.
\end{eqnarray}
In  eq.~(\ref{sfr}), $M_{cut}$ corresponds to the mass of the largest star that can
possibly form in the cloud. 
 Equivalently, according to the mass-scale relation given by eq.~(\ref{mass_rad}) or (\ref{mass_rad2}), one can define $y_{cut}=R/R_c$, which denotes the largest size fluctuations that can turn
unstable in the cloud.
The value of $M_{cut}$ is crucial in setting $SFR_{ff}$.
Indeed, as emphasized in paper I (\S5.1.3), the integral
$V_c \int _0 ^ \infty M {\cal N}(M) dM$
is equal to the mass of the system itself because integrating 
up to infinity implies that {\it all pieces of gas}, including the very diffuse ones,
are Jeans unstable.
Therefore, restricting the integration to 
a finite value implies that any piece of fluid which is not dense enough, 
i.e. whose density corresponds to a Jeans length larger than a significant fraction of the 
cloud's size, is excluded from the mass spectrum of collapsing structures. The immediate consequence of such a truncation of the integral
is obviously to reduce the SFR. 

If $R_c$ is the clump's radius, more generally its characteristic scale, 
the question is thus to determine up to which fraction of this 
scale or up to which fraction of the clump mass, 
the integration should be performed ?
A similar question concerns the behaviour  of the  density power spectrum whose scale-dependence is characterized by $\sigma^2(R)$ 
(eq.~\ref{power_spec}).
While in the inertial domain, it is well established that the power spectrum 
of $\log(\rho)$ exhibits a power law behaviour (Beresnyak et al. 2005, Federrath et al. 2008, Audit \& Hennebelle 2010), 
this may not be the case 
when the scale $R$ is approaching the clump size $R_c$. This question is 
directly related to the very definition of the clump itself and how it 
connects to the surrounding medium. 
Unfortunately, these questions are far from being
settled. 
Finally, as discussed in paper I, the second term of the right hand side
in eq.~(\ref{sfr}) is important only when $y_{cut}\rightarrow 1$. Therefore, given all 
the uncertainties when approaching this limit and because the influence of this second term remains limited, 
we elected to drop it in the rest of the calculations. 
In that case, after proper normalisation, eqs.~(\ref{time}), (\ref{dens_crit}), (\ref{mass_spec}) and (\ref{sfr}) yield

\begin{eqnarray}
 SFR_{ff}^0 = -  \int _0 ^{\widetilde{M}_{cut}}
 \left( { \widetilde{M}_R^c \over \widetilde{R} ^3 } \right)
 {d \widetilde{R} \over d \widetilde{M}_R^c}  {d\, \delta_R^c\over d\widetilde{R}}  {1 \over \sqrt{2 \pi \sigma^2} } \exp \left( -{(\delta_R^c)^2 \over 2 \sigma^2} 
 - {\sigma^2 \over 8}  \right) d \widetilde{M},
\label{sfr_red}
\end{eqnarray} 
with
\begin{eqnarray}
SFR_{ff} = ({\epsilon \over \phi_{t}})\times SFR_{ff}^0.
\label{sfr_tot}
\end{eqnarray}
For the case of an isothermal eos, for which the mass-size relationship is given by eq.~(\ref{mass_rad}), this yields
\begin{eqnarray}
 SFR_{ff}^0 = -2  \int _0 ^{\widetilde{M}_{cut}}    \, { 1 \over \widetilde{R}^3} \,
{ 1 + (1 - \eta){\cal M}^2_* \widetilde{R}^{2 \eta} \over
[1 + (2 \eta + 1) {\cal M}^2_* \widetilde{R}^{2 \eta}] }
\times    \left( {\widetilde{M} \over \widetilde{R}^3}  \right) ^{-{1 \over 2 \sigma^2} \ln ({\widetilde{M} \over \widetilde{R}^3}) }
\times {e^{( -{\sigma^2\over 8} )   }\over  \sqrt{2 \pi}\,\sigma}\, d \widetilde{M}.\nonumber \\
\end{eqnarray}
 Conversely, the SFR can also be derived from eq.~(\ref{sfr}) 
(still dropping the second term) as:

\begin{eqnarray}
\nonumber
SFR_{ff}^0
 &=& \int ^\infty _{\widetilde{\delta}_{cut}} {\tau_{ff}^0 \over \tau_{R}(\delta_R) } 
e^{\delta_R} {\cal P}(\delta_R)\, d\delta_R \nonumber \\
 &=& \int ^\infty _{\widetilde{\rho}_{cut}} 
\widetilde{\rho}^{1/2}  {\cal P}(\widetilde{\rho})\, d\widetilde{\rho} \nonumber  \\
&=& {1 \over  2 } e^{({ 3 \sigma^2 \over 8  })}
\left[ 1 + {\rm erf} \left( { \sigma^2- \ln( \widetilde{\rho}_{cut} ) \over  
2^{1/2} \sigma }  \right) \right].
\label{sfr_simp}
\end{eqnarray}

We stress that {\it this is possible only because the second term in 
eq.~(\ref{sfr}) has been dropped}.
\noindent The value of $\widetilde{\rho}_{cut}$ in eq.~(\ref{sfr_simp}) can be derived from the collapse condition derived in our theory (see eq.(29) of paper I), namely $\widetilde{\rho}=e^{\delta_R}>e^{\delta_R^c}$, or similarly from the mass-size relations given by 
eq.~(\ref{mass_rad}) or (\ref{mass_rad2}). For the isothermal case, this simply yields

\begin{eqnarray}
\widetilde{\rho}_{cut}=\widetilde{R}_{cut}^{-2}\,(1+{\cal M}_\star^2\widetilde{R}_{cut}^{2\eta}),
\label{rhoc}
\end{eqnarray}
where $\widetilde{R}_{cut}=(y_{cut}R_c/\lambda_J^0)$. As clearly expressed by this equation, physically, $\widetilde{\rho}_{cut}$ is the minimum density for which a perturbation, whose associated {\it total} (thermal + turbulent) Jeans length is equal to a maximum fraction $y_{cut}$ of the clump's size, can lead to a gravitational instability. Indeed, eq.(\ref{rhoc}) can be rewritten:

\begin{eqnarray}
(\lambda_{J})_{cut}={ y_{cut}R_c\over [1+{\cal M}_\star^2\,({y_{cut}R_c\over \lambda_J^0})^{2\eta}]^{1/2} }
\label{lcrit}
\end{eqnarray}

\noindent It is important to stress that this procedure differs  not only quantitatively but also qualitatively from the ones defined in KMK
 and PN. In these author formalisms, the critical density corresponds to a {\it new}, arbitrary scale, respectively a sonic scale or a shock  scale, characteristic of a process supposed to be necessary for the collapse. In contrast, in our formalism, the only relevant scale  is the size of the cloud itself. So, even though this can be expressed as a density in eq.~(\ref{sfr_simp}), the proper way to look at it, as clearly expressed by eq.~(\ref{lcrit}), is in terms of the
maximum size of the fluctuations, for a given cloud's size and density, which are able to grow and lead to gravitational instability. 
 Moreover, in the KMK and PN formalisms, the critical density is proportional to ${\cal M}^2$, implying that only very dense structures will lead to star formation. In contrast, in our formalism, any structure of any density can collapse, if its gravitational energy dominates over all sources of support, as long as the associated perturbation can grow and become unstable. In practice, our results do not depend significantly on the value of $y_{cut}$, except when $y_{cut}\rightarrow 0$ and for low Mach numbers (see next section). Indeed, $y_{cut}$ only affects the limit of the integration, which corresponds to a regime where the (log normal) PDF is small.

As defined above, $SFR_{ff}^0$ represents the SFR calculated 
from our theory for a core-to-star mass conversion efficiency $\epsilon=1$ during the collapse,
and for a characteristic time within which new mass reservoirs $\widetilde{M}$ of scale $\widetilde{R}$ become gravitationally unstable, i.e. new cores are produced,
equal to one free-fall time {\it at the core's density}, i.e. $\tau_R=\tau_{ff}(R)$ ($\phi_t=1$).
In that sense, $SFR_{ff}^0$ represents the "core formation rate" per free-fall time. As already mentioned, observations and simulations point to
$\epsilon \simeq 0.3-0.5$ while our estimate of 
$\tau_R$ in \S\ref{crossing_sect} leads to $\phi_t \simeq 3$.
Consequently, we typically have
\begin{eqnarray}
 {SFR_{ff}\over SFR_{ff}^0 }\simeq 0.1-0.2.
\end{eqnarray}

\subsection{Results. Dependence upon clump properties}
\subsubsection{Models and assumptions}

A detailed comparison with the theoretical SFR's derived by Krumholz \& McKee (2005, KMK) and Padoan \& Nordlund (2011, PN)
has been presented in Hennebelle \& Chabrier (2011, HC11) and thus will not be repeated here. As mentioned in this paper, the key
differences between these two theories and the present one are twofold. 
First, both the KMK and PN theories assume that the SFR is determined by one single typical free-fall time, defined as the free-fall time at the cloud mean density, $ \tau_{ff}^0$. Given the very clumpy nature of molecular clouds and the enormous range of density fluctuations present within these entities, defining one single mean free-fall time for star formation seems to be hardly justified. This point becomes particularly accute given the fact that there is no evidence that most of the regions within molecular clouds are collapsing, and if so that they are collapsing at $\tau_{ff}^0$ (see e.g. Kennicutt \& Evans 2012).  
The second essential difference resides in the fact that both KMK and PN assume an {\it ad-hoc} critical density  $\rho_{crit}$, for star formation to occur (see HC11). Their SFR is thus simply obtained by  estimating the gas fraction with density larger than $\rho_{crit}$. In contrast, in our theory, the free-fall time density dependence of {\it any collapsing structure} of density $\rho_R$
is properly accounted for as the crossing time 
$\tau_R$, proportional to $\rho_R^{-1/2}$, consistently  varies with $M$ and $R$. Therefore, a dense core of density $\rho$ collapses in a {\it density-dependent} free-fall time $\tau_{ff}(\rho)$, which can differ significantly from the clump's mean free-fall
time. In our theory, there is 
{\it no particular scale or critical density}, as we sum up
over {\it all gravitationally unstable overdense regions}, whatever their scale or density (see  eqns.~(\ref{bal_mass})-(\ref{spec_mass1})).
This is indeed expected if the density fluctuations are induced by turbulence, which is by essence a multi-scale process,
and by the fact that any piece of fluid can  collapse if its  (turbulent+thermal) kinetic energy is dominated by its gravitational energy.

Comparisons between eqns.(\ref{sfr_red}), (\ref{sfr_simp}) and the SFR's of KMK and PN have been presented in HC11 and will not be repeated here. It was shown in this paper 
that the KMK SFR differs by  more than one order of magnitude from the other ones, underestimating the observed SFR by a similar amount, as already noticed by Heiderman et al. (2010). The PN SFR and the ones given by  eqns.(\ref{sfr_red}) and (\ref{sfr_simp}) agree within a factor $\sim$2-3, {\it provided} the (${\epsilon  /  \phi_t}$) factor is properly included in the PN relation (see HC11). 
The main reason why the present determinations yield a larger SFR than KMK\footnote{besides the fact that KMK further assume that the critical free-fall time is equal to the free-fall time of the cloud, which implies that $\widetilde{\rho}_{crit}=1$, a very consequential issue, as shown in HC11.}
 and PN is that, when properly  accounting for the density-dependence of the collapsing structure free-fall times, denser  fluctuations, in particular the ones denser than a given arbitrarily defined density $\rho_{crit}$, collapse faster than lower density ones, leading to a larger formation of low-mass cores.  Although qualitatively similar to the PN results, however, ours still differ from these latter  (see Fig. 1 of HC11). Indeed,
while  $SFR_{ff}^0$ tends to decrease steadily
with increasing virial parameter, $\alpha_{{\rm vir}}=2E_{{\rm kin}}/E_{{\rm grav}}$, in the PN theory, 
it exhibits a more flattish (nearly constant) behaviour before decreasing more steeply at high $\alpha_{{\rm vir}}$ in our case, as will be examined in details below. 
The reason stems from the gas more diffuse than 
$\rho_{crit}$ not being taken into account in the PN
model while being accounted for, if indeed collapsing, in  our model. As mentioned above, in our theory, if a
piece of fluid of even very low density (e.g. smaller than $\rho_{crit}$ in PN) has a size significantly larger 
than the (thermal or turbulent) Jeans length at its density, it is subject to collapse and thus
must  be taken into account. This point will be crucial when comparing to SFR observational determinations at low density (\S\ref{obs}). 
As mentioned earlier, this is a fundamental difference between our and both KMK and PN theories.

\subsubsection{Star formation rate: results}

We now examine the dependence of our theoretical SFR upon various clump 
characteristic parameters and physical properties.

\begin{figure*} 
\begin{picture} (0,15)
\put(0,0){\includegraphics[width=15cm]{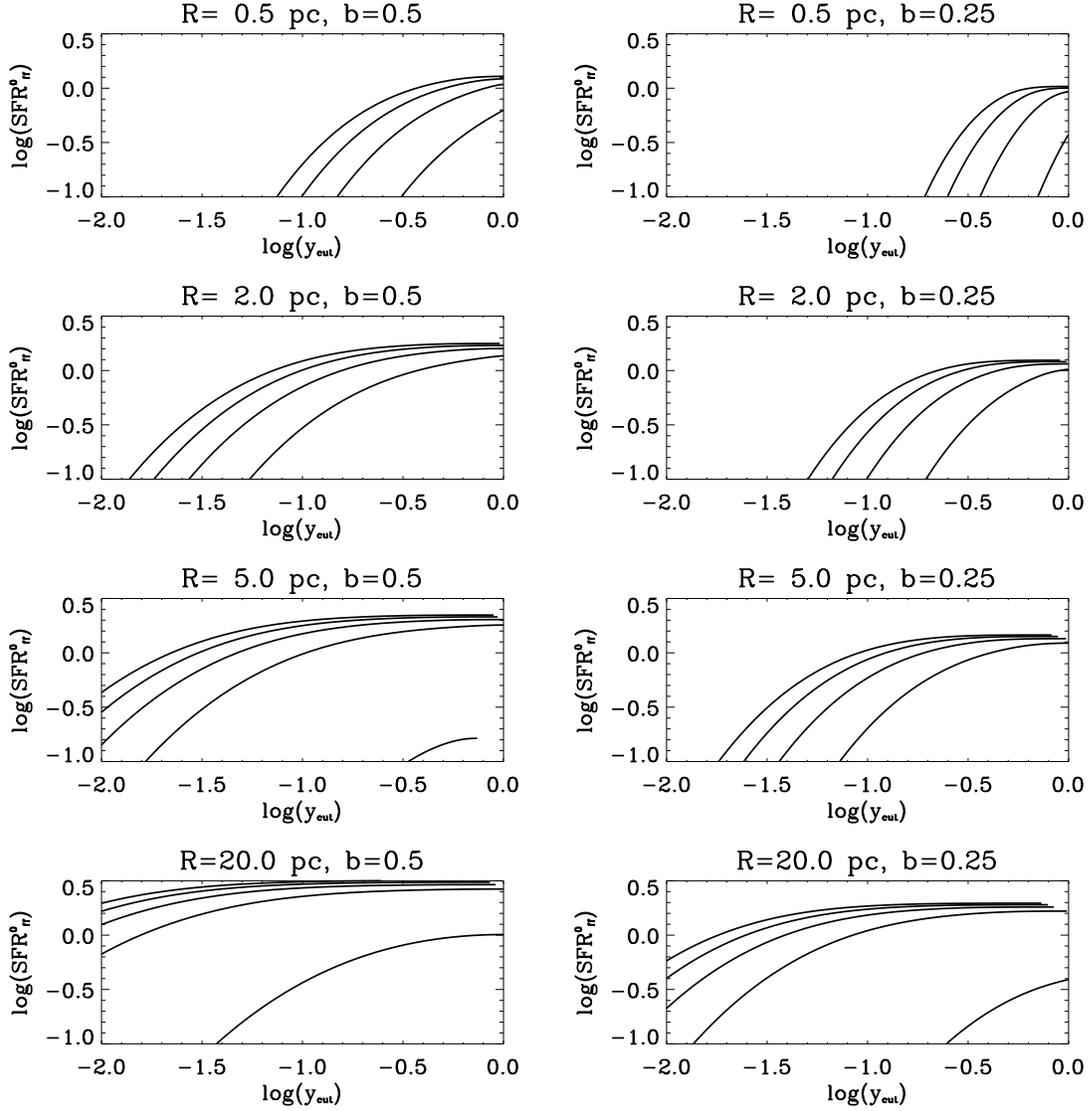}} 
\end{picture}
\caption{Value of $SFR_{ff}^0$ for various clump sizes and densities, namely 
 (from left to right) $d_0= 4.5, 3, 1.5, 0.7, 0.1$ in the non-isothermal 
case. Left column: $b=0.5$, right column: $b=0.25$.}
\label{fig_sfr_hym}
\end{figure*}

\subsubsubsection{(a) Dependence upon the clump's large-scale cut-off, 
size and density}
\label{cutoff}

Figure \ref{fig_sfr_hym} portrays $SFR_{ff}^0$, for the non-isothermal case, 
for $b=0.5$ (left column) $b=0.25$ (right column) 
in eq.~(\ref{base}), as a function of the cut-off parameter, $y_{cut}=R/R_c$,
 for various clump  
sizes and mean densities (this means that the parameter $d_0$,
which determines the density normalization at 1 pc in the Larson relation 
(eqn.~(\ref{def_dens})), is constant along each line but varies from one 
line to an other). 
As mentioned earlier and expected from the truncation of the integral in 
eq.~(\ref{sfr_red}), $SFR_{ff}^0$ is basically independent of $y_{cut}$ 
for $y_{cut}\ga 0.3$ but
quickly drops as this parameter approaches very small values. 
As also expected, the strong decrease of $SFR_{ff}^0$ occurs at higher  
 $y_{cut}$ values for
smaller, less massive clumps and the truncation becomes more drastic as 
density decreases. Indeed,  as already noticed in papers I and II 
(see Fig. 8 of paper II), 
the core mass spectrum is truncated both at high and small masses compared
 with the observationally determined CMF/IMF for small ($\la 0.5$ pc)
clumps, with a drastically reduced CMF for low-density clumps.
This stems from two reasons: (i) the scale-dependence, based on Larson's
 relations, of both the global and local Mach numbers, ${\cal M}$ and 
${\cal M}_\star$, which enter our formalism (see eqns.~(\ref{larson})
 and (\ref{mass_star})) and are responsible for generating and stabilizing
 the initial density fluctuations leading eventually to prestellar cores.
 The smaller the clump the smaller these values and thus the narrower 
the core mass spectrum (see paper I); (ii)  the fact that for small
 clumps or low-density clumps, the Jeans scale becomes comparable to 
or larger than the size of the clump itself (see eq.(\ref{ljeans}), 
inhibiting gravitational collapse. 

\noindent According to these results, star formation is thus predicted to occur dominantly in the largest, most massive clumps and/or in the densest (parts of) clumps, a conclusion indeed supported by observations. 
As seen in the figure, for  densities $\bar{n} \ga 2000$
cm$^{-3}$ and for $y_{cut}\ga 0.1$-0.3, depending on the clump's size, the SFR depends only weakly on $y_{cut}$ and reaches values in the range
$SFR_{ff}^0 \simeq 0.3$-$3.0$, depending on the size of the clump, for the  $b=0.5$ case. The fact that 
$SFR_{ff}^0$, the SFR per free-fall time,
 can be larger than 1 is due to its density dependence. 
 Indeed, $SFR_{ff}^0$ is defined with respect to $\tau_{ff}^0$, the free-fall time 
at the cloud's mean density, but fluctuations of 
size $R$ whose density is larger than 
$\bar{\rho}$ collapse in a shorter time. As already discussed, the value of $y_{cut}$ beyond which no star 
can form is ill-determined and may depend 
upon cloud parameters. In the rest of our calculations, we will assume 
$y_{cut} \simeq 0.1-0.3$ (see previous section), which means that only perturbations 
whose size is at most of the order of  about one tenth to one third of the cloud's size 
are relevant to produce gravitationally bound prestellar cores. As just mentioned, for larger values of $y_{cut}$, the SFR remains basically unchanged.

As seen from Fig. \ref{fig_sfr_hym}, low-density clumps ($\bar{n} \la 1500 $cm$^{-3}$) smaller than about $\sim 2$ pc
 yield quite small or even negligible values of $SFR_{ff}^0$, except for values of $y_{cut}$ approaching unity, a rather unlikely possibility. Phrased differently, at low density very large clumps, but {\it only} very large clumps still contribute appreciably to star formation. As mentioned in the previous section, this is one of the direct consequences of our theory and can not be the case in a theory arbitrarily defining a density threshold for star formation.
  A point of importance when we will compare with observational determinations (\S\ref{obs}).
As just mentioned, star formation should thus occur dominantly in dense enough (regions of) clumps or in very large and massive clumps and to increase with clump density and clump mass/size. It is worth stressing that this strong dependence upon density or mass for efficient star formation naturally arises in our formalism and does not stem from an ad-hoc threshold condition for star formation.

Finally, the figure shows the dependence of the SFR upon the coefficient $b$ entering eq.(\ref{base}),
which relates the rms Mach number to the width of the density PDF, $\sigma_0$ (see eq.(\ref{base})). Although the results are qualitatively similar, globally, $SFR^0_{ff}$ is  smaller by a 
factor of a few for large values of $y_{cut}$ and drops drastically
below  larger values of $y_{cut}$ for $b=0.25$ than for $b=0.5$. Indeed, as
shown in details in \S3.1 of   paper II, the mass spectrum peaks at larger 
masses and the density PDF gets narrower for smaller values of $b$, because the gas is less compressible. Since  the respective contributions of the  turbulence forcing mechanisms (and thus the value of $b$) in molecular clouds are
 rather ill-determined and may vary from cloud to cloud, depending for instance on the presence of expanding HII regions around massive stars or supernovae explosions,
this implies significant possible variations of the SFR, depending on the environment.

\subsubsubsection{(b) Dependence upon the virial parameter}
\label{vir}

\begin{figure*} 
\begin{picture} (0,15)
\put(0,0){\includegraphics[width=15cm]{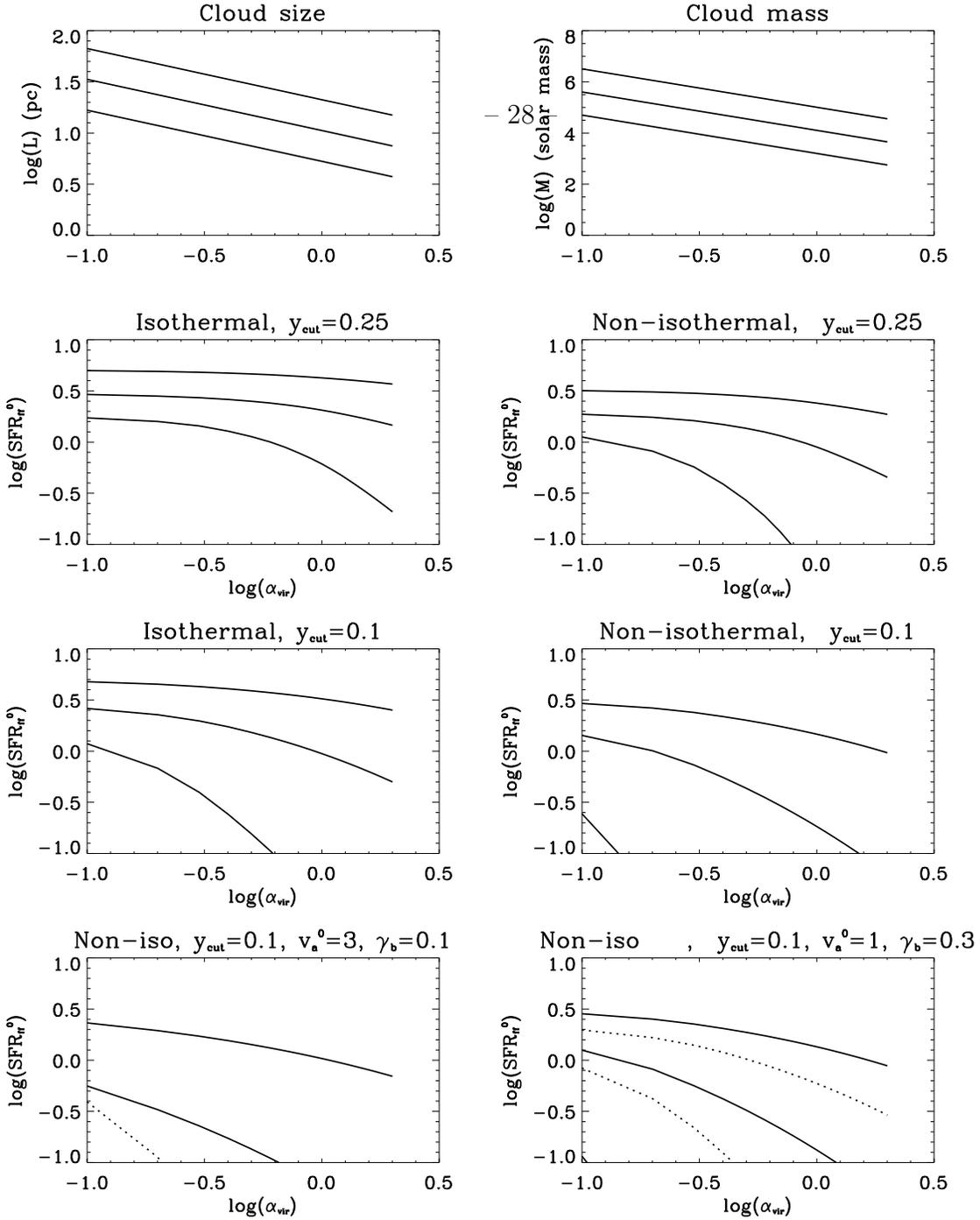}} 
\end{picture}
\caption{Values of $SFR_{ff}^0$ as a function of  $\alpha_{vir}$ 
corresponding to various cloud parameters and  ${\cal M}$=16 (top),
 9 (middle) and 4 (bottom), in the isothermal, non-isothermal, 
hydrodynamical and magnetized 
  cases, for 2 values of $y_{cut}$, as indicated on the figure, 
and our fiducial value $b=0.5$.  For the magnetized cases (right and left bottom panels), the solid lines correspond to the case where the magnetic field is not supposed to affect the density PDF, while the dotted lines display the results when the effect of the field on the width of the PDF is taken into account (see text).
}
\label{fig_sfr_comp}
\end{figure*}

The dependence of the SFR upon
the characteristic virial parameter $\alpha_{vir}$ of
 a  clump  of  radius $R_c=L_c/2$ and mean density ${\bar \rho}$,
$\alpha_{vir}=2E_K/E_G=(5/3\pi) V_{{\rm rms}}^2 / (\pi G {\bar \rho} L_c^2)$,
  which
measures  the ratio of turbulence over gravitational energy within the clump,
 has been examined in HC11 in the isothermal case and will be examined in the
 next subsection in the
non-isothermal one. It is illustrated in Fig. \ref{fig_sfr_comp}, which 
 displays the values of $SFR_{ff}^0$   for various clump sizes/masses, 
for three typical Mach numbers, ${\cal M}$=16 (top), 9 (middle) and 4 
(bottom). As a general trend, as discussed in HC11,
increasing $\alpha_{vir}$ leads to a decrease
of the SFR, with an abrupt reduction above some typical value of 
$\alpha_{vir}$, which depends on the Mach number. This stems from the
fact that as $\alpha_{vir}$ increases, the increasing contribution of kinetic
 energy over potential energy prevents gravitational collapse and thus
 inhibits star formation. As increasing $\alpha_{vir}$ implies decreasing 
the clump's mass/size, 
as $\alpha_{vir}\propto L_c^{-2}\propto M_c^{-2/3}$ 
(at fixed Mach number and density),
star formation is thus highest in the largest and most massive clumps, which
  correspond to the smallest values of $\alpha_{vir}$, in agreement with the
 conclusion reached in the previous subsection.
Giant molecular clouds in the
Milky Way, with masses  in the range $\sim 10^3$-$10^6\,\msol$, are generally
 weakly gravitationally bound structures, with virial parameters in the range
 $\alpha_{vir}\approx 0.3$-3, with slightly decreasing $\alpha_{vir}$ with
 increasing cloud mass, as just noted (Heyer et al. 2009, Murray 2011). 
Accordingly, the rather low observed SFR values
might partly stem from the fact that most molecular clouds have in general
 virial parameter values slightly above unity, i.e. depart from perfect
 virialization, being held together partly by the confining ram pressure
of turbulent flows in the ISM. 

\subsubsubsection{(c) Thermodynamics of the gas} 

 For the non-isothermal case, we need to specify
a temperature-density distribution (see eq.~(18) of paper II).
 We choose $n_0=200$ cm$^{-3}$ and 
$T_0=20$ K at $10^4$ cm$^{-3}$. Assuming that the velocity 
dispersion remains unchanged, we recalculate 
accordingly  the speed of sound
and the Mach number in the non-isothermal case (see eq.~(20) of paper II). 

As seen in Fig.~\ref{fig_sfr_comp},
taking into account the thermodynamics of the gas yields values of 
$SFR_{ff}^0$, for the same clump conditions,  smaller  by a factor of 
about $\sim$ 1.5-2 than the isothermal case in the nearly constant 
$SFR_{ff}^0$ region, which corresponds to small values of $\alpha_{vir}$, but
leads to a significantly steeper decrease for increasing values of the
 virial parameter, in particular for $\alpha_{vir}\gtrsim 1$. 
As explored in details in Paper II,
this is a direct consequence of  the density-dependence of the Mach 
number for non-isothermal gas.
Indeed, when taking into account the thermodynamics of the gas, the 
clump is warmer at low densities, which implies lower Mach numbers than 
for the isothermal case (see Paper II), decreasing the SFR, as seen in 
the figure. 

As seen in the figure, for low Mach numbers, which correspond to 
small-size clouds,  the value of $y_{cut}$ has a significant impact 
on $SFR_{ff}^0$, as already discussed in subsection (a), leading to 
substantial uncertainty on the SFR for such low-mass/small-size clumps. 
Fortunately,  such clumps are predicted to contribute almost negligibly 
to star formation, as mentioned earlier.

\subsubsubsection{(d) The role of magnetic field}

The two bottom panels of Fig.~\ref{fig_sfr_comp} show
the SFR in the two magnetic cases corresponding to $V_a^0=1$, $\gamma_b=0.3$
and $V_a^0=3$, $\gamma_b=0.1$, respectively, adopting $y_{cut}=0.1$. 

In order to decipher the  contributions of the various physical effects, we have calculated 
the SFR in the magnetized case first keeping the same PDF as in the hydrodynamical case (as
described by eq.~\ref{base}) but then we have also taken 
into account  the fact that the magnetic field 
can modify the width of the density PDF, as investigated for example in 
Molina et al. (2012). In a magnetized flow, the dependence upon the Alfv\'enic 
Mach number must be taken into account. 
 Based on their numerical simulations, Molina et al. (2012) propose an analytical relation to 
predict the dependence of the variance of the lognormal density distribution upon the rms Mach number in magnetized supersonic turbulent gas.
The result depends on the index $\gamma_b$. For $\gamma_b=0$, i.e. $B$ independent of the density, they recover the hydrodynamical dependence (eq. \ref{base}). When  $\gamma_b=$0.5, i.e. $B\propto {\sqrt \rho}$, they find that, in the super-Alfvenic case, the variance is 
given by

\begin{eqnarray}
\sigma_0^2 = \ln(1 + b^2 {\cal M}^2 {\beta \over \beta+1}),
\label{sigmag}
\end{eqnarray}
where $\beta=P_{th}/P_{mag}=2\, C_s^2 / V_a^2$.

The basic effect of the magnetic field on the SFR, when assuming 
no variation of the density PDF compared with the hydro case, is illustrated
by the solid lines in the bottom left and right 
panels, to be confronted with the SFR values displayed on the right third panel. As expected, the magnetic field tends to reduce the SFR by a factor 
of a few. The amplitude of this effect depends obviously on the field intensity, as explained in \S\ref{magfield}. 
For $V_a^0=1$, $\gamma_b=0.1$ (not displayed here), the SFR
is identical to the hydrodynamical case, while for $V_a^0=3$, $\gamma_b=0.1$
(left bottom panel), it is reduced by a factor of less than 2 for 
${\cal M}=16$ and  about 2 for ${\cal M}=9$. Stronger values of 
the magnetic field obviously lead to even lower SFR's. However, the values we used are within the range of the values typically 
measured in molecular clouds.

The dotted lines of bottom left and right panels show
the SFR when the influence of the magnetic field on the 
density PDF is taken into account, as described above. 
With our choice of parameters we find that at the mean
density $\beta=0.036$ for  $V_a^0=3$, $\gamma_b=0.1$
and $\beta=0.6$ for $V_a^0=1$, $\gamma_b=0.3$. The first value
in particular significantly reduces the width of the density PDF. 
Clearly the impact of the magnetic field is much more pronounced and the 
SFR  drops by a factor of several. We stress, however,
that, as discussed in Molina et al. (2012), the relation given by eq.~(\ref{sigmag})
only holds for $\gamma_b=0.5$ and is
is not working well at small and intermediate Mach numbers. Therefore,
there is still considerable uncertainties here due to the lack of 
exact knowledge of the density PDF in the presence of magnetic field and 
also of the precise gas density-magnetic intensity relation.

Finally, we also note 
that in the presence of a magnetic field, 
the coefficient $\phi_t$ could possibly be further increased, 
reducing further the SFR. Indeed, 
it is likely that the magnetic field and the velocity field 
of the perturbation that eventually gives rise to the gravitationally unstable 
perturbation must be sufficiently aligned. Otherwise, after a weak density enhancement, the magnetic 
pressure will probably stop the contraction.

\subsubsubsection{(e) The role of turbulence}

\begin{figure*} 
\begin{picture} (0,15)
\put(0,0){\includegraphics[width=15cm]{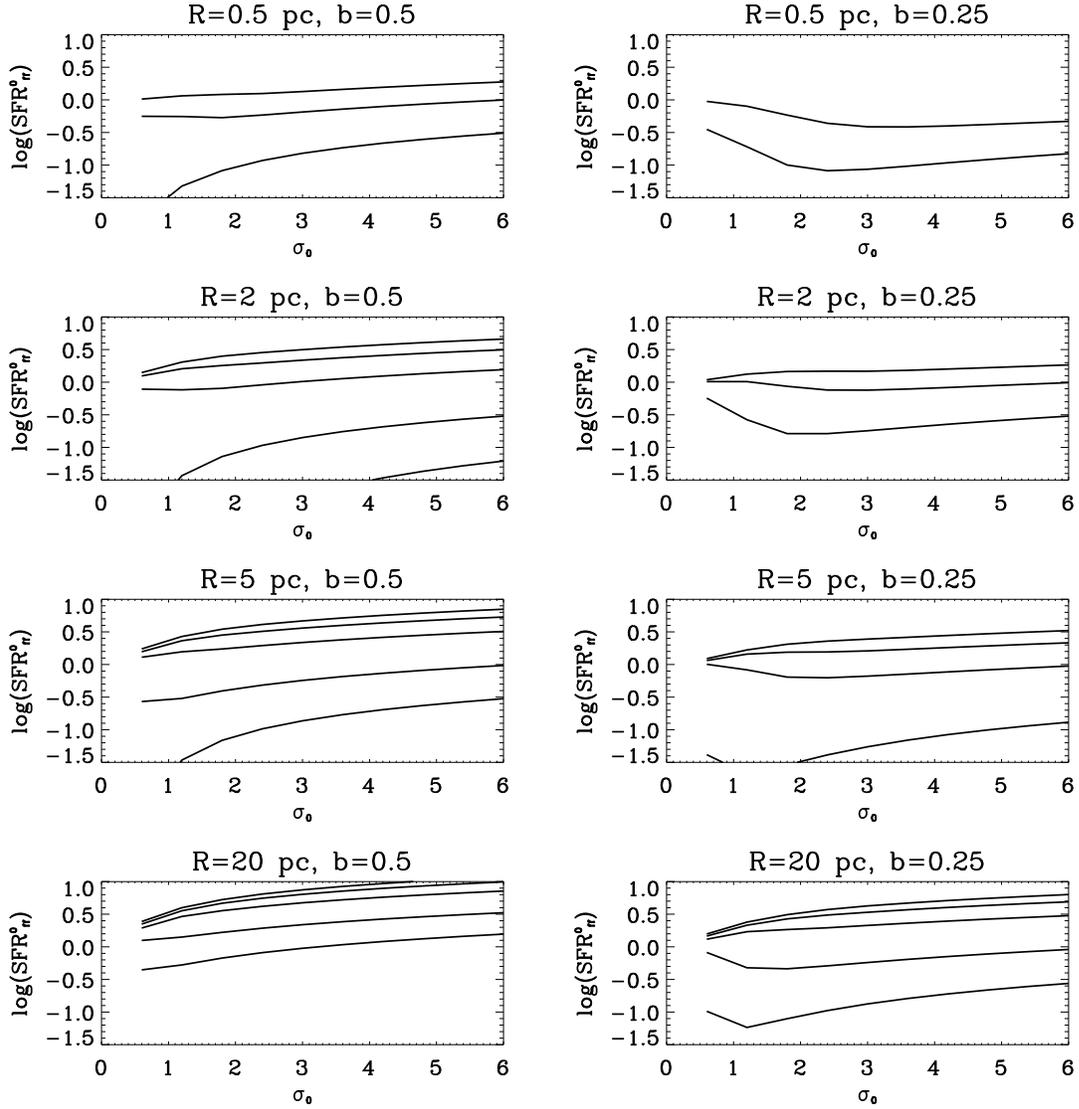}} 
\end{picture}
\caption{Value of $SFR_{ff}^0$ as a function of $V_0$ 
for various cloud sizes and densities, 
namely  (from top to bottom) $d_0=10, 5, 2, 0.5, 0.2$, in the non-isothermal 
case. Left column: $b=0.5$, right column: $b=0.25$.}
\label{fig_sfr_sig}
\end{figure*}

\begin{figure*} 
\begin{picture} (0,15)
\put(0,0){\includegraphics[width=15cm]{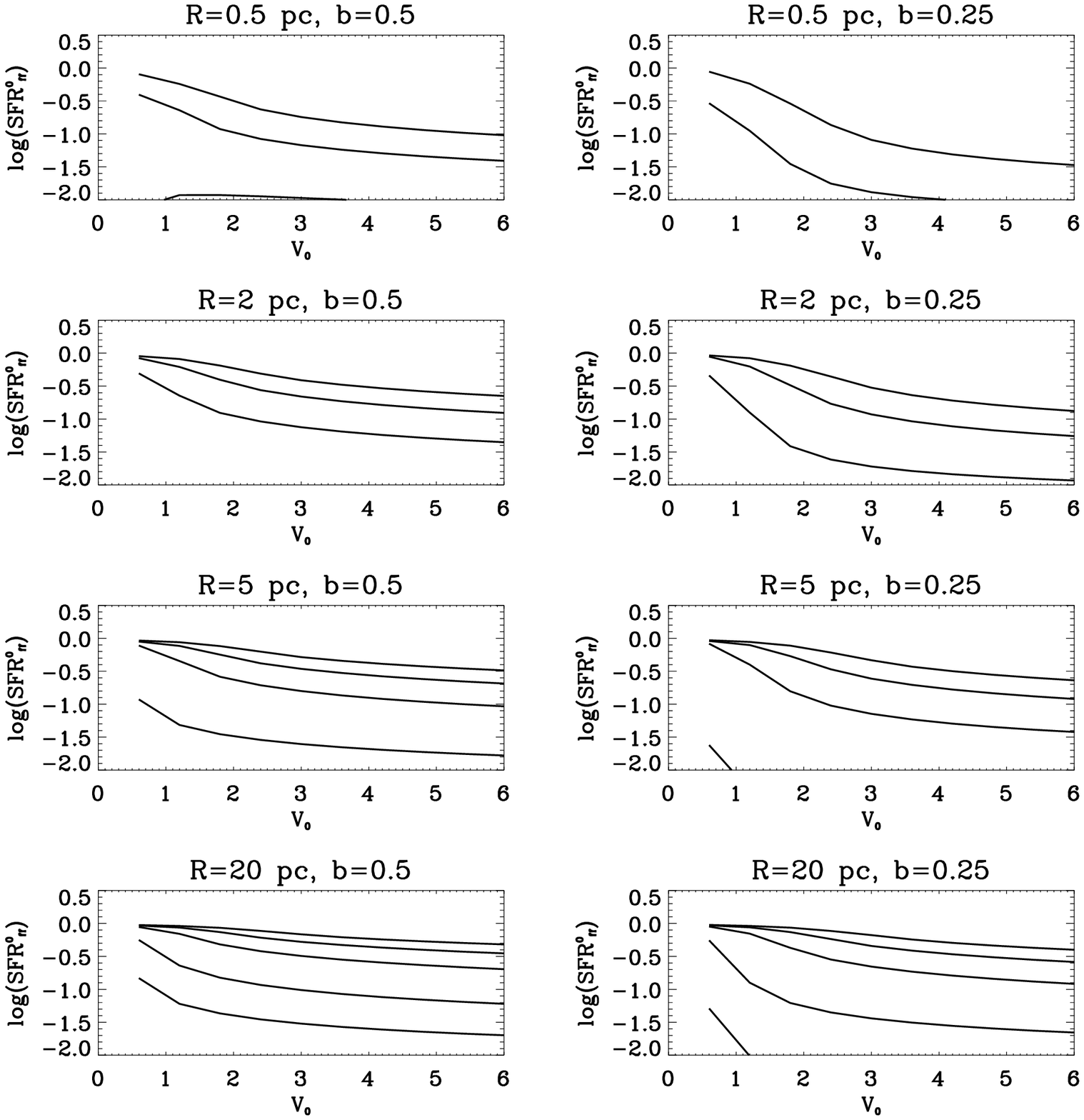}} 
\end{picture}
\caption{Same as Fig. \ref{fig_sfr_sig} when the freefall time dependence of the collapsing structures is not taken 
into account.}
\label{fig_sfr_sig_notime}
\end{figure*}

As in the theories of Krumholz \& McKee (2005), Padoan \& Nordlund (2011) 
 and Hopkins (2011), turbulence is at the heart of our star formation theory.
We have shown in Paper I (\S6.3) that, in the {\it  time-independent} version of our IMF theory, turbulence, globally, has a negative impact on star formation.
The stronger the turbulence, both at the global clump scale (as formalized by ${\cal M}$) and local (Jeans length) scale (as formalized by ${\cal M_\star}$), the smaller the global clump mass fraction  encapsulated into
gravitationally bound prestellar mass reservoirs. Indeed, although increasing the Mach number
increases the width of the PDF, promoting the formation of small-scale structures, it leads to a decrease of the number of collapsing structures around the peak of the CMF (see papers I and II).
As shown in the previous sections, however, star formation will keep processing during the clump's lifetime and time
dependence  changes the final clump's mass fraction ending up forming collapsing cores.
The impact of turbulence on $SFR_{ff}^0$ is  illustrated in Fig.\ref{fig_sfr_sig}. This figure portrays the
behaviour of $SFR_{ff}^0$ at {\it fixed cloud density} (${\bar n}$=constant along each  line) for different cloud sizes, as a function of $V_0$, thus of ${\cal M}$ (see eqn.(\ref{larson})), spanning a range ${\cal M}\simeq 4$-30.
The time-{\it independent} results are displayed  in 
Fig.~\ref{fig_sfr_sig_notime}. In this latter case, the SFR steadily 
decreases with increasing levels of turbulence, as mentioned above.
As shown in Fig.~\ref{fig_sfr_sig}, however,  
including time-dependence in our 
formalism significantly changes this behaviour, with the SFR now increasing,
 although modestly, with the level of turbulence.
 The reason is twofold. First, increasing the Mach number shifts the peak 
of the CMF towards smaller masses (see Paper I), a behaviour exacerbated when
 time-dependence is considered, as discussed in \S\ref{timedep}. Since
small-scale structures have shorter free-fall times
 ($\tau_R \propto \rho^{-1/2}$), this increases the number of small cores
 and thus globally increases the SFR (see eqn.~(\ref{sfr})).
Second,  massive mass reservoirs, considered to be stable in the
static case, can now fragment into small-scale structures as their internal 
turbulent motions lead to new small-scale overdense regions, globally 
increasing the number of collapsing cores. This positive impact of 
turbulence upon the SFR is in agreement with the
results of PN but contrasts with the ones of KMK. As demonstrated in HC11,
 this decreasing  dependence of the SFR with increasing Mach number in KMK
 theory (see their eq.(30)) stems from the missing 
$\widetilde{\rho}_{crit}^{1/2}$ term in   their eq.~(20), which stems from
 the fact that KMK assume that the critical free-fall time is equal to the
 free-fall time of the cloud. The dependence of the SFR upon $b$ illustrated
 in Fig.\ref{fig_sfr_sig} has already been addressed previously.

The fact that,
in the time-dependent formalism, turbulence is found to {\it enhance} 
star formation seems, at first glance, to be in contradiction with numerical simulations 
(see e.g. MacLow \& Klessen 2004). However, as discussed in 
HC2011, the behaviour we infer for the SFR is qualitatively similar to the 
one inferred by Padoan \& Nordlund (2011)
(see also  Federrath \& Klessen 2012).
 In their simulations, these authors find that, for a given $\alpha_{vir}$, the SFR 
increases with the Mach number.  
The critical point here is that when  turbulence is too high (i.e. 
$\alpha_{vir}$ is typically larger than 1), the kinetic energy 
will lead to an expansion
 of the clump, an effect which is neither included in the present
analytical calculations nor in the turbulence in the box numerical simulations
of Padoan \& Nordlund (2011).
We thus intuitively expect a non-monotonic behaviour 
of the SFR with turbulence. When $\alpha_{vir}$ is small, turbulence 
enhances star formation because of the presence of converging motions 
that lead to density enhancements. On the other hand, if turbulence becomes too
large   ($\alpha_{vir}\gtrsim 1$ or so), it triggers a fast expansion 
of the cloud which reduces the density and quenches
star formation.
(note that V\'azquez-Semadeni et al. 2003 do see such a non-monotonic
behaviour  with turbulence).

\subsubsubsection{(f) Summary of the results}

To conclude this section, we summarize the main results of our calculations at this stage.
As mentioned above and seen in Fig. \ref{fig_sfr_comp} and 
\ref{fig_sfr_sig}, for all typical clump conditions we have explored,
 which correspond to clump sizes $\sim$0.5-20 pc and masses 
$\sim 200$-$10^6\,\msol$, ${\cal M}=4$-30,  $SFR_{ff}^0$ lies in the 
range $\approx 0.01$-5 in the
non-isothermal case, even though the value of $y_{cut}$ significantly 
affects these values at low Mach. Lower values of $SFR_{ff}^0$
would imply either $y_{cut}$ of the order of 1/30 or even 1/100, which 
does not look very realistic, or very small clumps ($\la 0.1$ pc),
for which the mass spectrum is drastically truncated and   is completely
 incompatible
with the observed CMF/IMF (see Fig. 8 of paper II). 
According to
our calculations, the value $SFR_{ff}^0\approx 0.01$ thus seems to be a lower limit for realistic clump conditions. As shown above, the larger SFR values are reached for the densest clumps and the largest/most massive clumps, which correspond to the smaller values of the virial parameter and are thus likely to be gravitationally bound structures. Significant star formation, however, is still predicted to occur in low-density clumps, providing they are large enough (i) to generate enough turbulence levels (high enough Mach numbers) and (ii) to largely exceed their typical Jeans mass.
We must recall, however, that $SFR_{ff}^0$ is the SFR obtained assuming that self-gravitating 
fluctuations are replenished within one single free-fall time ($\phi_t=1$) and with a 100\% efficiency of initial mass reservoir to star conversion during the gravitational collapse ($\epsilon=1$).  
  $SFR_{ff}^0$ must thus be multiplied by a factor                   
 $\epsilon/\phi_t<1$ to yield the effective value $SFR_{ff}$.
 The dependence of these two parameters upon the local physical conditions such as magnetic
field intensity or radiative feedback remains very uncertain, but both processes tend to decrease the final star formation rate $SFR_{ff}$,
either by decreasing $\epsilon$ or by increasing $\phi_t$. As mentioned earlier, observational and numerical results suggest that $\epsilon/\phi_t\approx $0.1-0.2.
 Therefore, according to our calculations,
the star formation rate per free-fall time in dense molecular clumps with typical aforementioned properties
is estimated to lie in the range $SFR_{ff}\approx$0.001-1.0, for $y_{cut}=0.1$-0.3, for virial parameters in the range $0.3\la \alpha_{vir}\la 3$, in agreement with typical observed values (see e.g. Murray 2011 and \S\ref{obs} below). The calculations of a mean value of the SFR, integrating over the clump mass spectrum, will be addressed in \S\ref{clumpevol}.

\section{Comparison with observations}
\label{obs}

\subsection{Star formation rate versus gas surface density}

\begin{figure*} 
\begin{picture} (0,15)
\put(0,0){\includegraphics[width=15cm]{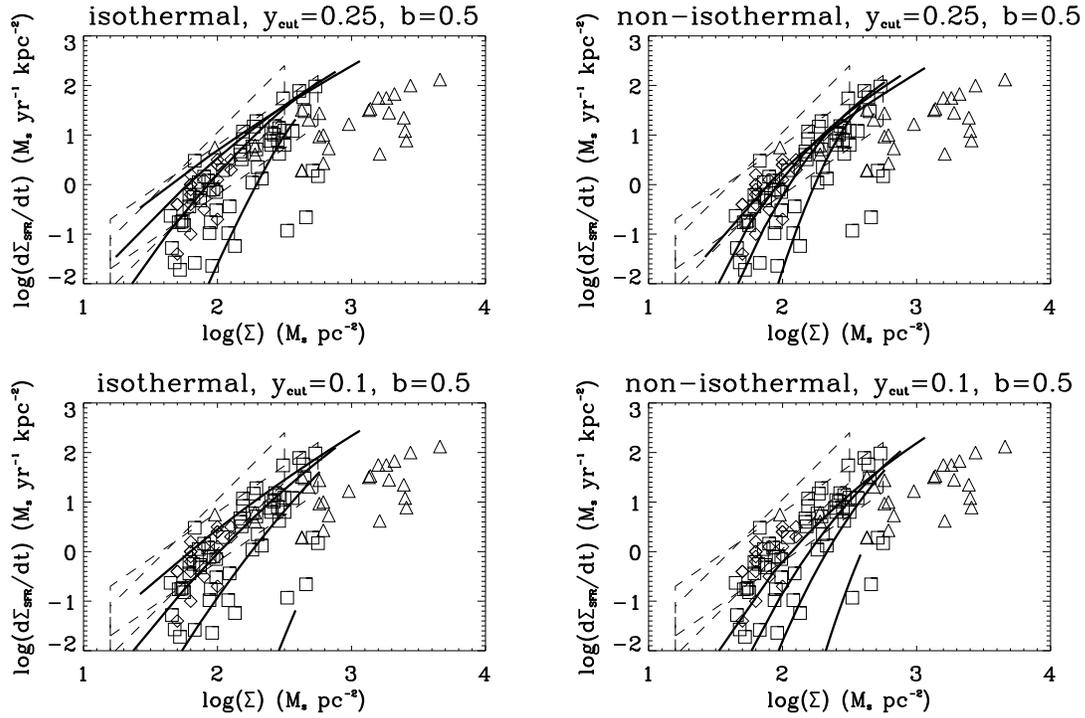}} 
\end{picture}
\caption{Comparison of the SFR per unit area, ${\dot \Sigma}_\star$, as a 
function of gas surface density, ${\Sigma}_g$, with various observational 
determinations: Heiderman et al. (2010) massive clumps (triangles) and 
molecular cloud YSOs (diamonds+squares), Gutermuth et al. (2011) 
(areas bracketed by dashed lines). The four solid lines correspond to four clump sizes, 
namely $R_c=0.5$, 2, 5 and 20 pc (right to left).}
\label{fig_sfr_obs}
\end{figure*}

As seen in the previous section, the SFR per free-fall time can vary by several orders of magnitude depending on the clump's conditions (mass, density, typical Mach number)
but also, unsatisfactorily enough, on the values of $y_{cut}$, $b$, $\phi_t$ and $\epsilon$. 
These large uncertainties prevent a precise theoretical determination of the SFR in a cloud. Nevertheless, it is instructive to compare our 
determinations with observational ones. Such a comparison has partly been done in HC11 and will be extended here.
The observed  star formation rate per free-fall time in star forming
molecular clouds in the Milky Way lies in the range $0.03\la   SFR_{ff}\la 0.3$ for clouds in the mass range $10^3\la  M_c/M_\odot \la 10^6 $, with ${\cal M}\approx 10$-15 and $\alpha_{vir}\approx 0.3$-2, with a mean value $\langle SFR_{ff}\rangle\approx 0.16$ (Murray 2011, Heyer et al. 2009). Looking at Galactic
nearby molecular clouds
 and massive star-forming dense clumps, Heiderman et al. (2010) find SFR's in the range $\approx 0.02$-0.12 for molecular clouds and $\approx 0.03$-0.5 for dense clumps, yielding a mean value $\approx 0.1$. These values are quite
consistent with the ones derived above from our theory, while
the typical rates obtained by Krumholz \& McKee (2005) for similar conditions are about an order of magnitude smaller, as noted by Murray (2011) and Heiderman et al. (2010), a point we addressed in the previous section. 

Traditionally, the observationally determined star formation rate for a given clump (or region of a clump)  of mass $M_c$, radius $R_c$, lifetime $\tau_0$ and mean dynamical time $\tau^0_{ff}$ is the {\it projected} SFR, i.e. the {\it SFR per unit area} :
\begin{eqnarray}
{\dot \Sigma}_\star={\Sigma_\star \over \tau_0}=SFE\,\times({\Sigma_g \over \tau_0})=SFR_{ff}\,\times({\Sigma_g \over \tau^0_{ff}}),
\label{sigma}
\end{eqnarray}
where ${\Sigma}_g=M_c/\pi R_c^2=(4/3){\bar \rho}R_c$ for a spherical clump is the clump gas surface density. The relationship between this projected SFR and the gas surface density  is generally found to 
 obey, within a large scatter, a power law relationship: 

\begin{eqnarray}
{\dot \Sigma}_\star =\Sigma_{100}\,({ {\Sigma}_g\over 100\,\msol\,{\rm pc}^{-2} })^{N}\,\,\,\,\,\msol\,{\rm yr}^{-1}{\rm kpc}^{-2}.
\label{KS}
\end{eqnarray}
Extragalactic studies yield the well known Kennicutt-Schmidt relation, with $\Sigma_{100}=0.16\pm0.6$ and $N=1.4\pm 0.15$.

Figure~\ref{fig_sfr_obs}  displays ${\dot \Sigma}_\star$  predicted 
by eqn.~(\ref{sfr_red}),
as a function of  surface density, $\Sigma_g$, for  four typical clump sizes,
 namely $R_c=$0.5, 2, 5 and 20  pc, both in the isothermal and 
non-isothermal case. 
The impact of the uncertainties due to $y_{cut}$ is shown by the two 
calculations with $y_{cut}=0.1$ and 0.25, respectively. 
The clumps are assumed to follow Larson's (1981) relations and thus have
 velocities and densities   given by eqn.~(\ref{def_dens}) and (\ref{def_dens2}), 
with $u_0=1$ and $d_0=0.1$-5, i.e. ${\bar n}\approx10$-$10^4\c3$.
The corresponding global and local Mach numbers entering our theory,
 ${\cal M}$ and ${\cal M}_\star$, are consistently derived from these values. 
We have taken $\epsilon/\phi_t=0.1$ in the figure. Also displayed on the 
figure are the data of
Heiderman et al. (2010), both for the class I and flat-SED YSOs and the 
massive clumps as well as the location of the observations of 
Gutermuth et al. (2011), as bracketed by the two large diamond areas. 
These latter authors determined the SFR in eight nearby 
molecular clouds, sampling gas surface density regions, inferred from
 near-IR extinction  mapping, 
ranging from $\sim 15$ to 300 $\mpc2$. Their analysis is similar to the 
one by Heiderman et al. (2010) but, as they examined larger clouds, put 
stronger statistical constraints 
on the SFR vs gas surface density determinations than that study. Moreover, 
these authors include pre-main-sequence Class II objects in addition to
 Class 0, I protostars, from {\it Spitzer} and {\it 2MASS} surveys.

\noindent Several conclusions can be drawn from the theory-observation comparison 
displayed in Fig. \ref{fig_sfr_obs}:

\indent (i) all calculations reproduce well the observed values, with the scatter, and exhibit a direct correlation between the density of star formation and the gas density, in  agreement with the observational determinations. 

Clearly, it seems difficult to provide a universal relation such as 
eq.(\ref{KS}) over the whole density range, given the strong dependence, 
both for the slope and the normalization, of the SFR upon the properties of the
 clump, i.e. its mass and the nature of turbulence forcing. What seems to be more robust is that, at very high density, 
 $\Sigma_g \gtrsim 300\mpc2$, the theoretical relations seem to merge
 towards a strongly super-linear behavior with an exponent $N \simeq 2$
 (an exact value is difficult to infer as it depends on the clump size) and a 
normalization $\Sigma_{300}\approx 32$.  
 Star formation is thus predicted to increase basically quadratically with 
density and thus to be largely dominated by the contribution of the densest
 (likely to be bound) clumps. Given the steeply decreasing number of clumps 
with increasing density, however, such dense clumps represent only a modest 
($\la 20\%$) mass fraction of the clouds (Kainulainen et al. 2011).
This slope becomes progressively steeper as density decreases, the steeper
 the smaller the clump, as expected from the analysis conducted in the 
previous section.

This $N\simeq 2$ value for the slope is in excellent agreement with the values $N=1.67$ to $2.67$ determined by Gutermuth et al. (2011) in eight nearby molecular clouds.
Lower values, including Kennicutt-Schmidt like values, $N\sim 1.4$, with a a star formation threshold are excluded by these authors.


\indent (ii) the dispersion in the observed SFR determinations is well explained by variations among the clump sizes, not mentioning possible variations of the ($\epsilon/\phi_t$) efficiency factor and of the contribution of different turbulence modes, as formalized by the factor $b$, between clumps.

\indent (iii) the  theory, as the observations, predict a severely decreasing SFR below a typical value $\Sigma_g \approx 100\mpc2$,  i.e. $N_{{H_2}}\approx 5\times10^{21}$ cm$^{-2}$, $\nbar \approx 2500\,(R/1\,{\rm pc})^{-1}\c3$, which corresponds to a visual extinction
$A_V\approx 7$ (Draine 2003), except for the largest (20 pc) clumps. 
We recall that such a drop in the SFR {\it naturally arises} from our theory, without invoking any density threshold.

\indent (iv) there is no real density {\it threshold} in star formation. Indeed, according to our calculations, star formation is predicted to keep processing, although with a steadily decreasing rate, even at densities lower than the aforementioned value, but {\it only}, at non negligible rates, in the largest clumps. 
This was expected from the discussion in \S\ref{cutoff}(a).
In this low-density domain, our values can be compared with the observations of Gutermuth et al. (2011). As mentioned earlier, 
these authors examined larger clumps than previous studies and found larger  SFR values than these latter 
at low density. The eight molecular cloud sample examined by these authors covers a range of clump sizes and surface densities $L_c\sim 4$-20 pc, $\Sigma_g\sim 22$-71$\mpc2$. In this range, they determine $\Sigma_\star/\Sigma_g^2$ in the range $(3\times 10^{-4})$-$(5\times 10^{-3})$ pc$^2\msol^{-1}$. These 
values can be directly confronted with the ones inferred from the two upper solid lines, which correspond respectively to $R_c=5$ and 20 pc, in the appropriate surface density range. This yields $\Sigma_\star/\Sigma_g^2\approx 2.0\times (10^{-5}$-$10^{-3})$ for the non-isothermal case and  values larger by about a factor 10 for the isothermal case, for our fiducial value of turbulence forcing, $b=0.5$.
Remembering that both observational and theoretical determinations are subject to significant uncertainties, the agreement can be considered as satisfactory.

This reinforces our suggestion that  the star formation rate smoothly 
decreases with decreasing gas density and that there is no real density 
threshold for star formation. The severe drop in the SFR, which is predicted 
to occur around $\Sigma_g \approx 100\mpc2$ for $\sim$pc-size clumps, simply 
reflects the basic 
mechanisms mentioned  in \S\ref{cutoff}:  star formation can still
 occur, although at significantly lower rates, in low-density clumps providing 
these latter, assuming they follow Larson's relation, are large enough to (i)
 significantly exceed their typical Jeans length and (ii) generate enough
 large-scale turbulence, i.e. have large enough Mach numbers,  to generate
 a large enough spectrum of density fluctuations, and thus of prestellar cores.
According to the observed clump size distribution, 
$dN_c/dL_c\propto L_c^{-1.9}$ (Kainulainen et al. 2011), however, 
such clumps are very rare.
{\it This stresses the need to observe very large areas at low-density 
in order to get statistically significant SFR detections. }

A threshold for core formation was also ruled out in the SCUBA survey of 
Perseus by Hatchell et al. (2005). These authors stress the existence of 
submm cores, identified from IRAS as likely Class I objects, down to $A_V=3$,
 i.e. $N_{H_2}\sim 10^{21}$ cm$^{-2}$, and demonstrate the steeply decreasing 
probability of finding a core with decreasing column density. This is also
 consistent with the analysis of Andr\'e  (2012, Fig. 3), who find a 
drastically decreasing but non-zero number of class-0 prestellar core 
candidates below about the aforementioned $A_V\sim 7$ value.

In contrast, several studies of various local clouds  (Onishi et al. 1998,
 Johnstone et al. 2004, Enoch et al. 2007, Lada et al. 2010, Andr\'e et al.
 2010) do find a  steep break in the SFR-gas density relationship around the
 aforementioned $\sim 100\mpc2$ value, that they identify as a density
 threshold for star formation.
It should be stressed, however,  that all these studies still find 
pre-stellar cores below this limit, suggesting that star formation continues,
 although with a much smaller efficiency, down to the limit of the surveys,
 i.e. $\Sigma_g\approx 10\,\mpc2$.
As mentioned above, a noticeable exception is the recent work by Gutermuth
 et al. (2011), who explore larger areas and who include pre-main-sequence 
Class II objects in addition to Class 0, I protostars, from 
{\it Spitzer} and {\it 2MASS} surveys, and who do not find evidence for a 
column density threshold for star formation from 15 to 300 $\mpc2$. 
These authors explain the different result by the very small number of 
Class 0, I protostars at low gas column densities, and thus the low 
statistics in previous surveys,  including the {\it Spitzer}-derived 
data of Heiderman et al. (2010).
Interestingly, by comparing the SFR derived from YSO surface densities
and the ones derived from far-IR luminosity, as commonly used
as a  proxy of the  SFR, they find that the $L_{{\rm FIR}}$-derived SFRs are systematically about an order of magnitude below 
the YSO-derived ones. This is confirmed for instance by the recent  analysis of  W40 and Serpens South with {\it Herschel}  by Maury et al. (2011), 
who have access to Class 0 objects and find SFRs an order of magnitude higher than the typical SFRs observed for embedded infrared clusters. This suggests that, if anything, SFR  determinations based only on Class 0 or Class I objects underestimate the real SFR by a fair amount. Therefore, the aforementioned identification of  a cut-off in the SFR below $\sim 100\,\mpc2$ might in fact be due to the {\it very small number of Class 0, I protostars at low gas  column densities and the too limited statistics
(too small fields of view) in the surveys}.
This is supported by the recent analysis of Bressert et al. (2010) who show that using only near-infrared detections and small fields of views allows identifications of young stars only in
the densest parts of the clouds, a bias which can be corrected with {\it Spitzer} data and larger surveys.

\bigskip

\subsection{ Molecular cloud vs galactic SFR}

As already found in the observations of Heiderman et al. (2010) and Evans et al. (2009), our SFR-gas relation, characteristic of Galaxy star-forming molecular clouds, lies above the Kennicutt-Schmidt relation by about a factor of 10 at high density.
It should be remembered,
however, that the Kennicutt-Schmidt relation derives from extragalactic determinations and is averaged
over much larger regions ($\sim$ kpc-size) than individual molecular cloud complexes ($\la 10$-100 pc). Such an average determination includes star-forming regions but also diffuse molecular gas or atomic gas that is not forming stars, leading to  overestimate  the amount of gas counted as star-forming gas (Heiderman et al. 2010). Recent SFR  determinations in
nearby galaxies  indeed show that the measurement size scale used changes significantly the SFR-gas surface densities relation (Liu et al. 2011). 
Moreover, SFR in GMC's are determined from inventoring the YSOs and assuming a star formation timescale of $\sim $Myr while galaxy-averaged SFR's are derived from conversion of a FIR flux into a mass growth rate assuming a timescale of 10 to 100 Myr, naturally leading to a smaller SFR for a given average gas density.
Since, as shown in the present study, the SFR depends on the clump's mean density and significantly decreases for low-density clumps, this provides a plausible explanation for the difference between GMCs and extragalactic SFR-gas quantitative determinations. In fact, a correct determination of the SFR-gas relation at extragalactic scales should take into account the density-dependence of the SFR-gas relation. The same explanation applies to the characteristic value of the SFR inferred for the Milky Way, found to be about an order of magnitude smaller than in GMCs.
A likely explanation is that most of
the mass of the clouds which compose the Galaxy is at low column density, with only a very small fraction in condense structures
at high enough density for efficient star formation (see e.g. Kainulainen et al. 2011). 
Indeed, large-scale clouds which compose the Galaxy have densities around $\sim 10^{21}$ cm$^{-2}$ (${\bar n}\lesssim 100$ cm$^{-3}$).
Therefore, most of the clouds in the Milky Way no longer produce stars.
On the other hand, our calculations suggest that the power dependence of the SFR upon the gas surface density in the (high-density) efficient star formation regime is not drastically different for  Galactic and extragalactic determinations, since our exponent at high density is $N\approx 2$ whereas the one characteristic of the Kennicutt-Schmidt relation is. $N=1.4\pm 0.15$ (see Fig.  \ref{fig_sfr_obs} ). 
Again, such a universal SFR-gas dependence naturally emerges from a picture of
 star formation being initiated by turbulence-induced fluctuations and the star formation rate being determined by the free-fall time of these fluctuations - not of the clump ! - once gravity takes over,
  leading to a strong dependence of the SFR upon the gas density.

\subsection{Other dependences of the star formation rate}

In the previous section, only the SFR dependence upon the column density was
examined. In the present section, we consider the SFR dependence upon the clump's
mass and upon the column and volume densities over free-fall time. The results are compared with the sample of Heiderman et al. (2010).

\begin{figure} 
\begin{picture} (0,8)
\put(0,0){\includegraphics[width=7.5cm]{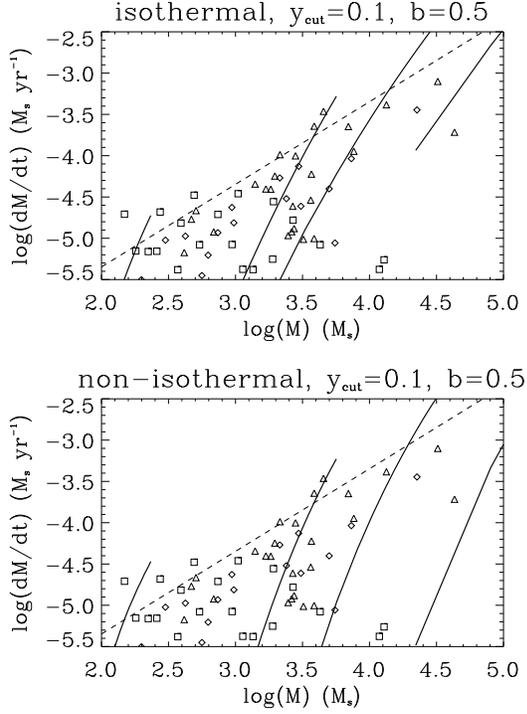}} 
\end{picture}
\caption{Star formation rate, ${\dot M}_\star$, as a function of the cloud 
mass, $M_c$, in  the isothermal and non-isothermal case. Each line corresponds
 to clump sizes $R_c=0.5$, 2, 5 and 20 pc, from  left to right, with
the normalization density $d_0$ increasing along the line from  $d_0=0.1$ to 
4.5. The squares, triangles and diamonds represent the 
data from Heiderman et al. (2010). The dashed line displays the fit of Lada et al. (2012).}
\label{fig_sfr_mass}
\end{figure}

\subsubsection{Star formation rate versus clump mass}
\label{clumpmass}

Figure~\ref{fig_sfr_mass}  portrays the star formation rate, ${\dot M}_\star$, 
as a function of the cloud mass, $M_c$, 
for the isothermal and non-isothermal case. Each line corresponds 
to the same clump sizes as in the previous figures, i.e. $R_c=0.5$, 2, 5 
and 20 pc, with
the normalization density $d_0$ increasing along the line from  $d_0=0.1$ 
to 4.5. We note the large scatter in the predicted ${\dot M}_\star$ for a
 given clump mass, in particular for the non-isothermal case. The data of
 Heiderman et al. (2010) are displayed in the figure for comparison. 
 The
 theoretical determinations agree very well with
the data, indicating that, within our theory, the range of clump characteristics we have investigated
reproduce well the observationally determined SFR values with the observed scatter.
The dashed line in th figure displays the relation derived by Lada et al. (2012), 
${\dot M}_\star = (4.6\times10^{-8})\,({M_c\over \msol})\,\msol {\rm yr}^{-1}$.
 The present calculations show that this relation, although broadly correct,
may not be an accurate 
 representation of the real star formation
 process, as this latter not only depends much more steeply on the clump's mass but also strongly depends on the clump's size.

\begin{figure} 
\begin{picture} (0,8)
\put(0,0){\includegraphics[width=7.5cm]
{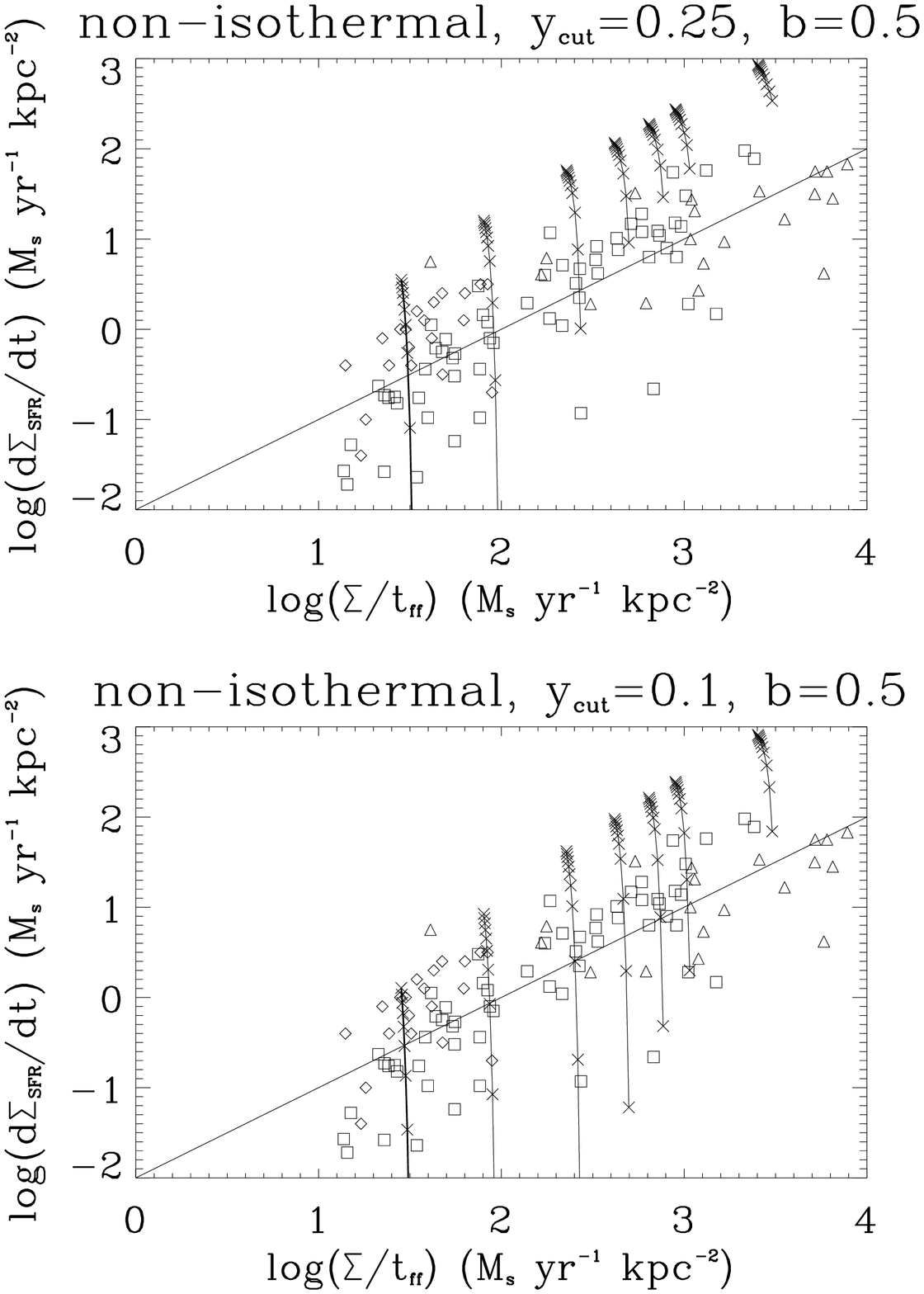}} 
\end{picture}
\caption{SFR per unit area, ${\dot \Sigma}_\star$, as a function of 
(${\Sigma}_g/\tau_{ff}^0$) for 7 values of $d_0=$0.5, 1, 2, 3, 4, 5, 10 (left to right) and
11 clump sizes  $R_c=$0.5, 1, 2, 4, 6, 8, 10, 12, 14, 16 and 18 pc (bottom to top on each line, as indicated by the crosses). Each curve corresponds to a constant $d_0$. 
The highest values of $d_0$ and $R_c$ correspond to the highest SFR's.}
\label{coldens_tff}
\end{figure}

\subsubsection{Dependence of star formation rate on free-fall time}

The values of star 
formation rates per mean free fall time for a sample of clumps can be 
directly inferred by determining the slope of the relation
 ${\dot \Sigma}_\star$ vs $({\Sigma_g / \tau^0_{ff}})$ (eq.(\ref{sigma})) or,
 equivalently, of the volume density relation 
${\dot n}_\star$ vs $({n_g / \tau^0_{ff}})$. A universal 
value of the SFR would correspond to  one single, constant value 
of the slope for the whole sample.

To confront such observational determinations with our theory, we first
vary the clump's size
for a fixed density normalization parameter $d_0$. Indeed,
assuming Larson's relation for our clumps, i.e.
$\rho \propto R_c^{-0.7}$, yields $\Sigma \propto 
 R_c^{0.3}$ and $\tau_{ff} \propto R_c^{0.35}$ for  the column density and free-fall time, respectively.
Therefore the column density over free-fall time, 
$\Sigma / \tau_{ff}$ is almost independent of the clump's size. The 
 corresponding ${\dot \Sigma}_\star$ vs $({\Sigma_g / \tau^0_{ff}})$ curves are thus almost parallel to the y-axis, making 
the dependence upon $d_0$ and $R_c$ straightforward. 
Seven values of $d_0$ are computed, namely 0.5, 1, 2, 3, 4, 5, 10 
as well as eleven sizes  $R_c=$0.5, 1, 2, 4, 6, 8, 10, 12, 14, 16 and 18 pc.
Figure~\ref{coldens_tff}  
 displays the theoretical predictions for
 ${\dot \Sigma}_\star$ vs $({\Sigma_g / \tau^0_{ff}})$ 
as well as the 
observational determinations of Heiderman et al. (2010).
As obvious in the figure, and as expected from the previous subsection,
the overall agreement between our predictions and 
the data is very good and the observational scatter is well reproduced by the explored variations of clump sizes and densities. Note that, given the large spread in the data 
and the uncertainties discussed previously in the theory, it is 
difficult to assess more than a qualitative agreement.

\begin{figure}[h!] 
\begin{picture} (0,8)
\put(0,0){\includegraphics[width=7.5cm]
{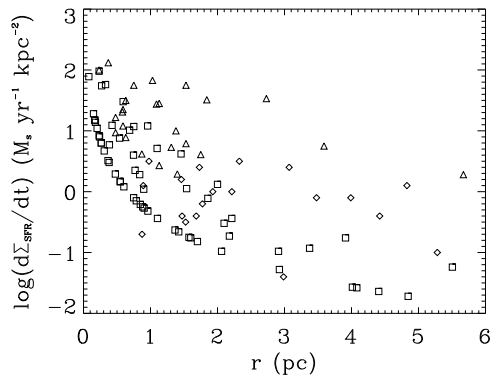}} 
\end{picture}
\caption{${\dot \Sigma}_\star$
as a function of  the clump radius, $r$. There is a clear 
trend of ${\dot \Sigma}_\star$ strongly increasing with decreasing clump's size, particularly if the massive clump determinations (triangles), 
which tend to differ from molecular clouds YSOs (diamonds
and squares), are excluded.}
\label{rad_sfr}
\end{figure}

Interestingly, the confrontation between our results and the observed SFR values
from Fig.~\ref{coldens_tff} 
leads to the conclusion that 
the size of star forming clumps should decrease with increasing column density over free-fall time. 
Typically, the clump's size should be less than about 2 pc  for
$\log (\Sigma / \tau_{ff}) \simeq 3$ and less than about 0.5 pc for 
$\log(\Sigma / \tau_{ff}) \simeq 3.5$
(with a dispersion of a factor 2-3). 
To verify whether this trend is indeed real, we have 
plotted  $\dot{\Sigma}_*$ as a function of  the clump's radius
(Fig.~\ref{rad_sfr}). 
Indeed, Fig.~\ref{rad_sfr} shows a clear trend
for the highest values of ${\dot \Sigma}_\star$ been obtained 
for the smallest clouds. The reason of this behaviour is not completely
clear but appears to be an important element to understand
the SFR dependence. 
A likely explanation is that large clouds never become dense enough, i.e. massive enough (see e.g. Fig. 8).
More precisely, as massive clouds collapse they undergo local gravitational fragmentation in dense
clumps where most of the star formation occurs. 

\begin{figure}[h!] 
\begin{picture} (0,10)
\put(0,0){\includegraphics[width=7.5cm]
{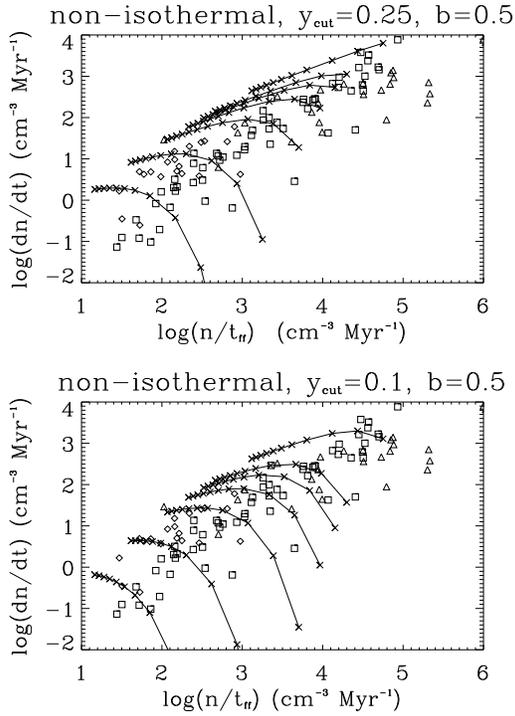}} 
\end{picture}
\caption{Volumic SFR density, ${\dot n}_\star$, 
as a function of $({n_g/ \tau^0_{ff}})$ for the same conditions as in
 Fig.\ref{coldens_tff}. Each line represents a fixed value of $d_0$.}
\label{dens_tff}
\end{figure}

\begin{figure}[h!] 
\begin{picture} (0,10)
\put(0,0){\includegraphics[width=7.5cm]
{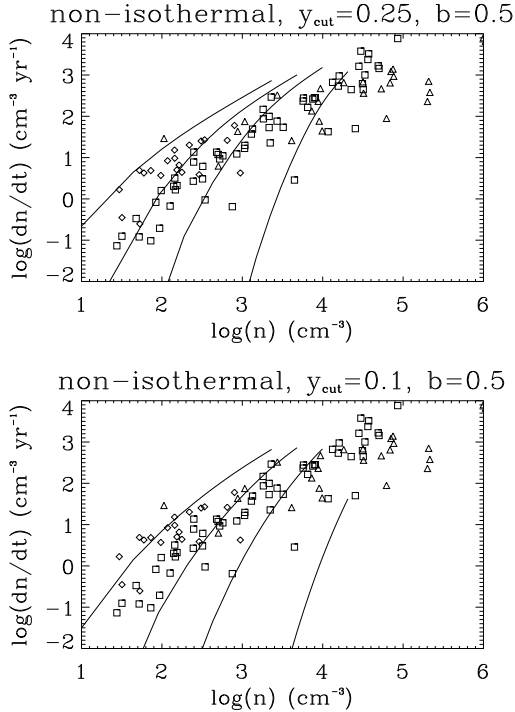}} 
\end{picture}
\caption{Volumic SFR density, ${\dot n}_\star$, 
as a function of $({n_g/ \tau^0_{ff}})$ for the same conditions as in
 Fig.\ref{fig_sfr_obs}. The four solid lines correspond to four clump sizes, 
namely $R_c=0.5$, 2, 5 and 20 pc (right to left).}
\label{dens_tff2}
\end{figure}

Finally, for completeness we also display ${\dot n}_\star$ vs $({n_g / \tau^0_{ff}})$ in
 Figure~\ref{dens_tff}  
  for the same cloud
parameters as in Fig.~\ref{coldens_tff} (the various 
curves correspond to a fixed $d_0$) while  Fig.~\ref{dens_tff2}
shows the results for the same cloud parameters as in Fig.\ref{fig_sfr_obs}.
For the data of Heiderman et al. (2010),
 we derived  the volume density from the surveyed area as 
(assuming spherical clumps of area $A$) 
${\bar \rho}=(3{\sqrt \pi}/4)(M_c/A^{3/2})$.

According to the present analysis, there is clearly no "universal" slope value 
neither for the  ${\dot \Sigma}_\star$ vs $({\Sigma_g / \tau^0_{ff}})$ 
nor for ${\dot n}_\star$ vs $({n_g / \tau^0_{ff}})$ relation. 
Instead,  these relations significantly
 vary  with the clump's size {\it and} density.
 Indeed, 
below $\Sigma / \tau_{ff} \lesssim 1000$ $M_\odot {\rm yr}^{-1} {\rm kpc}^{-2}$
there is between 1 and 2 orders of magnitude variations in the SFR. 
Even in dense regions, corresponding to 
$n / \tau_{ff} > 100$ ${\rm cm}^{-3} {\rm Myr}^{-1}$,
the mean value of the SFR, as measured by the slope of the illustrated relations, 
typically varies between $\sim 3\%$ and  $\gtrsim 10\%$. 
This analysis is in contrast with the claim by Krumholz et al. (2012) for 
 a universal SFR value,  of about $1 \%$ from clouds to starburst galaxies. 
As mentioned earlier, molecular clouds are very clumpy structures with orders of 
magnitude of 
density variations. Therefore, it seems difficult to invoke an average 
timescale, defined at the cloud's mean density, as the relevant timescale
 for star formation. 

\subsection{Discussion: The case of filaments}

According to various recent studies, in particular the ones conducted with {\it Herschel}, star formation is found to occur in filaments (Andr\'e et al. 2010, Men'shchikov et al. 2010, Ward-Thompson et al. 2010, Molinari et al. 2010, Hill et al. 2011). It is well known that (unbound or bound) molecular clouds exhibit a filamentary structure, as expected
from both hydrodynamical and MHD compressible (shock dominated) turbulence even in the absence of gravity (Padoan et al. 2001, Nakamura and Li 2008, Federrath et al. 2010, see also Banerjee et al. 2009).
 It is thus not surprising that the locations of
star forming populations tend to follow this morphology. We recall that in our formalism, the prestellar cores, progenitors of the protostars identified as Class 0 objects,
 take birth in turbulence induced clumps of enhanced density, and isolate themselves from the surrounding medium under the action of gravity (Hennebelle \& Chabrier 2008, Chabrier \& Hennebelle 2011). For sake of simplicity, 
the clumps and the cores in our theory are assumed to have a spherical geometry. Non-linear simulations of the collapse of filaments show that the resulting fragments are nearly spherical
(Inutsuka and Miyama 1997). It thus seems reasonable to assume spherical collapse for the {\it cores}.
 The reality might be more complex for the star-forming {\it clumps}, which may have a flattened, filament-like geometry, some of these filamentary clumps becoming 
themselves gravitationally unstable as they both accrete mass and dissipate turbulent energy, yielding eventually global collapse of the clump.
Interestingly, the very same  {\it Herschel}  observations reveal 
that protostellar cores seem to be present only in the gravitationally bound clumps (Andr\'e et al. 2010, Molinari et al. 2010, Arzoumanian et al. 2011), suggesting that star formation  takes place dominantly in such clumps,  prone to gravitational fragmentation. 
Gravitational contraction will thus cause  filamentary clumps to fragment along the axis.
  
This possible peculiar role of filaments in  star formation might affect our results by numerical factors of order of a few
but it does not fundamentally modify our theory of clump and  prestellar core formation, and of the resulting CMF\footnote{The characteristic Jeans mass for a nearly isothermal filament
($\gamma \sim 1$) differs only by a factor 0.6 from the one for a sphere (Larson 2003)} .
In our theory, the (scale-free) criterion for clump formation is simply set up by an arbitrary critical density threshold $\delta_c=\log( \rho_c/{\bar \rho})$  (see \S3 of paper I). The argument above thus suggests to choose for this threshold condition
 for the {\it star-forming} clumps - identified as the aforementioned filaments - the density at which these latter become
gravitationally unstable and thus exceed a critical mass per unit length $M_{crit}=2\langle \sigma \rangle^2/G$, where $\langle \sigma \rangle$ includes both the thermal and
non-thermal velocity dispersion contributions (Ostriker 1964, Larson 1985, Inutsuka \& Miyama 1992, Fiege \& Pudritz 2000). For a temperature $T\sim 10$ K, this corresponds to
 a mean column density $N_{H_2}\simeq 10^{22}$ cm$^{-2}$, similar to the value identified in \S4.1  for efficient star formation (point (iii)), about 5-10 times the
typical mean density of the surrounding {\it cloud}.
This particular choice for the {\it clump} formation in our theory will not affect the resulting {\it core} mass function (CMF), determined by the virial condition for spherical collapse, as mentioned above. 

Furthermore, a closer inspection 
of Table 1 and Fig. 8 of HC09 (see also Fig. 2) shows that small-size, low-density ($\lesssim 3000\c3$) clumps, for which the CMF is very narrow and thus star formation is  negligible, are
unbound or only marginally bound ($(E_{kin}+E_{th})\gtrsim |E_{grav}|$), whereas the ones above about this density are bound and thus prone to collapse. 
Therefore, our theory
 suggests that the clumps in which star formation is taking place dominantly should be denser than ${\bar n}\sim$ a few $10^3\,\c3$, i.e. $N_{H_2}\simeq 10^{22}$ cm$^{-2}$ for pc-size clumps,
 in fairly good agreement with the aforementioned critical value inferred for filaments.
This is consistent with the results found in \S\ref{vir}, that show that the SFR is highest in the largest and densest clumps, with $\alpha_{vir}\lesssim 1$.
Note that, given the scale-free nature of the turbulent flow responsible for the clump formation, such a minimum density for efficient star formation must apply at all scales and thus must vary accordingly  with the clump's size. The aforementioned value typically applies for a 1 pc size clump and should increase as $\sim R_c^{-0.7}$ for smaller sizes,  providing Larson's relations  apply.

Therefore, a picture where  star formation occurs {\it dominantly} in relatively dense, rather bound clumps, with mean density ${\bar n}\gtrsim 3000\,\c3$, in agreement with observational results, naturally emerges from our theory. The filamentary
nature of the clumps does not  significantly affect the results, except for possible numerical factors of order unity.


\section{Influence of the clump distribution and evolution}
\label{clumpevol}

\begin{figure} 
\begin{picture} (0,12)
\put(0,0){\includegraphics[width=8cm]{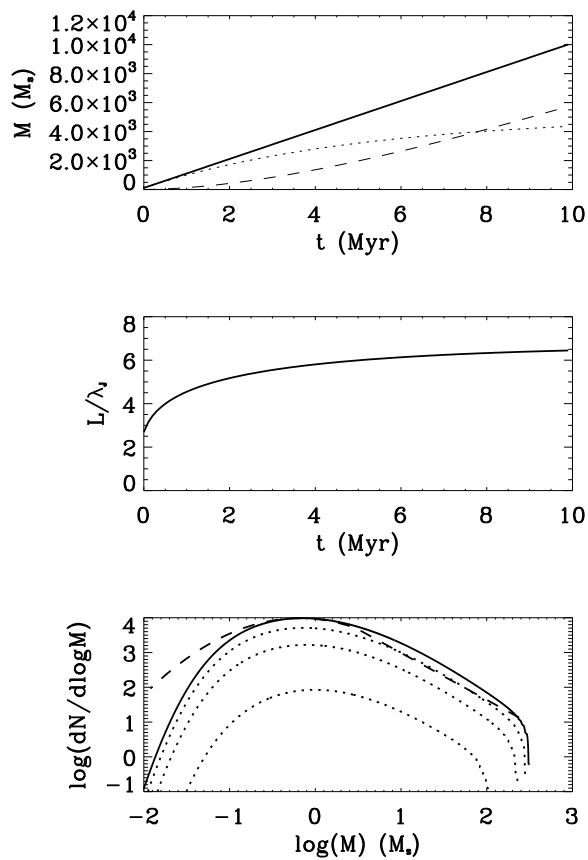}} 
\end{picture}
\caption{Clump evolution for $\dot{M}=10^{-3}$ M$_\odot$ yr$^{-1}$ and $\epsilon/\phi_t=0.1$.
Top panel: total cloud mass (solid line), gas mass (dotted line), 
star mass (dashed line).
Middle panel: Cloud radius over Jeans length. Bottom panel: mass spectrum
after different times: $t=$1, 4 and 7 Myr (dotted lines from bottom to top) and $t=10$ Myr (solid line).
The dashed line displays the SCIMF}
\label{imf_accret2}
\end{figure}

\begin{figure} 
\begin{picture} (0,12)
\put(0,0){\includegraphics[width=8cm]{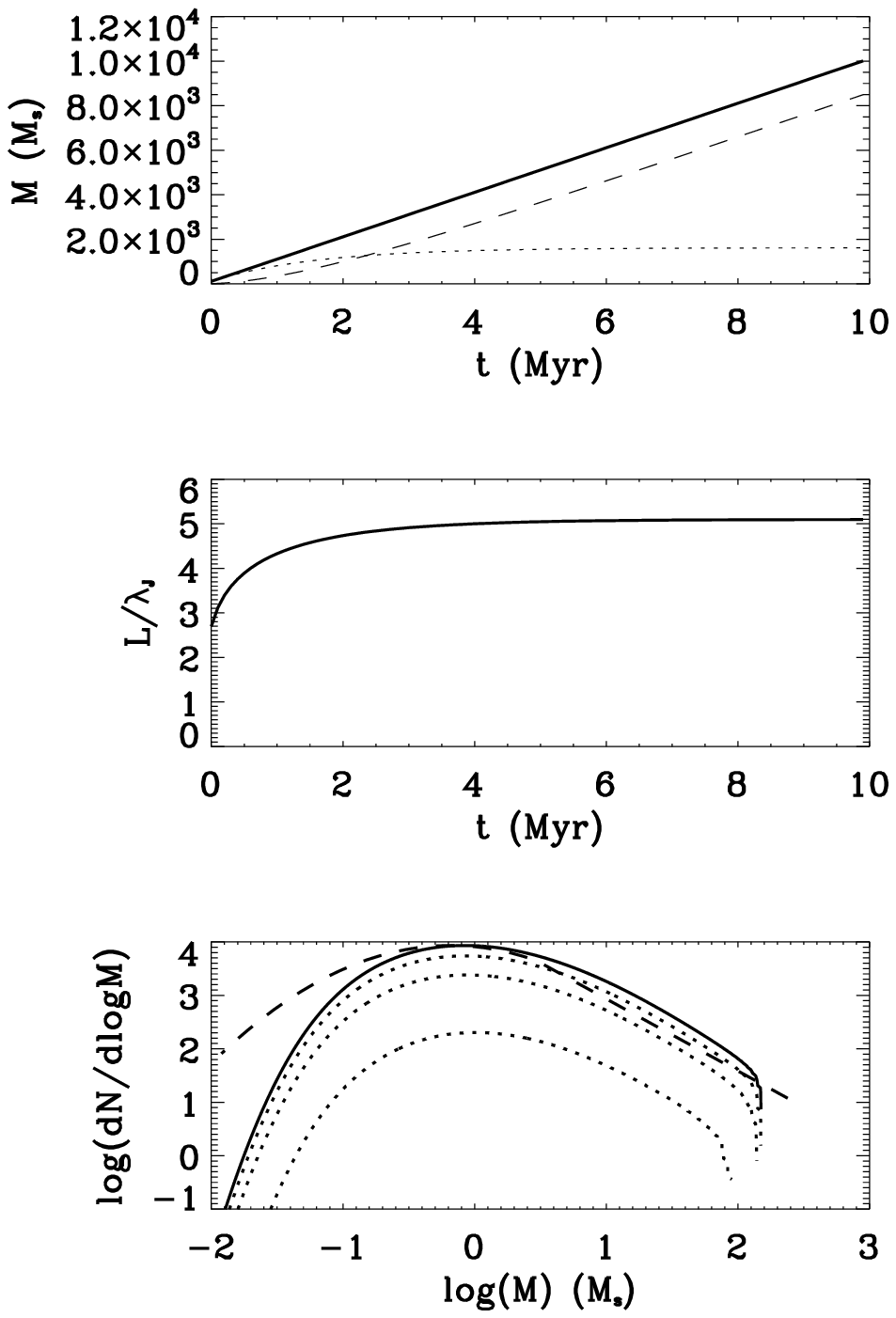}} 
\end{picture}
\caption{Same as Fig.~\ref{imf_accret1} for $\dot{M}=10^{-3}$ M$_\odot$ yr$^{-1}$ and 
$\epsilon/\phi_t=0.3$.}
\label{imf_accret3}
\end{figure}

\begin{figure} 
\begin{picture} (0,12)
\put(0,0){\includegraphics[width=8cm]{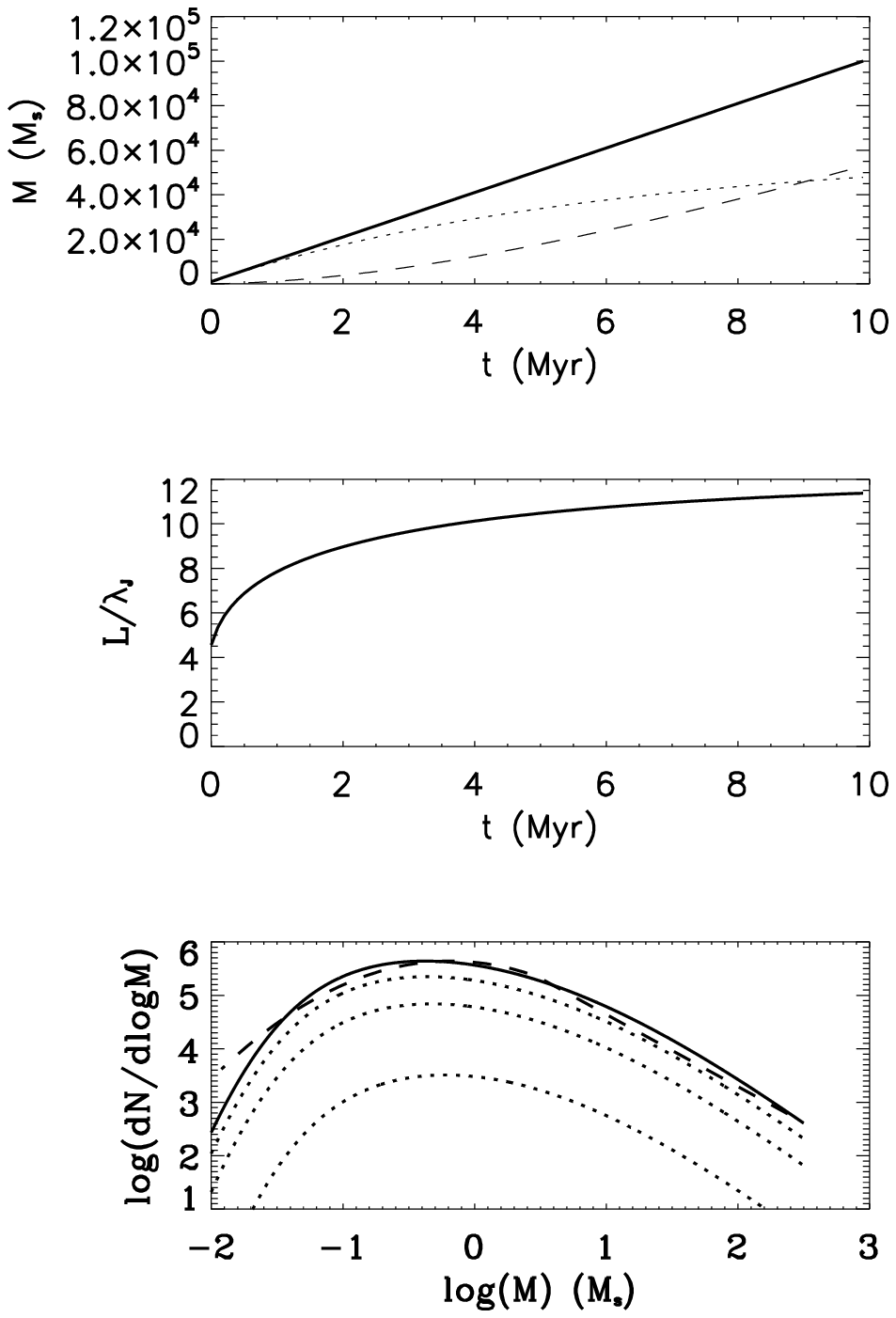}} 
\end{picture}
\caption{Same as Fig.~\ref{imf_accret1} for $\dot{M}=10^{-2}$ M$_\odot$ yr$^{-1}$ and $\epsilon/\phi_t=0.1$.}
\label{imf_accret1}
\end{figure}

\begin{figure} 
\begin{picture} (0,12)
\put(0,0){\includegraphics[width=8cm]{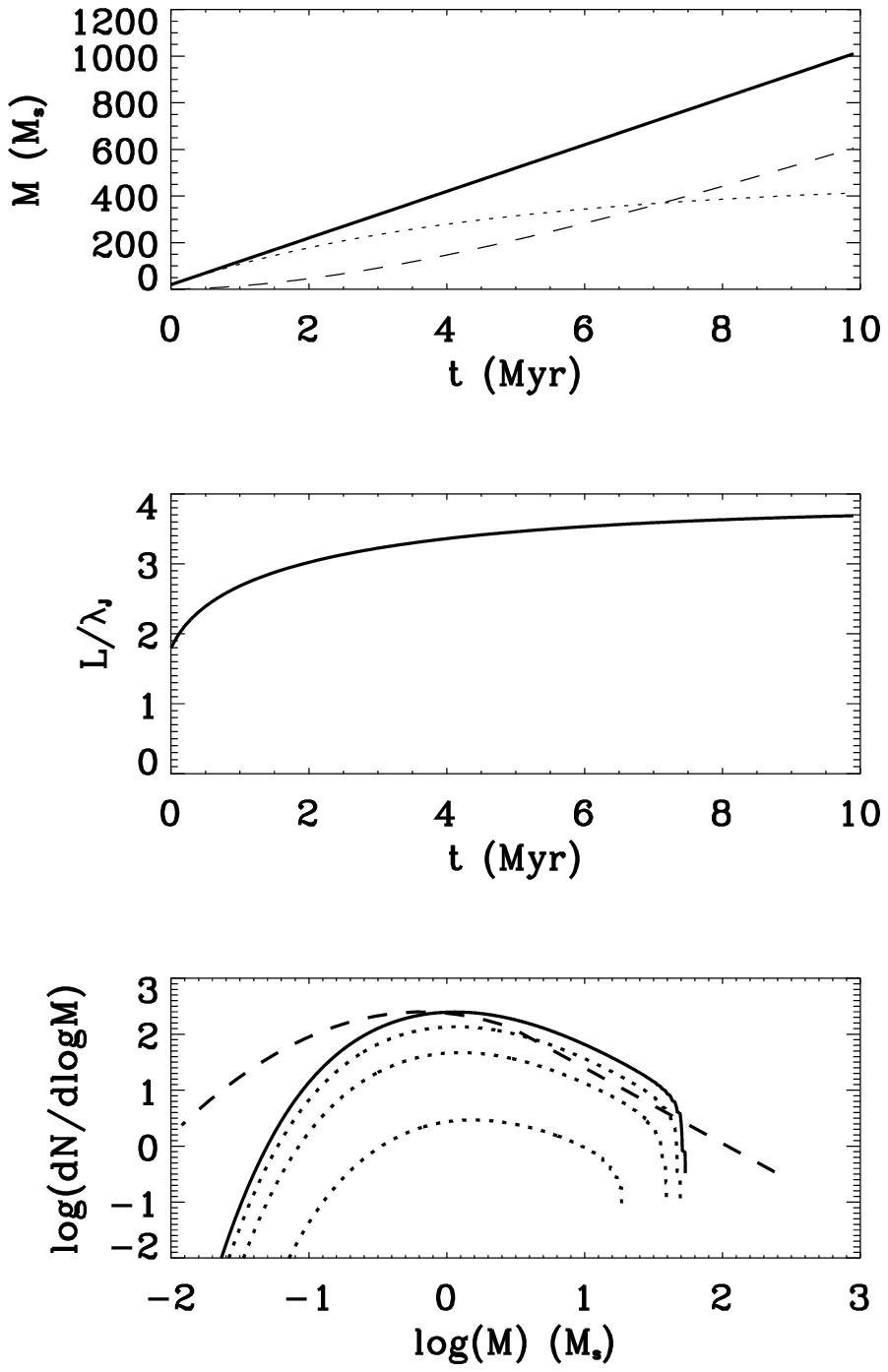}} 
\end{picture}
\caption{Same as Fig.~\ref{imf_accret1} for $\dot{M}=10^{-4}$ M$_\odot$ yr$^{-1}$.}
\label{imf_accret4}
\end{figure}

So far, we have considered the core mass spectrum produced 
within a single clump with fixed physical quantities (mass, size, density).
In real situations, there will be  a distribution of evolving clumps. In
this section, we investigate the impact  on the gravitationally bound core mass spectrum of the time dependence 
of the cloud parameters as well as having a clump 
distribution rather than a single clump. 
In order to distinguish the influence of these two effects, we first 
consider the time evolution of a single clump before investigating 
the effect of a distribution.

\subsection{A simple model of clump evolution}
One of the intrinsic difficulties in inferring the mass spectrum
of the self-gravitating fluctuations that may eventually lead to stars
is the choice of clump parameters (density, size, velocity dispersion...).
As seen from our calculations, these clump properties have a strong impact on the mass spectrum. 
The peak position, for instance, is determined by the clumps Jeans mass, thus mean density, temperature and  Mach number.
Clumps, however, are not well defined entities, in particular because 
they constantly grow in mass by accretion, so that their 
density and size evolve with time. Under these conditions, 
it is clear that the choice of a  fixed set of clump parameters is a 
simplification.
A proper investigation of the clump evolution would require to develop a fully self-consistent model for clump
formation and evolution, including growth by accretion, gravitational 
contraction or turbulence dispersion as well as spatial inhomogeneity. 
There is no such model of clump evolution yet; we thus consider  a very simple 
prescription, essentially inspired by observations.

As mentioned above, instead of setting a density and size, we assume that the clump 
is accreting at a fixed accretion rate, $\dot{M}$. Starting with a 
mass $M(t=0)\equiv M_0$, the clump mass grows as $M_c = M_0 + \dot{M} t$. This 
mass entails the mass of the gas, $M_g$, and the mass of the 
stars, $M_*$, that formed in the clump, $M_* = M_c - M_g$.
We further assume that the clump follows Larson-type relations as stated by 
eqs.~(\ref{def_dens}) with $d_0=3$ and $u_0=1$, which correspond to the 
canonical values of these relations. Then, knowing 
$M_g(t)$, we can infer the gas mean density, $\ng(t)$ and thus the cloud radius $R_c(t)$. 
Since at a given time, $t$, the clump's parameters are well determined, we can calculate 
the SFR from our CMF/IMF theory, as stated by eqs.~(\ref{sfr_red})-(\ref{sfr_tot}).
We can thus estimate the gas mass fraction converted into stars between 
$t$ and $t+dt$ as: $dM_*(t)=M_g \times SFR_{ff} \times dt$. The total 
mass of stars at time $t$, $M_*(t)$, is thus 
\begin{eqnarray}
M_*(t)=\int_0^t M_g(t) SFR_{ff}(t) dt.
\end{eqnarray}
We typically integrate from 0 to 10 Myr with time steps $dt=10^5$ yr.

The accretion rate onto the molecular clumps is a difficult quantity to 
infer observationally and  thus is not well known. Using observations 
of the LMC, Kawamura et al. (2009) and Fukui et al. (2009) propose that 
giant molecular clouds with a mass of $\simeq$a few $10^5 \, M_\odot$ are accreting 
at a rate $\dot{M}=$(1-5)$\times 10^{-2} \; M_\odot$~yr$^{-1}$. 
To estimate the accretion rate on smaller clumps, we simply use the Larson relations, 
\begin{eqnarray}
\nonumber
\dot{M} &\simeq& {M \over \tau_c} \simeq {M \over R_c/(\sigma/\sqrt{3})}  \\
& \simeq & M^{{1.3+\eta \over 2.3}} { u_0 \over \sqrt{3}} 
 \left( {4 \pi \over 3} d_0\right) ^{ 1 - \eta \over 2.3}
(1 \, {\rm pc})^{ 0.7 - 3 \eta \over 2.3} 
\label{dM_dt} \\
&\simeq&   10^{-3}  \, \left( {M \over 10^4 M_\odot} \right) ^{1.3+\eta \over 2.3}
({\sigma \over 0.8 \, {\rm km \, s^{-1}} } )\left( { {\bar n} \over 1 \, {\rm cm}^{-3}} \right)^{1 - \eta \over 2.3}\,\,M_\odot \, {\rm yr}^{-1},
\nonumber
\end{eqnarray}
where the numerical constant has been estimated for $\eta=0.4$. 
For a GMC of mass $10^5\,M_\odot$, this estimate yields
an accretion rate $\dot{M}  \simeq 5 \times 10^{-3}$ M$_\odot$~yr$^{-1}$, 
slightly lower but nevertheless close to the estimate 
of Fukui et al. (2009). 
For a cloud of $10^3$ M$_\odot$, we get
 $2 \times 10^{-4}$ M$_\odot$ yr$^{-1}$. We have thus explored three typical values for the 
accretion rate, namely
$\dot{M} = 10^{-2}, \;  10^{-3}$ and  $10^{-4}$ M$_\odot$ yr$^{-1}$, which correspond to clumps of approximately 
$10^5$, $10^4$ and $10^3$ M$_\odot$.

Figure~\ref{imf_accret2} portrays the results for $\dot{M} = 10^{-3}$ M$_\odot$ yr$^{-1}$ and $\epsilon/\phi_t=0.1$. 
The top panel shows the total mass of the clump (solid line), the  mass of gas (dotted line)
and the stellar mass (dashed line). While the total mass grows linearly with time, 
the mass of gas, which was  initially almost equal to the total mass, increases less and less rapidly as star formation is occurring. 
At the beginning of the process,  the mass of the cloud is small, 
and the amount of stars formed remains fairly limited. After $t \simeq 4$ Myr the stellar mass starts increasing significantly and after
$t \simeq 8$ Myr, the masses of gas 
and stars are roughly equal. After this time, the gas mass basically saturates,
as the gas added to the cloud is rapidly converted into 
stars. As a consequence, the size of the cloud (middle panel) increases rapidly at the beginning 
and much more slowly as star formation proceeds. The bottom panel displays the stellar 
mass spectrum
at different times, namely after 1, 4 and 7 Myr (dotted lines from bottom to top) and after
 $t=10$ Myr (solid line);
the dashed line correspond to the SCIMF. As 
seen in the figure, the {\it shape} of the mass spectrum remains about the same over time but
 of course
the integral, i.e. the total number of stars, strongly increases, a consequence of the growing 
size of the clump, as mentioned above.

Figure~\ref{imf_accret3} displays similar results for the same accretion rate, $\dot{M} = 10^{-3}$
 M$_\odot$ yr$^{-1}$,
but $\epsilon/\phi_t=0.3$. The behaviour remains qualitatively similar except that, as expected, 
stars form faster, since the SFR is three times larger
(see eqn.~(\ref{sfr_tot})), and thus
the gas mass remains smaller by a factor of about 2 compared with the previous case. Consequently, 
the size of the clump is slightly smaller and less small-mass objects form because the Mach number
 (${\cal M}\propto R_c^\eta$) is slightly smaller (see Paper I).

Figure~\ref{imf_accret1} displays the results for $\dot{M} = 10^{-2}$ M$_\odot$ yr$^{-1}$.
The SFR is slightly smaller than for
$\dot{M} = 10^{-3}$ M$_\odot$ yr$^{-1}$ because the clump grows more rapidly and thus is less 
dense.
Interestingly, the mass spectrum remains very similar, except for the slightly larger 
number of low-mass objects, a consequence of the larger Mach number.
Indeed, as discussed in paper I (eq.(47) of paper I), there is a partial compensation between 
the Jeans mass and the Mach number scale dependences as a function of the cloud's size/mass
 so that the location of the peak of the CMF remains almost unchanged. It is interesting to 
compare the top panel of this figure
with the top panel of Fig.~5 of V\'azquez-Semadeni et al. (2007), which 
shows the total gas+stellar masses in a simulation of colliding flows. 
The general behaviour and even some of the details of the simulation are very similar to the results portrayed in Fig.~\ref{imf_accret1}:
(i) the total mass of the cloud in both cases is a few $10^4$ M$_\odot$, 
(ii) the time at which stars and gas masses are equal agree within a factor of about 2
(particularly if we choose for the beginning of the cloud formation in the simulation the 
time $t \simeq 10$ Myr, which seems more accurate than $t=0$, (iii)
 when stars start forming efficiently, the mass of the gas remains nearly 
constant both in the simulation and in the present model.  The main differences appear in the evolution of the stellar mass,
which is more sudden in the simulations and faster at the beginning than 
in the later phases, in opposite to the present results. Moreover,
while in the model the mass of the gas always increases, it decreases in the 
simulation. This is likely due to gravity whose effects on the cloud dynamics are 
ignored in our model.

The last case we investigate is presented in 
Fig.~\ref{imf_accret4} which displays results for 
$\dot{M} = 10^{-4}$ M$_\odot$ yr$^{-1}$. 
The mass of the clump is only about 400 $M_\odot$.
The mass spectrum peaks at about the same mass, for the reason mentioned above but the mass spectrum 
is much narrower with a significant deficit of both massive and low-mass objects. This stems from the fact that for such small clumps, the Jeans length becomes
comparable to the size of the cloud, as discussed earlier, 
drastically decreasing star formation (see e.g. Fig. 2, and Fig. 8 of HC09).

\subsection{A distribution of time-dependent star forming clumps}
\begin{figure} 
\begin{picture} (0,12)
\put(0,0){\includegraphics[width=7.5cm]{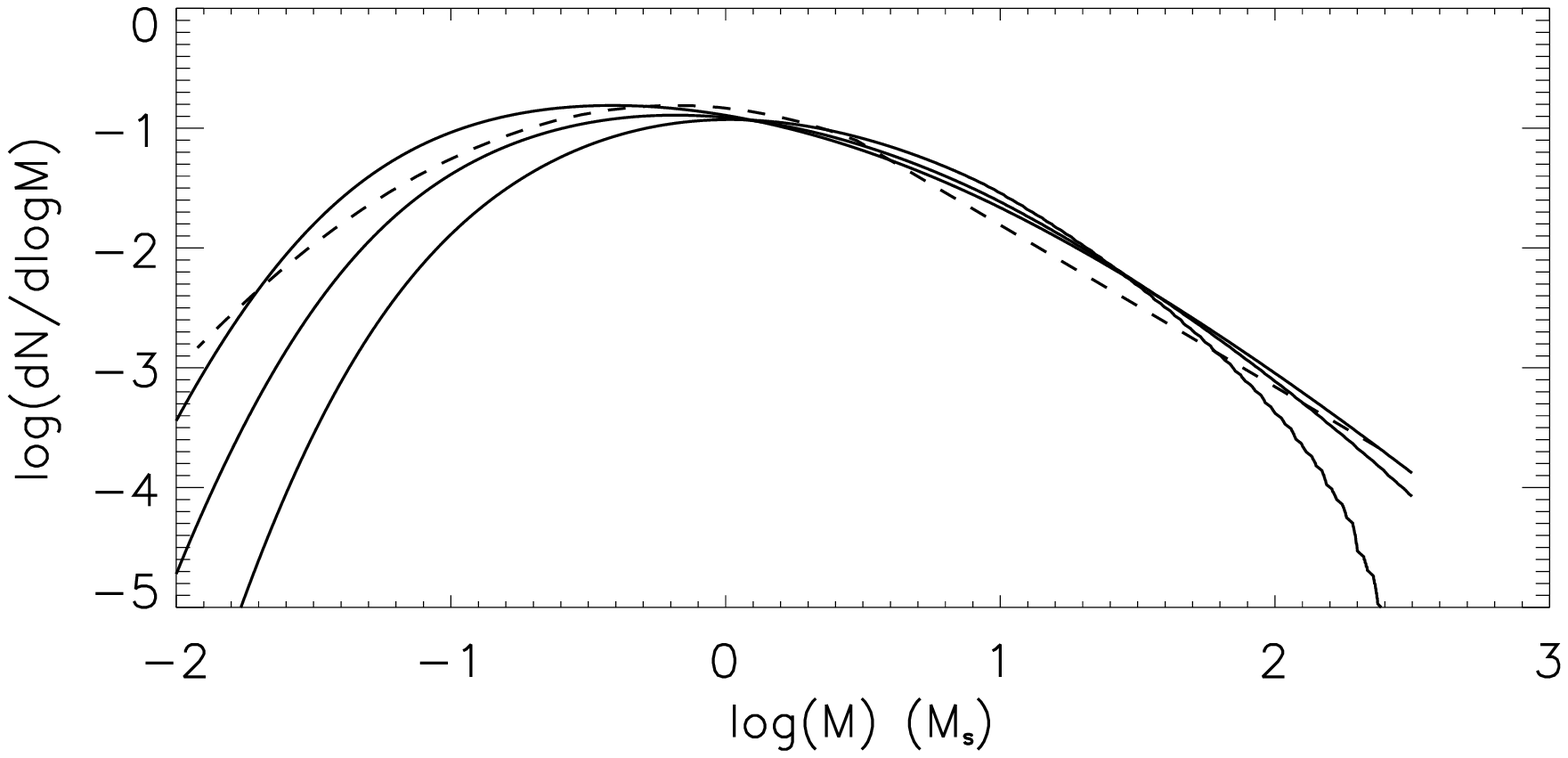}} 
\put(0,4){\includegraphics[width=7.5cm]{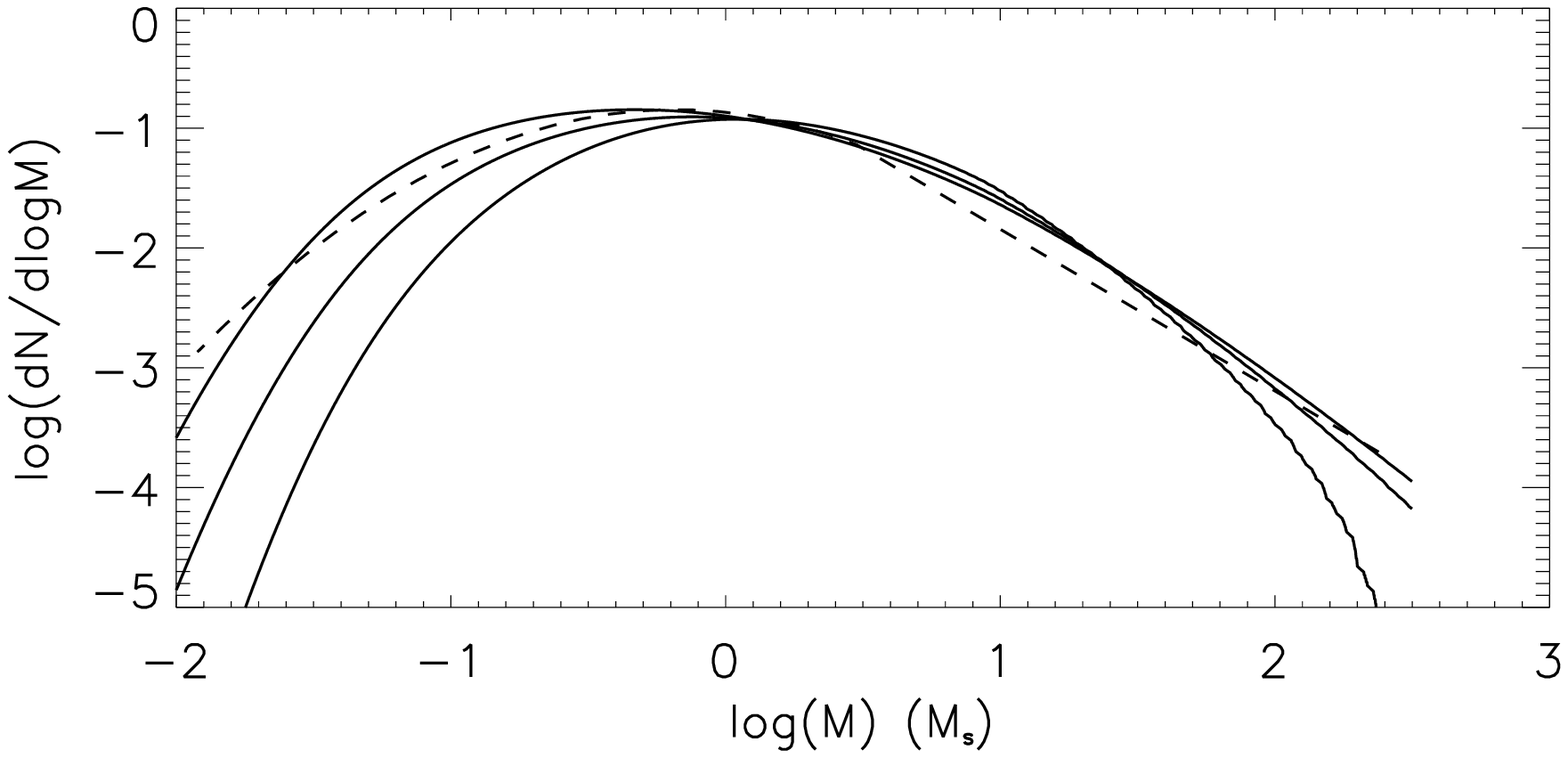}} 
\put(0,8){\includegraphics[width=7.5cm]{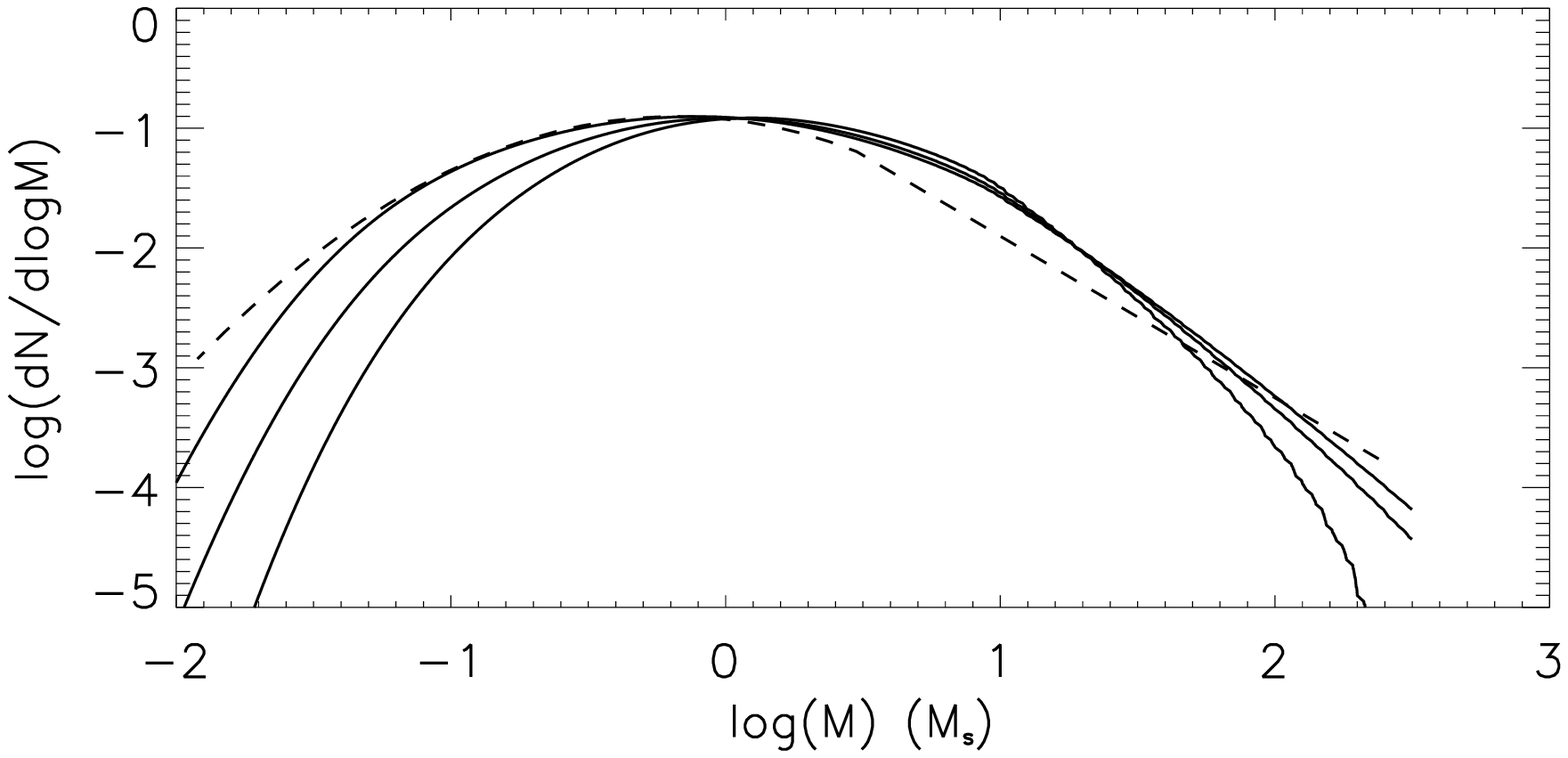}} 
\end{picture}
\caption{Integrated mass spectrum of self-gravitating 
objects (prestellar cores) for three values of $\gamma_{cl}$, the index 
of the mass spectrum of the star forming clumps. Top 
panel: $\gamma_{cl}=2$, middle panel: $\gamma_{cl}=1.7$, bottom
panel: $\gamma_{cl}=1.5$. The three solid curves correspond to
three different values for the upper value of the accretion rate, namely (from the leftmost line to the rightmost one): ${\dot{M}}_{sup}=$ 
$10^{-1}$, $10^{-2}$ and $10^{-3}$ $M_\odot$ yr$^{-1}$. The dash-line is the SCIMF.}
\label{distrib_accret_rate}
\end{figure}

It is now relatively well established that not only CO clumps (Heithausen et al. 1998, 
Kramer et al. 1998) but also infrared dark clouds (Peretto \& Fuller 2010) follow 
a mass distribution  ${\cal N}^{cl} = dN^{cl}/dM \propto M^{-\gamma_{cl}}$ with 
$\gamma_{cl}\simeq 1.7$.
It is important to recall that our theory predicts that this exponent, 
as well as the one for gravitationally bound cores when
a gravitational collapse condition is taken into account, is a direct consequence
of turbulence and is indeed related to the index $n'$ of the powerspectrum of 
$\log \rho$ by the relation $\gamma_{cl}=1+x=3-n'/3$ (see paper I). Therefore, the present theory 
naturally takes into account the clumpy structure of the gas and in principle no further 
calculations are needed. 
In other words, for spatially close enough clumps surrounded by sufficiently dense gas, 
there is no need to sum up over a clump population.
While strictly true as long as i) the gas 
is barotropic and ii) the density PDF is lognormal, this is no longer the case in real situations.
 Indeed, the diffuse ISM does not follow our eq.~(\ref{prescript}), as the atomic hydrogen
 which fills up the Galaxy can be as warm as $10^4$ K.  
Therefore, for clumps which are
spatially well separated, the  in-between gas is not isothermal and we must sum up over a clump
population. The scale at which this happens is not well determined. Numerical
simulations by V\'azquez-Semadeni et al. (2007),
Hennebelle et al. (2008), Heitsch et al. (2008), Banerjee et al. (2009)
suggest that the warm gas is tightly mixed with the cold gas at scales of about a few parsecs.

We calculate the mass spectrum resulting from the distribution of the 
time-dependent clumps described in the previous section. 
For that purpose, we proceed as follows. We perform 
a series of time-dependent clump calculations, choosing values of $\dot{M}$
between $10^{-1}$ and  $10^{-5}$ $M_\odot$ yr$^{-1}$ using logarithmic 
intervals. We stop the integration when $M_*=M_g$ in the clump
and we adopt for the mass of the clump the value $M_{cl}=M_*+M_g$.
Then we sum up the corresponding mass spectra after multiplying by
the aforementioned clump distribution ${\cal N}^{cl} (M_{cl}) \propto M^{-\gamma_{cl}}$.
The final core mass spectrum is given by 
\begin{eqnarray}
{\cal N}_{tot} = \int _0 ^{M_{sup}} {\cal N}_{M_c} ({M_{cl}}) V_{cl}\, {\cal N}^{cl}  (M_{cl}) dM_{cl}, 
\label{mass_distrib}
\end{eqnarray}
where ${\cal N}_{M_c} ({M_{cl}})=dn/dM_c$ is the mass spectrum of self-gravitating 
fluctuations (i.e. cores), i.e. the CMF, for a clump of mass $M_{cl}$ and volume 
$V_{cl}$, as given by eqn.(2) (see HC08 and HC09) and $M_{sup}$ is the 
largest mass in our sample which corresponds to the largest accretion rate.

At this stage, it is worth stressing that, since $\gamma_{cl}<2$, most of 
the mass is contained in the most massive clumps;  
$\gamma_{cl}=2$ then corresponds to a critical case. We have thus considered 
three values of $\gamma_{cl}$, namely 1.5, 1.7 and 2.
Figure~\ref{distrib_accret_rate} displays the results.

For $\gamma_{cl}=1.5$ and $\gamma_{cl}=1.7$,  the mass spectra are pretty similar to the results 
obtained in the previous section for a single clump. The reason is that most of the mass
is contained in the most massive clumps. Interestingly, varying the upper value of 
the accretion rate by two orders of magnitude has only a modest impact on the upper mass part of the
total mass spectrum while below about the mean Jeans mass, the low-mass part of 
the distribution quantitatively varies by orders of magnitude, the number of low-mass objects, in particular
brown dwarfs strongly decreasing with decreasing accretion rate. Values of ${\dot{M}}_{sup}\lesssim 10^{-3}$ $M_\odot$ yr$^{-1}$
yield a strong deficit of brown dwarfs compared with the SCIMF, representative of the observed population.
As mentioned above, this stems from the smaller cloud size, thus the smaller Mach number with decreasing accretion rate, yielding
a lack of overdense small-scale structures, progenitors of the brown dwarfs (see HC08 and HC09).

For $\gamma_{cl}=2$, the difference with the mass spectrum obtained for a single clump
is more pronounced. In particular at intermediate mass, the distribution tends to deviate 
from the SCIMF. This is due to the mass of the gas being equally distributed between
small and large clumps. As the CMF which results from small clumps peaks at 
larger masses than the CMF produced in massive clumps, this tends to create 
a small bump at intermediate masses ($M \simeq 10 M _\odot$).

\section{Conclusion. A paradigm for star formation and the initial mass function}

\subsection{Summary of the results}

In this paper, we have generalized our previous formalism and developed an analytical {\it time-dependent} 
theory of the stellar initial mass function. Although not very different from the time-independent prestellar mass spectra for given sets of cloud
initial conditions (temperature, density, Mach number), the time-dependent ones are not quite identical to the
former ones. The most significant differences are 
a slightly steeper slope at large masses, which arises from the 
time dependence of the characteristic Jeans mass, and the position of the peak of the CMF which is shifted 
towards smaller masses by a factor of the order of 2 to 3, depending on the initial conditions. This is 
a direct consequence of the small-scale structures being rejuvenated several times during the collapsing process of the
 larger ones. For all explored typical clump conditions, the resulting CMF, while slightly too narrow when considering an isothermal 
equation of state, agrees very well  with the 
shifted Chabrier  system IMF (i.e. the Chabrier IMF shifted by a factor of 
$\simeq 3$ to take into account the core-to-star mass conversion efficiency during the collapse) when including the thermodynamics of the gas, confirming the results obtained for the time-independent theory.

This time-dependent theory enables us to derive an expression for the star formation rate for molecular cloud conditions. Confrontations with previously published SFR theories show some
similar trends but quantitative or even qualitative differences. As mentioned in HC11, these differences arise essentially from the two essential characteristics of our theory, namely: (i)  each overdense region dominated by gravity collapses under its {\it own, density-dependent} dynamical time, (ii) there is no particular threshold density/scale for star formation; instead,
any overdense region produced by initial density fluctuations is susceptible to collapse, no matter its degree of internal turbulence, {\it if } dominated by gravity.

 We explore the dependence of the SFR upon the clump properties, namely the size, the level of turbulence  (through the virial parameter) and the 
magnetic field.
 We show that the exact value of the SFR depends significantly upon these clump parameters, which, unfortunately, 
are ill defined quantities, hampering precise theoretical estimates of the star formation rate in a clump/cloud in particular 
for magnetized clouds. 
We show that, when time-dependence is taken into account in the formation of collapsing dense cores, turbulence globally {\it favors} star formation, in contrast to the result obtained with the time-independent theory, in agreement with the results of Padoan \& Nordlund (2011). We also show that, when the clump is unbound, i.e. $\alpha_{vir}\gtrsim 1$, the star formation rate decreases drastically, as the dominant contribution of  kinetic (thermal+turbulent) over gravitational energy prevents gravitational collapse of the overdense structures, inhibiting star formation. The magnetic field also reduces the SFR when it is strong enough. Its 
exact influence depends on the magnetic-density relation and also on its influence on the density PDF none of these properties being known with 
great accuracy.

Our calculations show
 that star formation rate correlates with  both the gas surface density of the surrounding region and  the size of the clump
 and occurs dominantly in dense and/or massive - thus bound - clumps.
Such large clumps and dense regions, however, represent only a modest fraction ($\la 20\%$) of typical clouds.
In contrast,  for the smallest clumps, the SFR decreases drastically below a typical density $\Sigma_g\approx 100\mpc2$, i.e. a volume density ${\bar n}\approx 2500\,(R_c/1\,{\rm pc})^{-1}\c3$, i.e. a visual extinction $A_V\approx 7$,
in excellent agreement with  observational determinations in Milky Way molecular clouds. 
For clumps below about this density, star formation is still taking place, although at a much lower rate,
 only in very large ($\gtrsim 10$ pc) clumps, which are very rare. This stresses the need to explore large fields of view to accurately determine the SFR in low-density regions.
 We stress that this abrupt change in the SFR does not stem from an arbitrarily defined threshold for star formation, but naturally arises from the theory and reflects the fact that pc-size low density clumps barely form stars essentially because of their too large characteristic Jeans length compared with their size
and their too low level of turbulence.
 
 Interestingly enough, the aforementioned density corresponds to the density above which the cold gas in the ISM becomes essentially composed of molecular hydrogen, H$_2$. Indeed, simulations of compressible turbulence coupled with hydrogen chemistry (Glover \& MacLow 2007, see also Krumholz et al. 2009) suggest that (i) H$_2$ forms much more efficiently in (compressible) turbulent gas than in quiescent gas owing to the much shorter formation timescale and 
that (ii)  above $n\sim 300\c3$, the H$_2$ formation rate, which scales as the square of the number density, $n^2$, becomes larger than the photodissociation rate, allowing the efficient in-situ formation of molecular hydrogen. The gas becomes dominantly (resp. entirely)  molecular above about $5000 \c3$ (resp. $10^4\c3$) and remains "trapped" and shielded from external UV radiation in the collapsing structures. Our calculations thus support Glover \& MacLow's (2007) suggestion that dense (bound or unbound) star-forming regions {\it drive} H$_2$ formation, and not the opposite.
Phrased differently, the formation of molecular hydrogen is not a necessary condition for star formation but instead is a {\it consequence} of efficient star formation.
 
Finally, we derive a fully time-dependent calculation of the core
mass spectrum by considering a simple time-dependent clump 
model. That is, instead of assuming fixed clump conditions (density, mass, size), we consider an evolving, accreting
clump assuming a constant accretion rate. In order to explore the parameter space, we have conducted calculations for various accretion rates, typical of the observed determinations,
assuming that the clump parameters obey the standard Larson scaling relations.
The resulting core mass spectrum remains in very good
agreement with the observationally derived distribution and  reproduces
 various behaviors observed in simulations of converging 
flows. We also investigate 
the core mass spectrum that results from the contribution of 
a clump {\it distribution} such that $d N_c / d M_c \propto M^{-1.7}$. 
We show that the resulting core mass spectrum 
is dominated by the mass spectrum of the biggest clumps because 
they contain most of the mass. 

Our SFR model provides the general framework that includes the simplified model suggested by Lada et al. (2012), which relies on the fraction of dense (molecular) gas in the cloud and thus critically depends on a threshold density. 
As mentioned above (and shown in Fig. \ref{fig_sfr_obs}),  our theory naturally predicts a strong, roughly quadratic correlation between the SFR and the gas density, 
predicted to become dominantly molecular above  $\sim 5000\c3$.
As shown in \S\ref{clumpmass}, such a  correlation immediately implies a  similar correlation
with the mass of the clump/cloud itself, i.e. with the mass of molecular hydrogen above the aforementioned value.
The model suggested by Lada et al. (2012) is thus a direct consequence of our general theory of the IMF+SFR. 
As mentioned above, however, star formation is still predicted to take place, although at a much lower rate,
 in large, low-density clouds, which will be composed essentially of atomic hydrogen.

However, in contrast to what has often been claimed in the literature, we show that there is no universal value of the star formation rate, and that this latter does not simply correlate linearly with gas density. Indeed, as mentioned above, the SFR {\it strongly depends} not only on the gas density
but also on the clump mass/size, which leads to a large scatter in SFR values. 

\subsection{A paradigm for star formation and for the theory of the IMF}

The analytical theory described in our previous and present papers, which correctly reproduces various observational constraints, as discussed in these papers, suggests the following paradigm for star formation and for the resulting CMF/IMF:

1) compressible (shock dominated) large-scale turbulence in the cloud, due to various possible mechanisms such as accretion, converging cold and warm flows or star formation itself,
 generates a  field of density fluctuations at all scales in the cloud. The (nearly lognormal) PDF of these fluctuations  is entirely determined by the characteristic (universal)  log-density power spectrum index of turbulence. 
This PDF leads to overdense regions which correspond to the observed clump mass spectrum (see paper I). Observations suggest that the clumps have a filamentary structure.
 As they accrete mass and dissipate kinetic energy, the densest clumps become gravitationally unstable above a typical density ${\bar N}_{{H_2}}\approx 10^{22}\cc2$ (i.e. ${\bar n}\approx 10^4\,(R_c/1\,{\rm pc})^{-1}\c3$ for a spherical clump) for a temperature $T\sim 10$ K,  triggering the fragmentation into prestellar cores. 
Although probably quantitatively affecting our results by numerical factors of order unity, this filamentary nature of the clumps does not modify the general framework of our theory,
which indeed predicts a very narrow CMF below about this density for pc-size clumps, yielding a negligible number of prestellar cores.

2)  fragmentation then introduces  a {\it scale-dependence} (thus a density-dependence) in the processus, set up the virial criterion for gravitational collapse (see paper I).
As demonstrated in paper I and in  Chabrier \& Hennebelle (2011), we suggest that turbulence plays a crucial role in setting the massive initial mass reservoir distribution (thus the Salpeter slope of the IMF),  progenitors of massive stars, 
above about the mean thermal Jeans mass, at the early stages of star formation. {\it Not} by providing a pressure support, in a static sense, but by dispersing the gas within 
these structures, which would otherwise have collapsed, until they reach a mass that we identify as their {\it turbulent Jeans mass} (see eq.~\ref{mass_rad2}).

3) below about the typical density mentioned above, our theory predicts a drastically decreasing CMF and thus SFR, for the typical size of the dominant clump population, 
essentially because the characteristic Jeans length of the clump becomes comparable to or larger than its size, stabilizing the clump against fragmentation. For a filament, the same density corresponds to the critical mass per unit length below which the clump is gravitationally stable, yielding the same conclusion. 
We predict, however, that star formation can still occur below this density in very large ($\ga 10$)pc size) clumps,  large enough to significantly exceed  their typical Jeans length and
 generate large enough turbulence levels. This seems to be supported by the observed population of class II cores in low-density, large clumps (Gutermuth et al. 2011) and
 stresses the need to observe very large areas at low-density in order to get statistically significant core detections.
Such clumps, however, are very seldom, as inferred from the steeply decreasing clump size spectrum, and the probability to find pretellar cores in
low-density environments drops drastically, making their detection very difficult.

4) star formation is thus a continuous process and can  occur, statistically speaking, in any density environment, as any turbulence-induced overdense region can collapse, if dominated by gravity, 
and produce eventually a prestellar core. There is thus no real "threshold" for star formation. 
This leads to a direct correspondence between the star formation rate and the gas (atomic + molecular) density, as observed and indeed predicted by the theory, with star formation
occurring most actively in the densest regions of molecular clouds, which entails only a small ($\la 20\%$)  fraction of their mass. 
 However, as mentioned above, star formation is basically choked off 
below ${\bar n}\sim 1000 \c3$ for the typical ($\sim$pc) size of most clumps. 
Above this density, the H$_2$ formation rate,
which increases as the {\it square} of the gas density, dominates the photo-dissociation rate and  the gas quickly becomes dominantly and entirely molecular.
This drastically decreases the photoelectric heating efficiency, causing in turn a drop in dust and gas temperature (Tielens \& Hollenbach 1985, Glover \& McLow 2007). 
Star formation thus {\it promotes} H$_2$ formation, and not the opposite.

We stress, however, that, from the general point of view, star formation strongly depends on the clump characteristic properties (mass, size); therefore,  there is no "universal" relation between star formation rate and gas density, but instead large variations, depending on the clump's environment.

This paradigm, which relies on the results presented in the present and former papers, shows that a star formation rate determined at the early stages by turbulence-induced fluctuations provides quite a consistent picture of  star formation in Milky Way molecular clouds, with star formation occurring dominantly in the densest regions of the cloud. 
This gravoturbulent picture of star formation correctly predicts the observed CMF over the entire mass range from brown dwarfs to massive stars and naturally leads to 
a star formation rate vs clump masses/sizes and gas density correlations in very good agreement with observational determinations. 
We stress, however, that ill-defined quantities such as the size of the clump or the dynamical efficiency of core-to-star mass conversion, which may well vary from clump to clump, make the determination of the SFR uncertain by a factor of a few. 

\acknowledgments
The research leading to these results has received funding from the European Research Council under the European Community's Seventh Framework Programme (FP7/2007-2013 Grant Agreement no. 247060. This work was initiated while G.C. was a visitor of the Max Planck Institute for Astrophysics, where he benefited from numerous discussions with various colleagues.

\end{document}